\documentclass[a4paper,twocolumn,11pt,accepted=2025-08-29]{quantumarticle}
\input{setup}

\begin{document}

\title{Bayesian Quantum Amplitude Estimation}

\author{Alexandra Ramôa}
 \email{alexandra.ramoa@inl.int}

\author{Luis Paulo Santos}%

\affiliation{International Iberian Nanotechnology Laboratory (INL), Portugal\\
High-Assurance Software Laboratory (HASLab), INESC TEC, Portugal\\
Department of Computer Science, University of Minho, Portugal}%

\maketitle
\begin{abstract}
    We present BAE, a problem-tailored and noise-aware Bayesian algorithm for quantum amplitude estimation. In a fault tolerant scenario, BAE is capable of saturating the Heisenberg limit; if device noise is present, BAE can dynamically characterize it and self-adapt. We further propose aBAE, an annealed variant of BAE drawing on methods from statistical inference, to enhance robustness. Our proposals are parallelizable in both quantum and classical components, offer tools for fast noise model assessment, and can leverage preexisting information. Additionally, they accommodate experimental limitations and preferred cost trade-offs. We propose a robust benchmark for amplitude estimation algorithms and use it to test BAE against other approaches, demonstrating its competitive performance in both noisy and noiseless scenarios. In both cases, it achieves lower error than any other algorithm as a function of the cost. In the presence of decoherence, it is capable of learning when other algorithms fail. 
\end{abstract}
\section{Introduction}
\label{sec:intro}

Quantum amplitude estimation (QAE) is a routine for estimating the measurement probability associated with a given subspace for some wavefunction \cite{brassard2002}. Notable applications are Monte Carlo integration \cite{montanaro2015} and estimation tasks \cite{knill2007}, with applications in finance \cite{woerner2019, gomez2022, rebentrost2018, miyamoto2022, stamatopoulos2020, gomez2022}, chemistry \cite{kassal2008, katabarwa2021, johnson2022}) and machine learning \cite{wiebe2016, wiebe2015, kerenidis2019, wiedemann2022}.

The original proposal for QAE achieves a quadratic quantum advantage by performing phase estimation on a quantum amplitude amplification (QAA) operator \cite{brassard2002}. However, the complexity of the corresponding circuit is prohibitive for the currently available quantum devices, due to their susceptibility to noise and limited size. 

Alternatives have been proposed in the literature, often hybrid quantum-classical algorithms where simpler circuits are embedded in a classical feedback loop \cite{aaronson2020, wie2019, suzuki2020, wang2017, nakaji2020,grinko2021}. The simpler circuits typically consist of a sequence of $m$ non-controlled applications of the amplification operator, where $m$ is an experimental degree of freedom controlled by the classical processing unit, possibly in an adaptive fashion.  

In this formulation, QAE is largely similar to common characterization tasks associated with superconducting qubits \cite{ramoa2024}. These tasks are related to precession dynamics, such as Larmor \cite{malcomlevitt2008}, Rabi \cite{merlin2021} and Ramsey \cite{ramsey1950} oscillations. Similarly, in photonic quantum computing, Mach-Zehnder interferometry gives rise to a similar framework \cite{pezze2014}. 

All of these dynamics admit squared-sinusoidal descriptions, just like Grover circuits. Techniques applied to one of them likely transfer to the others, extending the applicability of QAE algorithms to tasks such as sensing or quantum gate implementation and calibration; and inversely, QAE can borrow from algorithms devised for those tasks.

Importantly, unlike in many hybrid near-term quantum algorithms, such schemes can provably achieve the full quadratic advantage for QAE \cite{aaronson2020, fukuzawa2023}. However, the implementation details remain an open question. They are determinant for performance factors such as the quantum cost offset, classical processing cost, parallelism, and noise resilience. 

While the aforementioned alternative algorithms require simpler circuits, most of them still assume ideal executions, considering at most sampling noise as a limitation. What is more, their performance is often assessed under that assumption, rendering their behavior unpredictable in the presence of noise. 

One framework capable of generalizing to the noisy case is Bayesian inference, which applies naturally to the iterative formulation of QAE. However, the involved processing costs are high. The circuit length is expected to increase exponentially with the iteration number \cite{ferrie2012, suzuki2020}, with higher precision requiring more iterations. In a naive strategy, the searching range and thus the optimization cost per iteration are expected to increase exponentially with the total iterations. 

On the other hand, the cost of a naive (deterministic) statistical approximation increases exponentially with the number of parameters. More efficient approximations are vital to ensure tractability; importantly, care must be taken lest they jeopardize scalability or correctness. In \cite{wang2021}, a noise-aware approach to QAE based on Bayesian inference is proposed under a normality assumption. While this is a cost-effective approach for simple cases, it is not expected to generalize well to more complex ones, namely problems involving more elaborate noise models. Furthermore, this proposal requires generalizing the Grover operator and optimizing over large numbers of parameters, incurring a significant cost overhead.

\subsection{Contributions}

In this work, we propose BAE, an approximate Bayesian algorithm for quantum amplitude estimation that is capable of handling noise without making restrictive assumptions or incurring prohibitive costs. Problem tailored heuristics are used to cut down optimization costs while preserving the quantum advantage, bringing the optimization cost per iteration down to constant. 

By employing scalable and highly parallelizable statistical methods and other approximations, we further lower the classical processing costs without affecting the quantum advantage. The employed  quantum circuits are of the simplest type presented in the literature; they do not require controlled or generalized versions of the Grover operator, nor any other qualitative complication as compared to Grover search.

Our method offers quantifications of model merit with little extra cost, can incorporate available information (such as past experimental data), and is flexible with respect to the cost trade-offs involved. For instance, one may trade in part of the quantum advantage for a lower classical cost, or for increased parallelism in the quantum sub-routine.

To benchmark our algorithm, we numerically test it along with those of other QAE algorithms presented in the literature, and propose thorough, problem-agnostic and cost-effective methods to evaluate and analyze their performance. BAE achieves Heisenberg-limited estimation, has the lowest cost offset among all tested algorithms, and is capable of learning in noisy scenarios where other algorithms fail. 

Additionally, inspired by a widely used method in statistics, we propose aBAE, a variant of BAE where the quantum circuits are guided based on measures of statistical efficiency not directly related to the error or uncertainty. This further highlights the importance of this component, and offers a robust and systematic way to work with challenging models or data. We call this variant \textit{annealed} Bayesian amplitude estimation, after the method it is inspired in, annealed likelihood estimation \cite{south2019}. 

The code and datasets used for this paper are available at \cite{ramoa2024a}.

\subsection{Document structure}

The rest of the paper is organized as follows. Section \ref{sec:background} presents relevant background, including the original quantum algorithm, the classical counterpart, and alternative hybrid algorithms. BAE is introduced in section \ref{sec:BAE}. Section \ref{sec:methods} describes  the numerical methods used for benchmarking. The results of these tests are shown in section \ref{sec:results}. Finally, section \ref{sec:conclusion} discusses and summarizes the results, as well as directions for future research. Supplementary details about the algorithm, data processing and background are given in the appendices.  
\section{Background}
\label{sec:background}

\subsection{Amplitude estimation}
\label{sub:ae}

Quantum amplitude estimation (QAE) \cite{brassard2002} is a process that estimates the parameter $a$ in a wavefunction of the form
\begin{equation}
\label{eq:psi}
     \ket{\psi} = \op{A}\ket{0}^{\otimes n} = \sqrt a\ket{\psi_1} + \sqrt{1-a} \ket{\psi_0},
\end{equation}

\noindent with $a \in [0,1]$ and $\ket{\psi_1}$, $\ket{\psi_0}$ projections of $\ket{\psi}$ into two different subspaces. $\op{A}$ is a problem-dependent initialization operator.

Despite being a probability, the parameter $a$ is commonly termed \textit{the amplitude} in the literature, as we work in a domain where their relation is bijective. We will follow this convention. 

It is often convenient to rewrite equation (\ref{eq:psi}) in terms of an angular parameter $\theta$, which we call the Grover angle:
\begin{equation}
\label{eq:theta_def}
    \theta = \arcsin(\sqrt{a}).
\end{equation}

The wavefunction then becomes:
\begin{equation}
\label{eq:psi_theta}
    \ket{\psi} = \sin(\theta)\ket{\psi_1}\\
    + \cos(\theta) \ket{ \psi_0}.
\end{equation}

This type of wavefunction arises most commonly in the context of Grover search \cite{grover1996}, where given a function $f(x): \{0,1\}^n \rightarrow \{0,1\}$, we want to find the pre-image of $1$ under $f$, i.e. $x$ s. t. $f(x)=1$. The operator $\op{A}$ is known; we can prepare a state of the form (\ref{eq:psi}), where $\ket{\psi_1}$ is the subspace spanned by solution states and $\ket{\psi_0}$ is its orthogonal complement in the Hilbert space. For that reason, $\ket{\psi_1}$ is often called the \textit{good} subspace, and $\ket{\psi_0}$  is called the \textit{bad} subspace. If no prior knowledge is available, the initialization could be a Hadamard transform, assigning equal probability to all possible states. The probability of finding a solution upon measuring state (\ref{eq:psi}) is then $a=2^{-n}$. 

In the quantum framework, the ability to identify a solution can be formulated as having access to a phase oracle operator $\opsub{U}{f}$ that marks solution states by assigning them a $\pi$ phase,
\begin{equation}
	\label{eq:oracle}
    \opsub{U}{f} = (-1)^{f(x)} \ket{x},
\end{equation}

 or, since it flips the sign of the good subspace and leaves the bad one unchanged,

\begin{equation}
	\label{eq:oracle2}
    \opsub{U}{f} = \op{I} - 2 \ketbra{\psi_1}{\psi_1},
\end{equation}

\noindent where $\op I$ is the identity operator.

The Grover operator is given by:
\begin{equation}
\label{eq:groverop}
    \op{G} = -\op{A}\opsub{U}{0} \op{A}^{-1} \opsub{U}{f},
\end{equation}

where $\opsub{U}{0}$ reflects the all-zero state, leaving all others unaltered. Note that the sequence $A\opsub{U}{0}A^{-1}$ reflects the initial state $\ket{\psi}$.

 It can be shown that the effect of applying operator (\ref{eq:groverop}) $m$ times on state (\ref{eq:psi}) is:
\begin{align}
    \op{G}^m\ket{\psi} = \sin((2m+1)\theta)\ket{\psi_1} \nonumber \\
    + \cos((2m+1)\theta) \ket{ \psi_0}. \label{eq:psi_amp}
\end{align}

For small enough amplitudes and low enough $m$, the effect of this operator is to increase the amplitude; for this reason, it is also called the \textit{amplitude amplification} operator. In Grover search, $m$ is chosen to maximize the probability of obtaining a \textit{good} state upon measurement,
\begin{equation}
    P(\psi_1 \mid m) = \lvert \bra{\psi_1} \op{G}^m \ket{\psi} \rvert^2.
\end{equation}

More generally, this can be seen as an \textit{amplitude oscillation} operator; once that optimal $m$ is exceeded, the probability will decrease, and then change in a periodic fashion:
\begin{equation}
\label{eq:p1}
    P(\psi_1 \mid m) = \sin^2((2m+1)\theta).
\end{equation}

For our purposes, the measurement outcomes are binary: the only relevant feature of a quantum state is which of the two subspaces it belongs to, regardless of the Hilbert space dimension. We can then say that the result of measuring the amplified state $\op{G}^m \ket{\psi}$ is a datum $D \in \{0, 1\}$. 

\subsection{Quantum algorithm}
\label{sub:qalg}

Amplitude estimation is the task of learning $\theta$ or $a$. It can be shown that $\pm 2\theta$ are eigenphases of the operator defined in equation (\ref{eq:groverop}). Hence, the problem can be solved by applying quantum phase estimation (QPE) to this operator. 

The QPE protocol requires ladders of controlled applications of $\op{G}$, followed by the inverse quantum Fourier transform (QFT) \cite{brassard2002}. The output precision is determined by the number of qubits used in the auxiliary register for the QFT, whereas the main register is problem-dependent. 

The phase measurement made by the QPE circuit collapses the wavefunction of the readout register into an encoding of the one of eigenphases, allowing for the calculation of $\theta$. 

In general, QPE acts according to:
\begin{equation*}
    \text{eigenvalue is } e^{i2\pi \phi} \rightarrow \text{QPE outputs } r=\phi * K ,
\end{equation*}
\noindent where $K=2^k$ is the QFT order, with $k$ the number of auxiliary qubits. This result is deterministic if $r$ can be represented exactly with $k$ bits and probabilistic otherwise. In the latter case, the most likely output is the closest integer to $r$.

In the case of QAE, we have the eigenphases $\phi=\pm \theta/\pi$. The subspace spanned by $\ket{ \psi_0}$ and $\ket{ \psi_1}$ finds an orthonormal basis in the two corresponding eigenvectors of the Grover operator. The initial state for QPE can be $\ket{\psi}$. The measured QPE outcome $r$ will be given by:
\begin{gather}
     \left( r =  K\frac{\theta}{\pi} \right) \vee  \left( r= K-K\frac{\theta}{\pi}\right).
\end{gather}

This allows us to compute an estimate for the amplitude, $\tilde a$:
\begin{equation}
    \tilde a = \sin^2(\pi \frac{r}{K}).
\end{equation}

The result is the same for the two eigenvalues, since the sine function is symmetric about $\pi/2$. This is a crucial point, as the measurement would otherwise be ambiguous. Also note that while the minus sign on the operator of equation (\ref{eq:groverop}) is irrelevant for quantum searching, such is not the case for amplitude estimation, where its absence would have one calculating the complement of $a$ rather than itself.

A key theoretical result of \cite{brassard2002} states that estimating $a$ in the manner described above uses $\mathcal O (K)$ oracle queries, and achieves an error upper bounded as:
\begin{equation}
    |\tilde a -a| \leq 2\pi s \frac{\sqrt{a(1-a)}}{K}+s^2\frac{\pi^2}{K^2} 
\end{equation}

\noindent for any positive integer of choice $s$, with probability $8/\pi^2$ if $s=1$ and $1-1/(2(s-1))$ otherwise. 

The major takeaway is that the error in estimating $a$ scales as: 
\begin{equation}
\label{eq:qae_scaling}
    \epsilon \in \mathcal O(N_q^{-1}),
\end{equation}
\noindent where $N_q$ is the number of oracle queries (equal to the Fourier order $N_q=K=2^k$). We have also defined the error $\epsilon \equiv |\tilde a -a|$.

\subsection{Classical algorithm}
\label{sub:calg}

We now want to compare equation (\ref{eq:qae_scaling}) to the best performing classical algorithm. 

Classically, we estimate $a$ by sampling from the original distribution and evaluating the function $f$ for each sample. One sample or function evaluation corresponds to one query. We then average these evaluations to get an estimate for the expected value of $f(x)$, which is the amplitude. This corresponds to the problem of estimating the parameter of a Bernoulli distribution $\mathcal B(a)$  by sampling from it, and fits into the sample mean framework. 

The associated error is thus the standard error of the mean, which is equal to $\sigma/\sqrt N_q$, with $\sigma$ the standard deviation of the original distribution - in this case, a binomial distribution - and $N_q$ the number of samples/queries to $f$. The error thus scales as:

\begin{equation}
    \epsilon \in \mathcal O(N_q^{-1/2}).
\end{equation}

Hence, formula (\ref{eq:qae_scaling}) represents a quadratic speedup: the error in estimating $a$ decreases quadratically faster with the number of queries in the quantum case.

\subsection{Alternative quantum algorithms}
\label{sub:altqalg}

The quantum algorithm described in section \ref{sub:qalg} offers an optimal and straightforward solution to amplitude estimation. However, the required quantum circuits are deep and complex, involving ladders of controlled operations and large numbers of qubits. This renders it unfeasible for current quantum devices, which suffer from noise and - relatedly - limited sizes and continuous computation times.

In particular, this algorithm employs phase estimation, just as Shor's algorithm \cite{shor1999} - despite offering a much smaller complexity advantage (quadratic rather than exponential). This prompts the question of whether phase estimation is truly necessary to solve this problem. This question has been answered negatively by \cite{aaronson2020}, where a simplified approach to amplitude estimation is proposed. The alternative algorithm achieves the same (optimal) asymptotic complexity of \cite{brassard2002} but replaces the QPE circuit by an ensemble of simpler quantum circuits combined with classical processing and control.

These consist of amplification circuits identical to those used for Grover's quantum searching, where the Grover operator acts $m$ times on an initial state. This parametrized family of circuits has an output distribution given by equation (\ref{eq:p1}), where doing $m=0$ recovers the classical case. The extra degree of freedom of the quantum case can be used within a classical processing framework to attain the full quadratic speed-up.

The only employed quantum resources are Grover measurements. We can think of the quantum device as a black-box that receives an integer $m$ and outputs a binary outcome $D$ according to equation (\ref{eq:p1}). In this context, amplitude estimation can be seen as a metrological problem. We can say that $m$ is an experimental control and $D$ an experimental outcome, or datum (or alternatively, absorb $m$ into $D$). 

The black-box is the source of quantum advantage. In the classical case, we are restricted to sampling from $\mathcal B (\sin^2(\theta))$; whereas using quantum resources, we can sample according to the conditional distribution:

\begin{equation}
     \mathbf{P}(\theta \mid m) = \bra{\psi} \op{G}^m \ket{\psi}= \sin^2(r_m\theta),
\end{equation}

with $r_m=2m+1$, for $m$ any odd integer, using a number of queries in $\mathcal O (r_m)$. This extra freedom can be used to unlock otherwise unachievable learning rates (up to quadratically faster).

It is then clear that the quantum advantage stems from the amplification process, and can be distilled down to this fundamental motif. However, how to best employ it is an open question. While the algorithm of \cite{aaronson2020} has optimal complexity, it incurs an impractical cost offset. Furthermore, while the employed circuits are \textit{simpler}, they are still expected to be \textit{perfect} - not only for the correctness of the results, but even to assure termination. In particular, a certain degree of amplification is required to complete the algorithm as proposed, which may be infeasible in the presence of decoherence.

\begin{figure}[!htb]
    \centering
    \includegraphics[width=\linewidth]{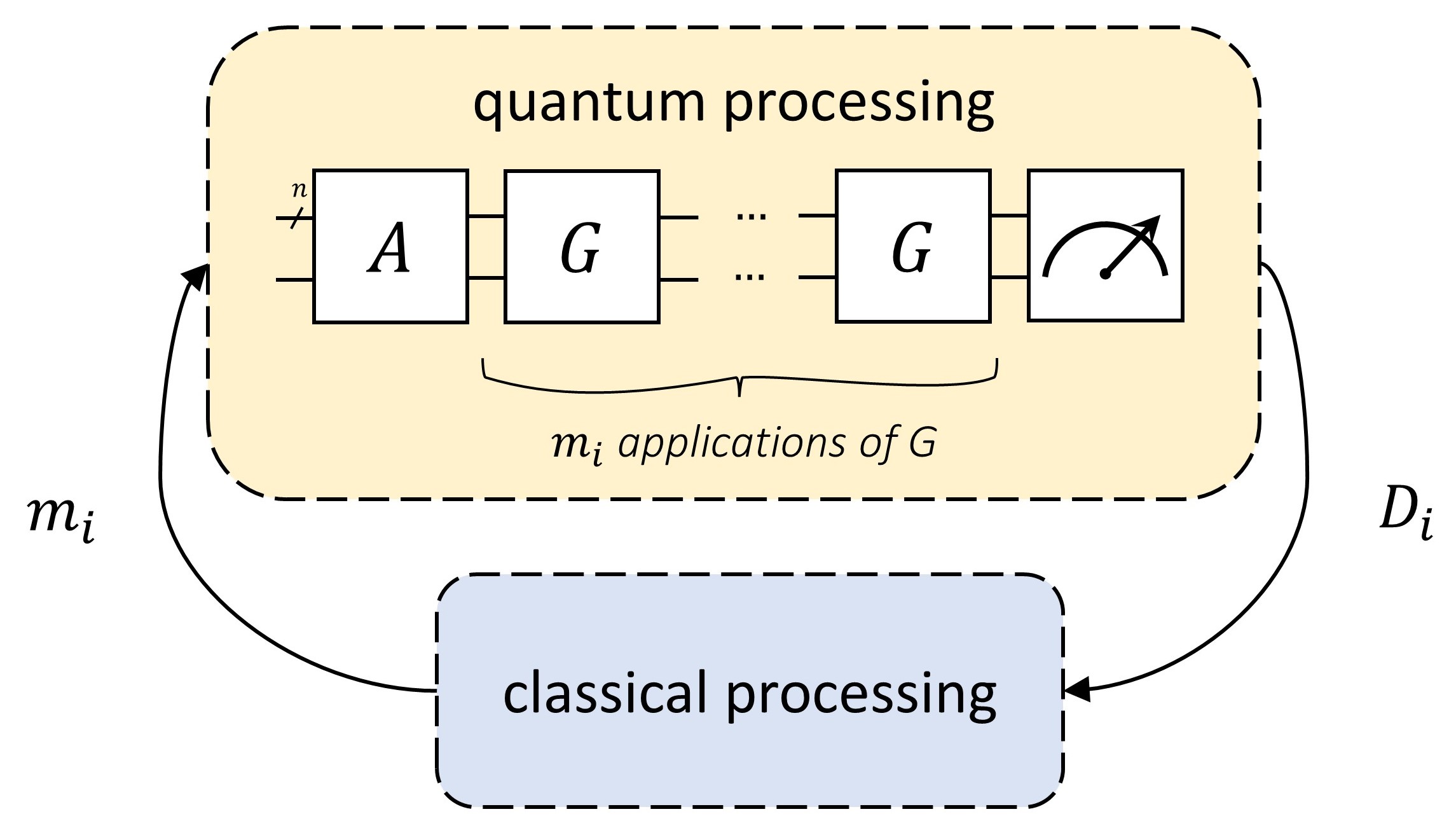}
    \caption{Typical structure of hybrid quantum-classical algorithms for amplitude estimation. Orange and blue are used for quantum and classical processes respectively.}
    \label{fig:hybdrid}
\end{figure}

Other alternatives have been proposed in the literature, most of which use the same family of circuits. Most often, these consist of hybrid quantum-classical algorithms where the quantum circuits are embedded in an iterative scheme with classical feedback. That is, the parameter $m$ specifying the quantum circuit is chosen adaptively at each iteration based on previous steps. Figure~\ref{fig:hybdrid} shows the usual workflow for this type of algorithm.

The classical processing component is generally the distinguishing factor, and the one responsible for the differences in performance between algorithms. The most commonly used metric of success is the complexity advantage, studied theoretically and/or numerically. However, in practice, other features can be relevant - namely the maximum circuit depth, the classical processing cost, the quantum cost offset, the noise resilience, the online processing cost, and the parallelism.

A few notable algorithms that follow this structure are phase estimation-free amplitude estimation, or maximum likelihood amplitude estimation (MLAE) \cite{suzuki2020}, faster amplitude estimation (FAE) \cite{nakaji2020}, iterative amplitude estimation \cite{grinko2021}, and robust amplitude estimation (RAE) \cite{wang2021}. Section \ref{sub:litoverview} provides an overview of these algorithms. 

\subsection{Quantum enhanced estimation}
\label{sub:quantum_enhanced}

When considering alternative algorithms for QAE, it is useful to view the problem as rooted in metrology, in order to properly assess their merits. In this section we briefly discuss quantum enhanced estimation \cite{toth2014, zwierz2010}.

For the scope of this work, we are interested in expressions relating the estimation error to the resource count, to be called $\epsilon$ and $N_q$, respectively. The former is the root mean squared error, which quantifies the uncertainty, and can be estimated by the standard deviation. The latter quantifies a cost; for instance, a number of measurements, probes or queries, or a (cumulative) evolution time.

Classically, the fundamental limit for the behavior of the error under an optimal strategy for amplitude estimation is given by the standard quantum limit (SQL):
\begin{equation}
    \label{eq:sql}
    \epsilon^{(SQL)} \propto N^{1/2}
\end{equation}

This limit is due to sampling noise: the resolution of any measurement is limited by the number of shots, or measurement repetitions. For this reason, this is also called the shot noise limit. Even though it is called ``noise'', we do not throughout this paper consider it to be so, instead using "noise" to mean deviations from ideal behavior (extrinsic noise). 

Although this is the best achievable performance given independent measurements, one may exploit quantum effects to improve upon equation (\ref{eq:sql}) when characterizing a system of quantum mechanical nature. A correlation between measurements can be introduced in space (entanglement) and/or time (adaptivity) to enhance the estimation precision. 

More specifically, quantum control allows for attaining the Heisenberg limit:
\begin{equation}
    \label{eq:hl}
    \epsilon^{(HL)} \propto N^{1}
\end{equation}

Equation (\ref{eq:hl}) represents the ultimate bound of metrology, and the gold standard for estimation tasks.

When working with algorithms that are not necessarily optimal, these bounds provide an insight into the performance, namely how much of the quantum advantage they preserve. This can be observed as a scaling that sits in-between the classical and quantum bounds (equations  \ref{eq:sql}  and \ref{eq:hl}, respectively).

\subsection{Literature overview}
\label{sub:litoverview}

\begin{table*}[t]
\small
\resizebox{\textwidth}{!}{
\begin{tabular}{ |c|c|c|c|c|c|c| } 
     \hline
      & \textbf{Key idea} & \makecell{\textbf{Paralle-} \\ \textbf{lizable}} & \textbf{Circuits} & \textbf{Strength(s)} & \textbf{Weakness(es)} & \textbf{Complexity} \\
     \hline

     \makecell{\textbf{QAE} \\ \cite{brassard2002}} & \makecell{QPE \\on Grover \\operator} & no \par  & \makecell{QAA \\ + QPE} \par & \makecell{provably \\ optimal \\ complexity} & \makecell{circuit depth \\and width; \\noise oblivious} & 
     \scalebox{0.8}{\footnotesize $N_q \in \mathcal O (\frac{1}{\epsilon} \log\frac{1}{\alpha} \cdot  a^{-1})$}  \par \\
     \hline

     \makecell{\textbf{MLAE} \\ \cite{suzuki2020}} & \makecell{heuristic \\ circuits \\ $\rightarrow$ statistical \\estimation} & fully \par & \makecell{QAA} \par \par & \makecell{simplicity; \\ numerical \\performance} & \makecell{limited formal \\guarantees; depth \\ grows indefinitely} & 
     \makecell{\fontsize{6.5}{1}\selectfont  best case $\mid$ observed: \\
     {\footnotesize  \scalebox{0.8}{$ N_q^\text{(LIS)} \sim \epsilon^{-0.75} \mid \epsilon^{-0.76} $}} \\
     {\footnotesize \scalebox{0.8}{$N_q^\text{(EIS)} \sim \epsilon^{-1} \mid \epsilon^{-0.88}$}}}\\
     \hline

     \makecell{\textbf{V-MLAE} \\ \cite{plekhanov2022}} & \makecell{MLAE with \\ variational \\ approximation} & partially \par & \makecell{QAA} \par & \makecell{numerical per-\\formance; limits \\ circuit depth} & \makecell{limited formal \\ guarantees; \\ cost overhead} & 
     \makecell{{\scriptsize similar to MLAE}\\ { \scriptsize except for EIS-} \\ { \scriptsize observed (untested)} }\\
     \hline

     \makecell{\textbf{AES} \\ \cite{aaronson2020}} & \makecell{rough estimate \\ $\rightarrow$ exponential \\ refining} & partially \par & \makecell{QAA} \par & \makecell{provably \\ optimal \\ complexity} & \makecell{large quantum \\ cost offset; \\ noise oblivious} & \scalebox{0.8}{\footnotesize $N_q \in \mathcal O (\frac{1}{\epsilon} \log\frac{1}{\alpha} \cdot a^{-1}$})  \par \\
     \hline

     \makecell{\textbf{SAE} \\ \cite{wie2019}} & \makecell{amplification \\ $\rightarrow$  probability \\ inversion} & partially  \par & \makecell{Hadamard \\ tests}  \par & simple  \par & \makecell{non-essential \\ circuits; incon-\\clusive analysis} & 
     {\footnotesize $N_q \in \mathcal O (a^{-1})$}  \par \\
     \hline

     \makecell{\textbf{IAE} \\ \cite{grinko2021}} & \makecell{watchful \\ optimization \\ of Fisher inf.} & partially \par & \makecell{QAA} \par & \makecell{provably \\ near optimal \\ complexity} & \makecell{depth grows \\ indefinitely; \\ noise oblivious} & 
     \scalebox{0.8}{\footnotesize $N_q \in \mathcal O (\frac{1}{\epsilon}\log(\frac{1}{\alpha}\log\frac{1}{\epsilon}))$}  \par \\
     \hline

     \makecell{\textbf{M-IAE} \\ \cite{fukuzawa2023}} & \makecell{IAE but \\ distribute shots \\ more favorably} & partially \par & \makecell{QAA} \par & \makecell{provably \\ optimal \\ complexity} & \makecell{same as IAE}  \par & 
     \scalebox{0.8}{\footnotesize $N_q \in \mathcal O (\frac{1}{\epsilon} \log\frac{1}{\alpha})$}  \par \\
     \hline

     \makecell{\textbf{FAE} \\ \cite{nakaji2020}} & \makecell{complementary  \\ measurements \\ $\rightarrow$ invert \\ probability} & partially \par & \makecell{QAA} \par & \makecell{provably \\ near optimal \\ complexity} & \makecell{depth grows \\ indefinitely; \\ noise oblivious} & 
     \scalebox{0.8}{\footnotesize $N_q \in \mathcal O (\frac{1}{\epsilon} \log( \frac{1}{\alpha} \log\frac{1}{\epsilon})$}  \par \\
     \hline

     \makecell{\textbf{BAE} \\ \footnotesize{(this paper)}} & \makecell{problem \\ tailored \\ Bayesian \\ inference} & partially \par & \makecell{QAA} \par & \makecell{optimal observed \\ complexity; \\ noise resilient} & \makecell{limited formal \\ guarantees; \\ classical cost} & 
     \makecell{$N_q \in \mathcal O (\epsilon^{-1})$  \\
     {\fontsize{6.5}{1}\selectfont (observed) }}\\
     \hline
\end{tabular}}
\caption{Table summarizing the characteristics of a selection of QAE algorithms. All circuits are based on Grover operations, but differ in how they're structured. Complexities were demonstrated analytically unless otherwise stated, and denote the number of queries $N_q$ necessary to estimate an amplitude $a$ to error at most $\epsilon$ with probability at least $1-\alpha$.}
\label{tb:algs_table}
\end{table*}

As mentioned in section \ref{sub:altqalg}, the first QFT-free algorithm for QAE was proposed in 2019 by Aaranson and Rall \cite{aaronson2020}, who replaced the QFT by a sequential scheme relying on pure amplitude amplification - that is, on Grover-type circuits, where the only variable is the number of Grover iterations. They call the algorithm "quantum counting, simplified"; more generally, we use the term"amplitude estimation, simplified" (AES). 

Even though this algorithm accomplishes the intended speed up, a large constant factor is involved; meaning that even though the uncertainty in the amplitude shrinks as fast as desired, it starts out unfavorably. What is more, precise measurement results are necessary for the algorithm to function as intended, making it especially susceptible to noise. 

In the same year, other authors had set forth their own alternatives, trading off rigor for practicality. In \cite{suzuki2020}, Suzuki and co-authors proposed a maximum likelihood approach to amplitude estimation (MLAE). Their strategy relies on heuristic sequences of simple Grover circuits, and infers the amplitude based on the measurement data extracted from them. 

Two heuristics are proposed: LIS and EIS; where the $m$ increase linearly and exponentially respectively. The authors prove lower, but not upper, bounds for the estimation error. The lower bound for EIS is the Heisenberg limit, whereas LIS has a worse (but still quantum-enhanced) best case performance. A numerical analysis shows good performance, although neither of the strategies saturates the lower bound. Some authors have studied the behavior of this algorithm in the presence of noise, and suggested improvements \cite{brown2020, tanaka2021, callison2023, Tanaka2022}.

Two years later, \cite{plekhanov2022} considered reworking this maximum likelihood algorithm from another perspective, by periodically replacing chunks of Grover iterations with variational approximations in order to reduce circuit depth. They demonstrate interesting numerical results, despite incurring a cost overhead due to the variational circuit optimization. 

In the same year, \cite{giurgica-tiron2022} reworked the scheme of \cite{suzuki2020} to cover the ground between the classical and quantum approaches - the goal being to mindfully exploit the limited \textit{quantumness} of near-term devices. On top of that, the authors developed another algorithm achieving the same feat while offering stronger formal guarantees. Both approaches demonstrate robust numerical performances.

Shortly after \cite{suzuki2020}, \cite{wie2019} introduced a straightforward approach based on Hadamard tests, which the authors termed simpler quantum counting (or simpler amplitude estimation, SAE). After several executions, the model for the outcome distribution is inverted to obtain the parameter of interest. However, the theoretical and  numerical analyses don't directly address the performance metrics of interest. 

Not much later, \cite{grinko2021} came up with iterative amplitude estimation (IAE), an algorithm combining formal rigor, a solid numerical performance, and a modest cost offset. Although it couldn't attain the ideal asymptotic complexity, it came close, with only a double-logarithmic factor separating the two. What is more, the (noiseless) experiments demonstrated its competitiveness, which was not surpassed by any other considered algorithm.

This algorithm drew the attention of other authors, prompting modified versions. In particular, \cite{fukuzawa2023} enhanced it through a rearrangement of its failure probabilities across iterations - managing to shave off the unwanted logarithmic factor to get an optimal asymptotic performance. This development further consolidated the significance of this algorithm as a both rigorous and practical approach. Yet its noise-obliviousness may render it impractical for near-term use, where more heedful strategies may bring an advantage.

Another algorithm was later proposed in \cite{nakaji2020}. Again it falls short of achieving Heisenberg scaling, but again it comes close, the difference being yet again a double-logarithmic factor. It relies on straightforward  inversions of circuit until they are barred by redundancy; at which point it changes into more involved inversions requiring additional measurements  for disambiguation. As in \cite{grinko2021}, numerical tests show quite satisfactory results, although with a less favorable offset. 

All of these algorithms are considered hardware-friendly - in the sense that the required circuits are shallower than in the original algorithm of \cite{brassard2002} - but assume fault tolerant computations. Put differently, the circuits are simpler, but still expected to yield ideal results upon execution. 

Not only does this underpin most or all proofs of correctness, it is also assumed when constructing measurement schedules. More specifically, to refine knowledge, algorithms usually rely on circuits of progressively increasing depth. If this carries on indefinitely, it is unavoidable that the times required for execution exceed the device's coherence time, which is always finite in practice. 

As a result, we will eventually be measuring classical noise at the output; clearly, in this noisy regime, less ambitious measurements could be more informative. What is more, the overreliance on exact outcomes makes noise oblivious algorithms incapable of recovering from aberrant measurements.

These points highlight the need for amplitude estimation algorithms that can adapt to noisy scenarios. In 2021, \cite{wang2021} gave a step in that direction, with an algorithm termed robust amplitude estimation (RAE). Instead of relying on rigid schedules and meticulous calculations underpinned by a rigorously crafted analytical backbone, it relied on a more flexible framework based on Bayesian inference \cite{koh2022}. Not only is this framework capable of noise mitigation \cite{katabarwa2021}, these capabilities can be combined with or enhanced by complementary techniques \cite{dalal2023}. The richness of this take immediately opens a multitude of interesting paths to pursue in an attempt to bring QAE closer to practical applications \cite{johnson2022}.

RAE relies on Bayesian inference with \textit{engineered likelihood functions}, which are achieved by generalizing the Grover operator. A similar idea was proposed in \cite{brassard2002} to make Grover search deterministic. In this case, the goal was to further customize the circuit model to be used for data collection. Moreover, a simple noise model is incorporated into the likelihood, which further upskills the algorithm.

The main downside of this approach is the trade-off between computational cost and scalability. To make the problem tractable, the authors work under a Gaussian assumption. This is shown not to significantly affect the numerical results for the case tested therein. While such approximations are often agile for a few parameters, as is the case, they don't scale well for higher dimensions, which arise when considering more complex noise models. 
The statistical representation is arguably the most critical implementation detail in Bayesian inference. It is then important to find methods that are efficient, general and scalable. Failing to do so is bound to frustrate the inference process, affecting not only the acuity of optimality, but correctness itself.

Additionally, RAE requires generalizing the reflections of the Grover operator (\ref{eq:groverop}) into arbitrary rotations, whereas originally they are fixed and equal to $\pi$. This  increases the  classical optimization cost, since we are now working with $2m$ parameters (with $m$ the number of circuit layers/applications of the Grover operator) rather than $1$ ($m$ itself). 

Furthermore, working with $m$ different operators consisting of two continuous rotations brings added complexity to the experimental set up as compared to a single constant operator consisting of two reflections. Such customization makes the circuit harder to implement, calibrate, and possibly compile; applying error correction becomes more costly. 

For these reasons, tailoring the Grover operator can be considered a disadvantage as compared to other hybrid QAE schemes.

In this paper, we propose an approach that does not generalize the Grover operator, instead working with standard fixed-operator amplitude amplification. We leverage QAE-specific insights to alleviate classical processing costs while retaining the quantum advantage, and employ cost-efficient statistical methods that generalize well to noisy scenarios.

Table \ref{tb:algs_table} presents a summarized overview of the discussed algorithms.

\section{Bayesian amplitude estimation}
\label{sec:BAE}

\begin{figure*}
    \centering
    \begin{subfigure}[b]{0.48\linewidth}
        \centering
        \includegraphics[width=0.6\linewidth]{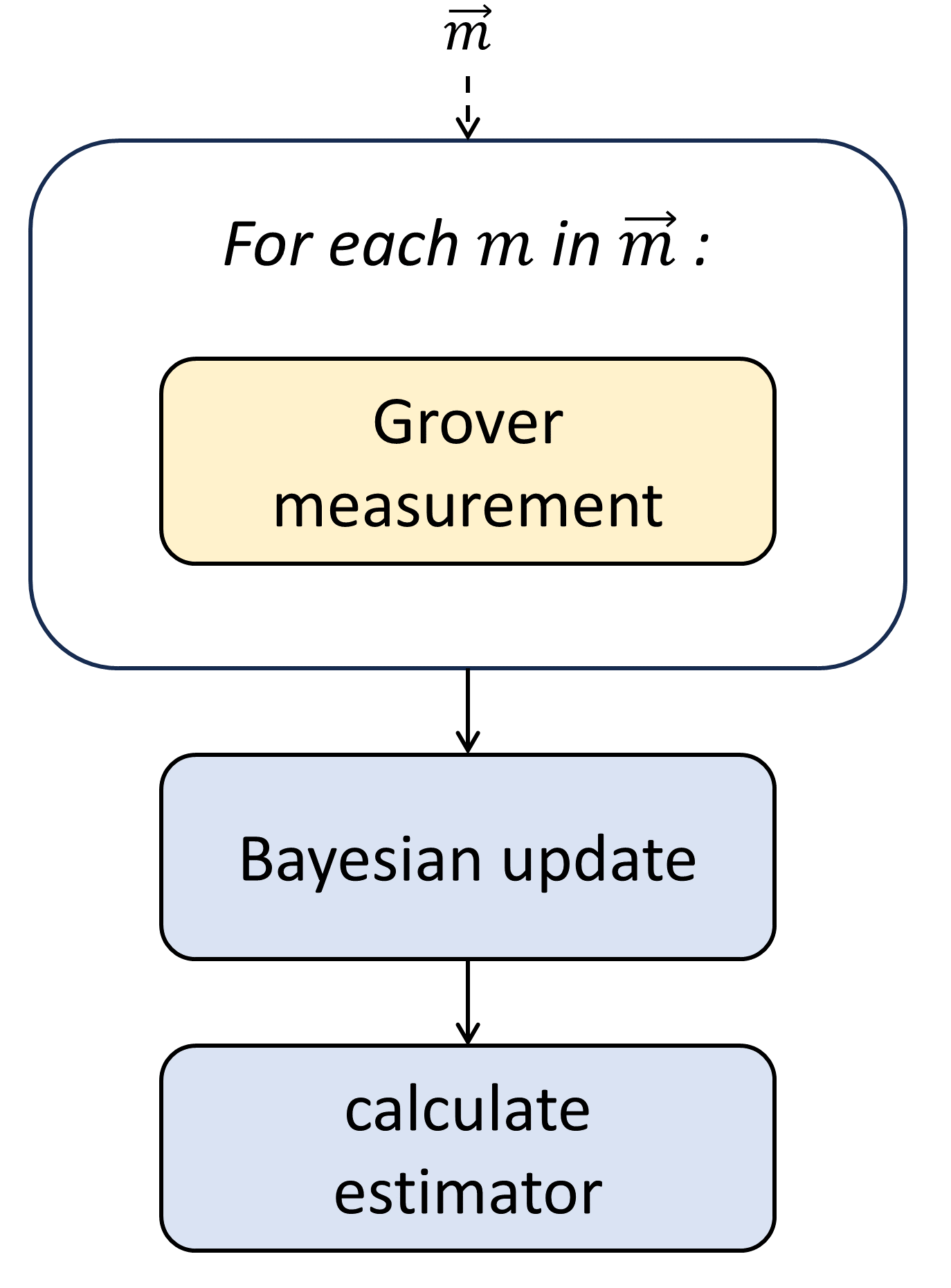}
        \caption{Non-adaptive Bayesian algorithm for quantum amplitude estimation.}
        \label{fig:inference}
    \end{subfigure}
    \hfill
    \begin{subfigure}[b]{0.48\linewidth}
        \centering
        \includegraphics[width=\linewidth]{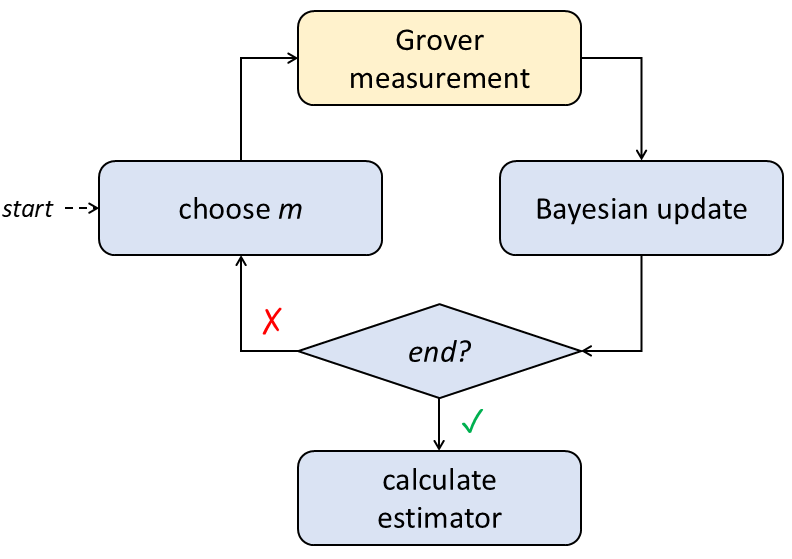}
        \caption{Bayesian quantum amplitude estimation (BAE). Pseudo-code is provided in Algorithm~\ref{alg:adaptive_inference}; blocks \textit{choose m} and \textit{Bayesian update} refer to algorithms \ref{alg:optimize_control} and \ref{alg:SMC_update}, respectively.}
        \label{fig:update}
    \end{subfigure}
    \caption{Diagrams of Bayesian algorithms for amplitude estimation.}
    \label{fig:diagrams}
\end{figure*}

We propose Bayesian amplitude estimation (BAE), a Bayesian inference based algorithm following the hybrid scheme of Figure~\ref{fig:hybdrid}. In each iteration, a number of Grover steps $m$ is chosen classically, after which the corresponding quantum circuit is executed. The measurement outcome is fed into the classical processing unit, and this cycle repeats as necessary to achieve the intended uncertainty. 

The Bayesian framework allows us to refine our (possibly null) initial knowledge about the amplitude, representing it by probability distributions. This description allows us to evaluate the expected utility (variance) of any $m$ in a look-ahead fashion, and thus optimize $m$ or sequences $\vec m$. We opt for a greedy algorithm - unitary look-ahead, where a single $m$ is optimized at a time -, to minimize optimization costs. For related reasons, we perform a classical warm-up phase, in the initial stage where information is scarce. 

The workflow is represented diagrammatically in Figure~\ref{fig:update}, while pseudo-code is provided in algorithm \ref{alg:adaptive_inference}. Overviews of Bayesian inference and Bayesian experimental design are provided in Appendices~\ref{app:Bayesian_inference} and \ref{app:expdesign} respectively. The Bayesian updates and estimator calculations are the backbone our inference-based protocol; for efficient numerical representations of the probability distributions, we use sequential Monte Carlo (SMC) with Markov Chain Monte Carlo transitions, detailed in appendix \ref{app:integration}. 

The crux of the algorithm is the \textit{choose m} sub-routine. A greedy strategy would involve discrete optimization over an infinite domain, as $m$ is unbounded above. To remedy this, we develop a heuristic strategy to define the search range adaptively, based on window expansions (Appendix~\ref{subapp:window}). We start with a low upper bound for the search range; each time one of the highest possible controls allowed by the window is chosen in the optimization, we register a \textit{hit}. When the number of hits achieves a pre-determined threshold, a window expansion is triggered: the former upper bound becomes the new lower bound, while the upper bound doubles. We then reset the hit counter to zero, and start again. This relies on the fact that Heisenberg-limited QAE algorithms have amplifications schedules where $m$ grows exponentially with the iterations \cite{aaronson2020, fukuzawa2023, suzuki2020}. While the window width still grows exponentially, we achieve constant cost by having the evaluations within be exponentially sparse.

Importantly, this Bayesian framework can easily be adapted to a noisy scenario, by adapting the generative model. We extend the Bayesian framework to learn noise parameters beyond the amplitude, obtaining an algorithm that can characterize device noise and adapt in real time. Details about the noise tailoring are provided in Appendix~\ref{app:noise}. 

Our algorithm is capable of exploiting prior knowledge, e.g if the amplitude is known to be small (Appendix~\ref{app:Bayesian_inference}); can evaluate noise models for negligible extra cost (Appendix~\ref{subapp:model_evidence}); is well-suited for parallel execution, with largely independent tasks that can be distributed across both quantum and classical devices (Appendix~\ref{app:parallelism}); amd features adjustable parameters that can be adapted based on practical trade-offs (Appendix~\ref{app:hyperparameters}).

\begin{figure*}
    \centering
    \begin{minipage}{0.9\textwidth}
        \begin{algorithm}[H]
        \caption{Algorithm for adaptive inference. Non trivial functions are defined in Appendix~\ref{app:SMC}.}
        \label{alg:adaptive_inference}
            \begin{algorithmic}[1]
            \Parameters \texttt{wNshots}, \texttt{Nshots}
            \Require   \texttt{prior} \Comment{Distribution over parameter.}
            \Ensure ${\hat \theta}$ \Comment{Parameter estimate.}
            \State \Gets {\vec{m}} {\Callt{repeat}{0, wNshots}}
            \State \Gets {\vec{o}} \Callt{measure}{$\vec{m}$} \Comment{Classical warm up.}
            \State \Getst d \Callt{batch\_bayesian\_update}{prior, $\vec{m}$, $\vec{o}$} \Comment{Update distribution.}
            \While {not  \ttt{termination\_criterion}} \Comment{E.g. target uncertainty.}
            \State \Getst m \Call{optimize\_control}{\ttt{d}}
            \State \Gets {\vec{m}} {\Callt{repeat}{m, Nshots}}
            \State \Gets {\vec{o}} \Callt{measure}{$\vec{m}$} 
            \State \Getst d \Callt{batch\_bayesian\_update}{d, $\vec{m}$, $\vec{o}$}
            \EndWhile
            \State \Gets {\hat \theta} \Callt{estimator}{d} \Comment{E.g. mean.}
            \State \Return ${\hat \theta}$ 
            \end{algorithmic}
        \end{algorithm}
\end{minipage}
\end{figure*}

\subsection{Annealed Bayesian amplitude estimation}
\label{sub:abae}

BAE is an inference algorithm that greedily minimizes the variance. In this section, we propose a different approach - annealed BAE (aBAE), borrowing from annealed importance sampling \cite{neal1998, south2019}, a well-known algorithm in the statistics literature.

This algorithm has the goal of sampling from a complex probability distribution $\mathbf{P}(\theta)$, possibly a posterior distribution. It does so by traversing a sequence of annealing coefficients, or temperatures, $\beta_1, \dots, \beta_N$, where $\beta_1 = 0$, $\beta_N=1$, and the coefficients are increasing. This effectively raises the distribution up from a uniform distribution into the target one in gradual increments. 

The successive powers of the distribution form a sequence of distributions that can be sampled from using sequential Monte Carlo. Note that previously, the sequence of distributions was given by cumulative datasets of increasing size. The SMC algorithm can be applied to any sequence of distributions. Refer to Appendix~\ref{app:SMC} for details.

The choice of the sequence of coefficients determines the performance of the algorithm. One option is to choose the coefficients adaptively to maintain the effective sample size (ESS) around a target value. The ESS is a measure of particle degeneracy in SMC. A set of $K$ samples may correspond to less than $K$ effective samples due to correlation; if dealing with weighted "grid points", as in SMC, uneven weights mean low representativeness. The ESS quantifies to how many uniform samples a set of weighted samples corresponds. 

Thus, keeping the ESS around a target value helps assure a stable representation. Intuitively, assuring that the ESS is not too low guarantees that the information is captured properly; whereas assuring it is not too high assures that a significant amount of information is taken in. 

In standard BAE, we control the ESS by refreshing the point locations when it falls below a certain threshold. For annealed BAE, we instead choose at each iteration the experimental control that minimizes the expected distance to the target ESS. Note that the Bayesian and SMC frameworks are unchanged, except for the choice of the utility function.  The ESS expectation can be calculated in a look-ahead fashion, just like the variance or other utility functions. 

The experimental design then takes the role of the choice of coefficients. While not as well-behaved, they can achieve a similar effect due to the structure of the likelihood function. 

The advantage of this method is statistical robustness, which we expect to be especially notable in higher dimensions, where adequately exploring the space is more challenging. Additionally, in multi-dimensional estimation,the variance cost function must be generalized to a scalar representation of the covariance. This raises new points to consider, such as the relative scales of the parameters. Using the ESS eliminates these problems. Lastly, the complexity of calculating covariance-related metrics scales with the dimension, whereas the complexity of calculating the ESS does not.

\section{Methods}
\label{sec:methods}

\subsection{Processing and benchmarking}

To asses the merit of a QAE algorithm one must find where results fall relative to the fundamental limits of metrology. 

For a graphical depiction of this assessment, we will represent the root mean squared error (RMSE) as a function of the number of queries $N_q$. In previous QAE proposals, tests are performed for a fixed amplitude \cite{grinko2021, nakaji2020, wie2019} chosen by the authors. Such choices may lead to biased assessments, as well as overfitting. 

Instead, we test the algorithms in a problem agnostic and general way, by sampling amplitudes at random and taking a normalized average of the normalized values of the RMSE, NRMSE:

\begin{equation}
    \text{NRMSE} =  \sqrt {\frac{1}{N}\sum_{i=1}^N \left(\frac{a_i-\hat{a}_i}{a_i}\right)^2}.
\end{equation}

, where $a_i$ and $\hat a_i$ are the real and estimated amplitude values, and $N$ is the number of random amplitudes used. 

Similarly, we use a normalized version of the average standard deviation. This allows for a more thorough performance assessment, reveals behavior irregularities in some algorithms, and allows for universal tuning of the algorithm hyperparameters.

Importantly, we use a different set of random amplitudes for each plot presented here: an amplitude $a \in [0,1]$ is sampled afresh for each run. We did not reuse amplitudes in any two executions, nor during the development of the algorithm and tuning of the hyperparameters. As such, BAE is in principle well-fitted to all possible amplitude estimation problems, given its robust performance across hundreds of amplitudes picked at random; and all the demonstrations shown here hold for arbitrary amplitudes. Accordingly, we always assume the absence of prior knowledge about $a$ for BAE, putting it on equal footing with the other algorithms. 

As for the queries, they could be applications of the Grover operator as a whole, of the oracle, or of the initialization operator. We consider the latter. The queries to these two operators are nearly proportional, with the initialization operator $\op{A}$ being applied twice within each oracle application (once forwards and once backwards, i.e. the inverse) and once more for state preparation.

Note that the cost of an algorithm is calculated as the total number of queries used throughout its execution. These queries may be split among multiple circuits. This comprehensive definition of cost allows for a direct comparison with the limits of quantum metrology (section \ref{sub:quantum_enhanced}), which is not concerned with the specifics of the resource allocation. 

Nonetheless, this division affects the behavior of the algorithm, as the number of queries in a circuit is proportional to its depth, which determines the degree to which the outcomes are affected by decoherence. This is reflected in the performance of the algorithms under the presence of noise.

This ends the discussion on the quantities to be plotted. We now discuss \textit{how} to plot them. Firstly, we will use a double logarithmic scale. We represent the limits in equations (\ref{eq:sql}) and (\ref{eq:hl}) for reference; they take the form of straight lines with $-0.5$ and $-1$ slopes respectively, facilitating visual assessments.

Secondly, we customize the y intercept to the datasets to facilitate a visual analysis. Details are provided in Appendix~\ref{app:intercept}. Note that the vertical scale offset should still be heeded: it is a relevant cost metric in practice, depending on the precision regime. It may be undesirable to have algorithms requiring a higher quantum resource count from the outset, even if the scaling is Heisenberg-limited. Eventually, the performance of such an algorithm will beat that of another with a smaller offset but a less favorable evolution pace, but whether this happens depends on the target resolution. 

Another challenge arises from the use of adaptivity. In quantum metrology, the performance metrics tend to respect \textit{average} results, due to statistical noise. These averages are straightforward in the case of deterministic algorithms, but not with adaptive ones. In that case, due to a mismatch in x-coordinates, brute force averages would require discarding most data points and using a disproportional amount of executions, growing exponentially with the maximum query number. To avoid this, we find and employ a good approximation that allows us to use all data points while still depicting the statistic reliably. We analyzed multiple possible strategies, and found the best performing one to be binning the points according to their x-coordinates and averaging both their coordinates independently.

\begin{figure}[!htb]
    \centering
    \includegraphics[width=\linewidth]{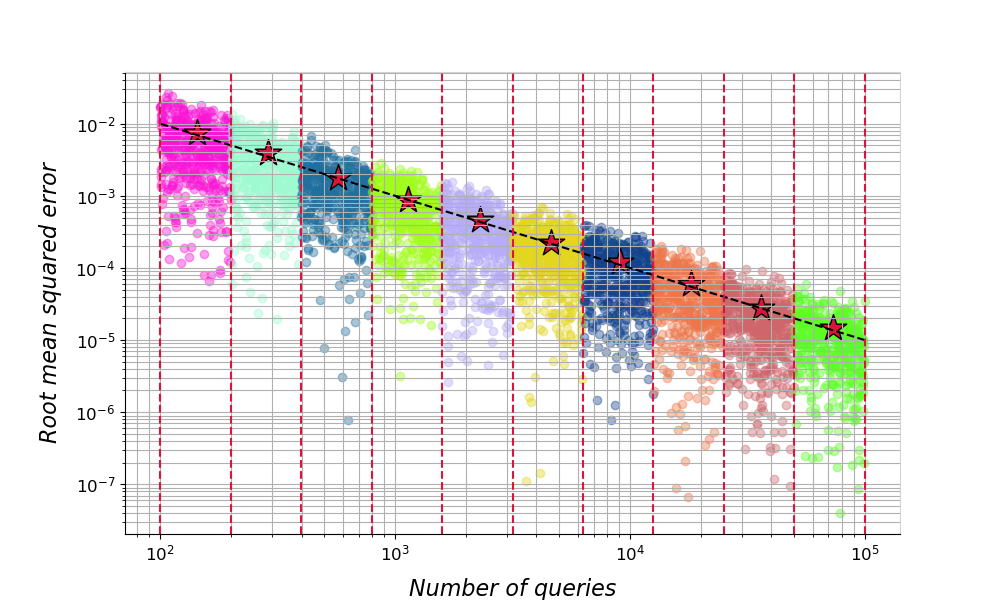}
    \caption{Behavior of the adopted strategy to process adaptive algorithm data: binning and averaging. Here it is applied to dummy data, which is grouped into bins (different colors) and then summarized (star markers). It comes close to ideal behavior - markers lying on the line.}
    \label{fig:averaging_best}
\end{figure}

Results for mock data are displayed in Figure~\ref{fig:averaging_best}. These data were generated to reproduce ideal Heisenberg-limited behavior, so that the results can be verified. By means of the aforementioned strategy, the same-colored points contained in each bin are summarized by a single point: the star markers. Ideal processing would have these markers lying unbiasedly on the dashed line. Details can be found in Appendix~\ref{app:proc_adapt}; Appendix~\ref{app:otherstat} shows the results using other strategies, demonstrating that our results hold for any processing strategy that behaved well for dummy data.

\subsection{Quantum simulation}
\label{sub:qsim}

We want to numerically simulate BAE, other hybrid approaches to QAE, and the canonical QAE algorithm (to be used as a reference). For testing purposes, the quantum parts of each of these algorithms can be simulated efficiently using analytical calculations and multinomial sampling. This does not mean that the circuits can be efficiently simulated by a classical computer, as generating these data requires knowing the value of $a$. Refer to Appendix~\ref{app:num_qae} for more details. 

In addition to the ideal behavior, we want to observe the behavior of the algorithms under the influence of extrinsic noise (i.e. not shot noise). We thus augment equation \ref{eq:p1} with an extra parameter $T$ - a coherence time -, and assume an exponential decay affecting the basis states equally. The factor that is independent of $\theta$ assures this symmetry (and proper normalization). The result is expression \ref{eq:p1_Tc}.

\begin{gather}
\label{eq:p1_Tc}
    P(\psi_1 \mid m) = e^{-\sfrac{m}{T}}\sin^2((2m+1)\theta) 
    \nonumber \\
    + \frac{1-e^{-\sfrac{m}{T}}}{2}.
\end{gather}

This model has been used in other works, and similar exponential decays arise when considering common sources of noise in quantum devices: depolarization, dephasing, energy relaxation, and gate miscalibration \cite{granade2012, wiebe2016a, wang2021, katabarwa2021}.

For practicality, we define a time unit as the time taken by one application of the Grover operator. $T$ can then be expressed using these units. Having a closed expression given by equation \ref{eq:p1_Tc}, sampling noise can then be introduced as usual.
\section{Results}
\label{sec:results}

This section graphically presents the results of testing our BAE algorithm as described in section \ref{sec:methods}. The statistics are taken over 100 executions. In the comparative plots, we consider the evolution of the median errors to facilitate the visual analysis. Average results are provided in Appendix~\ref{app:otherstat} for completeness.

Shot noise is present in all tests; in subsection \ref{sub:noiseless_sim} no other sources of noise are considered, whereas in section \ref{sub:noisy_sim} finite coherence times are imposed. Section \ref{sub:comparative} presents a comparative analysis, benchmarking BAE against the state-of-the-art. 

The code and datasets used for these simulations are publicly available on GitHub \cite{ramoa2024a}, along with a demonstrative notebook.

\subsection{Noiseless performance of BAE}
\label{sub:noiseless_sim}

\begin{figure*}[!htb]
    \centering
    \begin{subfigure}[b]{0.41\linewidth}
        \centering
        \includegraphics[width=\linewidth]{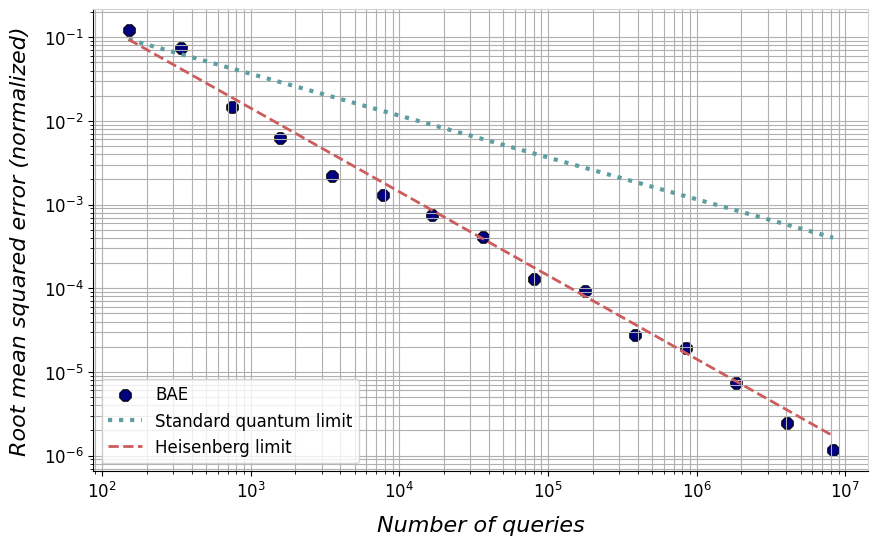}
        \caption{Baseline.}
        \label{fig:res1}
    \end{subfigure}
    \hfill
    \begin{subfigure}[b]{0.48\linewidth}
        \centering
        \includegraphics[width=\linewidth]{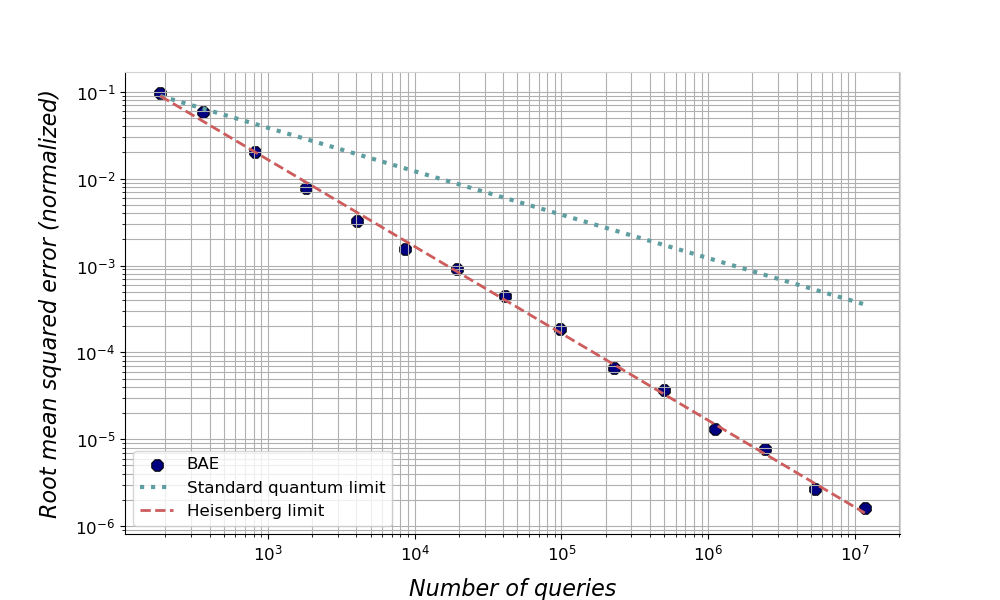}
        \caption{With additional cost-sparing measures.}
        \label{fig:res2}
    \end{subfigure}
    \caption{Performance of Bayesian amplitude estimation in ideal conditions (shot noise only). We observe Heisenberg-limited estimation. We find extra cost-sparing measures that barely affect performance, while decreasing classical processing costs and increasing quantum parallelism.}
    \label{fig:res_combined}
\end{figure*}

Figure~\ref{fig:res1} shows the numerical simulation results for BAE. We can see that the evolution of the error is smooth and parallel to the Heisenberg limit.

Figure~\ref{fig:res2} shows the results of employing the cost sparing measures mentioned in Appendix~\ref{app:expdesign}: the resampling steps of sequential Monte Carlo were suppressed for utility calculations, and $10$ shots were realized for each $m$. The offset is only slightly worse, and we still observe Heisenberg-limited estimation. The runtime for the classical processing was cut by a factor of 5. In all other simulations, we employ these cost sparing measures.

Note that a slightly better cost offset may be obtained in exchange for less regular behavior, by reducing the number of warm-up shots and shots per measurement, or relaxing the criteria for triggering window expansions (appendices \ref{app:expdesign} and \ref{app:hyperparameters}).

\begin{figure}[!htb]
    \centering
    \includegraphics[width=\linewidth]{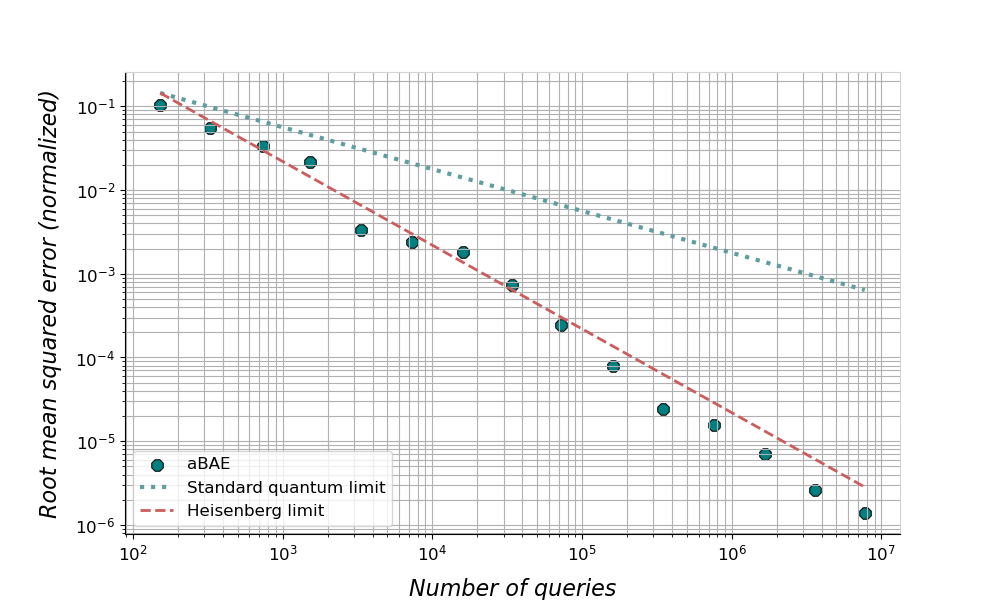}
    \caption{Performance of annealed Bayesian amplitude estimation in ideal conditions (shot noise only). This variant, despite optimizing statistical representativity rather than uncertainty, yields similar results to the original, and can replace it when variance-based methods are problematic.}
    \label{fig:ESS}
\end{figure}

We also test annealed BAE, as per subsection \ref{sub:abae}. Figure~\ref{fig:ESS} shows the results, with the estimation still parallel to the Heisenberg limit. The cost offset is similar to BAE under the same circumstances and for unidimensional estimation. This is promising, given that its advantages are expected to show for higher dimensional problems. For instance, multi-parameter noise models may give rise to multimodality, in which case this approach is expected to outpeform the original BAE, which is variance-based.

\subsection{Performance of BAE with decoherence}
\label{sub:noisy_sim}

To demonstrate BAE's potential for estimation in the presence of noise, we run the same simulations considering a depolarizing channel with a finite coherence time.  For these tests, we assume random coherence times $T_c \in [2000,5000[$, with the time units described in section \ref{sub:qsim}. The time is picked at random for each run and assumed constant within it. BAE uses $500$ shots to estimate the coherence time.  

\begin{figure}[!htb]
    \centering
    \includegraphics[width=\linewidth]{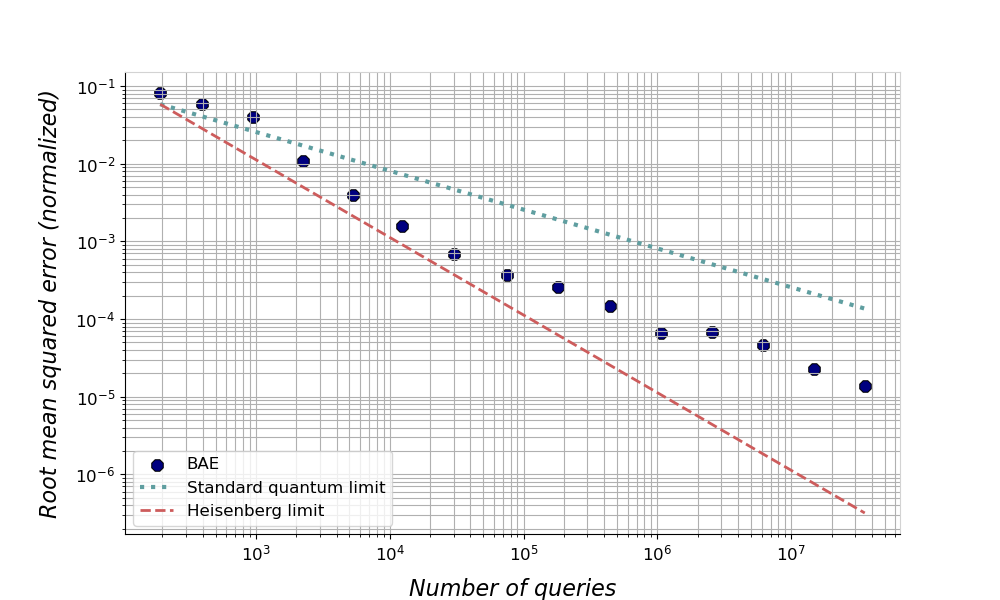}
    \caption{Performance of Bayesian amplitude estimation in the presence of decoherence. The estimation is initially Heisenberg limited and then slows down, continuing at a pace between the quantum and classical limits. }
    \label{fig:noisy_bae}
\end{figure}

Figure~\ref{fig:noisy_bae} shows the estimation process. We can see that the evolution is initially parallel to the Heisenberg limit, and slows down as the number of queries increases, achieving an intermediate slope between the quantum and classical limits. 

In the noiseless case, as the error shrinks, deeper circuits are chosen to improve the estimation error optimally. In this case, where the coherence is limited, there is a trade-off: deeper circuits are more affected by decoherence. Thus, BAE chooses the circuit depth that achieves the best balance between these two considerations, retaining part of the quantum advantage despite the noise.

\subsection{Comparative analysis}
\label{sub:comparative}

\begin{figure*}[!htb]
    \centering
    \begin{subfigure}[b]{0.48\linewidth}
        \centering
        \includegraphics[width=\linewidth]{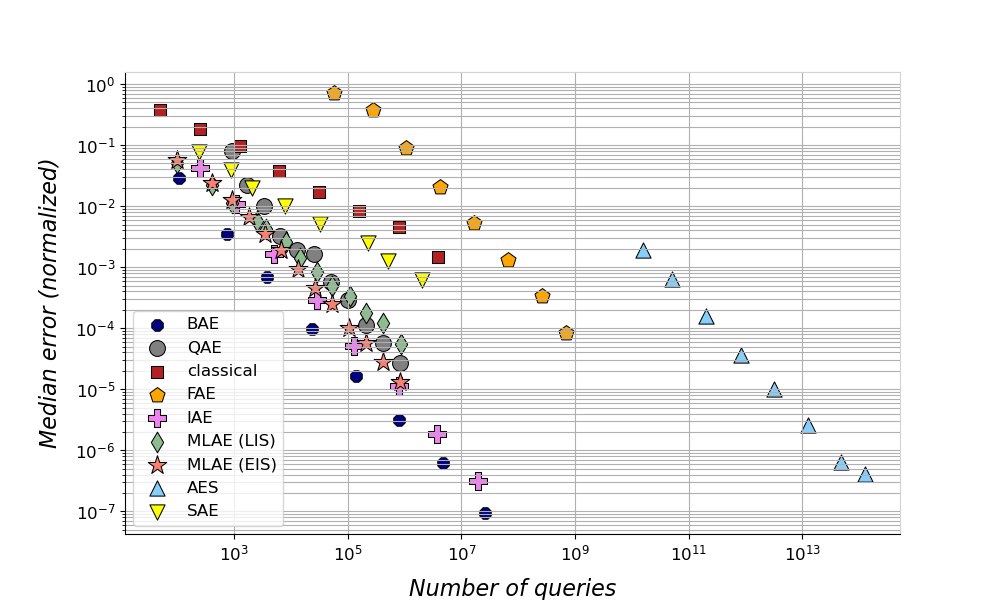}
        \caption{Ideal case (shot noise only). BAE achieves the lowest cost for any error, and the lowest error for any cost. }
        \label{fig:noiseless_median}
    \end{subfigure}
    \hfill
    \begin{subfigure}[b]{0.48\linewidth}
        \centering
        \includegraphics[width=\linewidth]{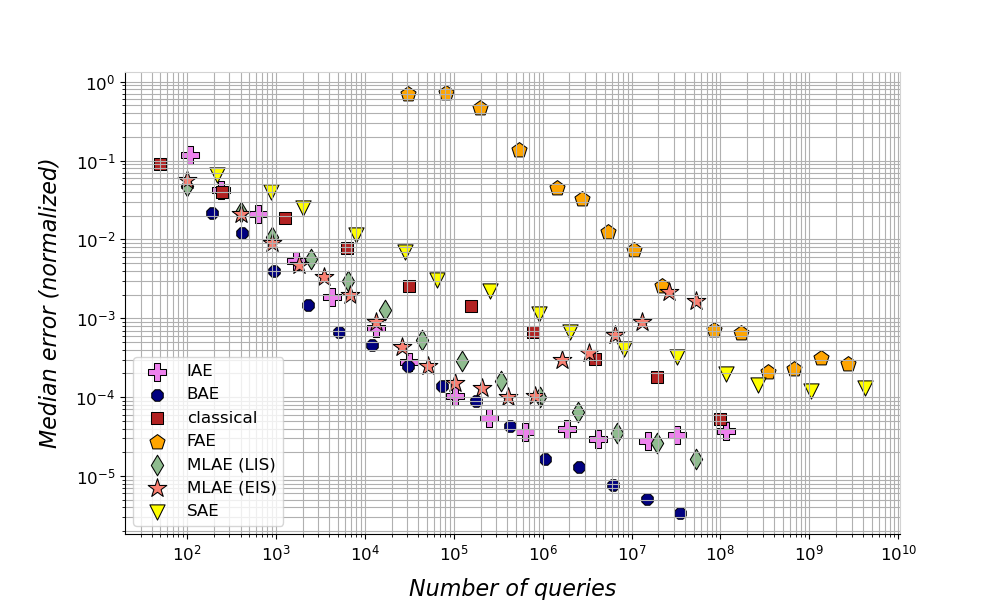}
        \caption{Noisy case (decoherence). BAE achieves the lowest error, as the other algorithms stagnate due to noise.}
        \label{fig:noisy_median}
    \end{subfigure}
    \caption{Performance of quantum amplitude estimation algorithms. Bayesian amplitude estimation (BAE) outperforms all others in both cases. Acronyms are described in Section~\ref{sub:litoverview}; methodology and units are discussed in Section~\ref{sec:methods}.}
    \label{fig:comparison}
\end{figure*}

Finally, we test other algorithms proposed in the literature and compare them with BAE. The results are presented in Figure~\ref{fig:comparison}. 

Starting by the ideal case (Figure~\ref{fig:noiseless_median}), we observe that BAE has competitive performance in terms of both complexity and offset: its learning rate (slope) matches that of the best-performing methods, while its offset is better than any other. BAE not only saturates the Heisenberg limit, it also manages to shrink the error faster than any other algorithm as a function of the resource count. This illustrates how even though it can adapt to noisy devices, BAE is not exclusively a NISQ algorithm: it is capable of achieving full quantum advantage in a fault-tolerant scenario. 

Turning to the noisy case (finite coherence - Figure~\ref{fig:noisy_median}), BAE remains the best performing algorithm\footnote{QAES is not represented because it doesn't terminate (as discussed in section \ref{sub:altqalg}). The original algorithm for QAE is not represented due to the runtime, but it is easy to see that it wouldn't fare well. Reaching $N_q=10^6$ with the phase estimation based circuit would require a Fourier order $m=19$, and thus $2^{18}$ consecutive applications of the (controlled) Grover operator followed by the QFT. This far exceeds the coherence times.}, with an even bigger separation. 

Among all tested algorithms, BAE achieves the lowest estimation error for any given number of queries. In fact, it achieves the lowest estimation error regardless of cost, as the other algorithms do not seem capable of achieving error below $10^5$ due to stagnation in the learning rate. BAE performs reliably even for higher numbers of queries, corresponding to longer executions and higher precision, whereas others with similar costs come to a stall or erratic behavior due to the noise.

Of the other algorithms, some only reveal  this trend for higher numbers of queries: FAE, SAE and MLAE (LIS). This is due to their larger offsets and/or slower learning rates; they use shorter circuits than would be optimal, which inflates the quantum cost and delays the effect of decoherence. This does not constitute an advantage: we are interested in the effect of noise for a given magnitude of the error, not of the cost. (Otherwise, one might "improve" an algorithm's noise resilience by inflating the cost for no benefit, causing a shift to the right in the plot.) For similar reasons, MLAE-LIS runs into runtime and numerical stability issues for high query numbers, due to the optimization of products envolving large numbers of small terms (likelihoods for tens of thousands of circuits).

Relatedly, we note that the total number of queries may not be straightforwardly related to the impact of decoherence, nor is it the sole predictor of unstable behavior in the presence of noise. These queries are cumulative, and not necessarily proportional to the maximum circuit depth (although they are approximately so for some algorithms, such as the geometric progression circuits of MLAE-EIS).

These are the main results of our work. We have demonstrated that BAE:

\begin{enumerate}
    \item Achieves Heisenberg-limited estimation in the absence of noise, with a slope as good as the best algorithms; 
    \item Has a reduced offset as compared to all other tested algorithms; 
    \item Can, unlike the other algorithms, adapt to noisy scenarios. 
\end{enumerate}

The last point is related to other perks coming from the pliable and wide-ranging framework: BAE is capable of accomodating experimental limitations or preferences, assessing the merit of noise models, leveraging previously available information, and exploiting various trade-offs.
\section{Conclusions and future work}
\label{sec:conclusion}

We propose BAE, a QAE algorithm that is capable of attaining Heisenberg-limited estimation; is highly customizable and capable of negotiating trade-offs across the multiple costs involved; is parallelizable and scalable; and is resilient to noise. Numerical simulations show that our algorithm has a robust performance as compared to state-of-the-art algorithms, both in the presence and in the absence of noise. This makes it especially interesting in the transition between the NISQ and fault tolerant eras, as an algorithm that can interpolate between those regimes.

In particular, BAE is capable of characterizing noise and self-adapting accordingly. This makes it better suited to faulty quantum devices than algorithms that fail to account for extrinsic noise sources. While noise can still slow down the learning rate, properly treating it can minimize this slowdown while safeguarding correctness. We observe that our algorithm continues to learn even after others saturate due to noise. 

We additionally propose aBAE, an annealed version of BAE which guides the inference based exclusively on a measure of statistical degeneracy. This makes the cost of evaluating the utility function independent of the number of parameters, and helps assure a stable numerical representation in multi-dimensional scenarios, as those brought about by detailed descriptions of noise. 

Relatedly, an interesting direction for future work is to test the algorithm using more complex noise models. Such models are expected to benefit the most from the robustness of the presented methods, namely the efficient multi-dimensional sampling and the annealed variant of BAE. This improvement is a key step towards another target for future work: performing quantum amplitude estimation on real quantum devices, rather than relying on numerical simulations. 

Other possible research directions include modifications to specific parts of the BAE algorithm, namely the utility function, the control optimization routine, and the numerical representation. This may further improve the performance or reduce classical costs. 

Finally, the BAE algorithm could be directly applied to the characterization tasks mentioned in the introduction, with numerous potential applications in quantum technologies - namely superconducting and photonic quantum computing, and sensing.

\begin{acknowledgments}
  The authors thank Ernesto Galvão, Antonio Molero and Bruna Salgado for helpful comments. This work is funded by national funds through FCT – Fundação para a Ciência e a Tecnologia, I.P., under the support UID/50014/2023 (\href{https://doi.org/10.54499/UID/50014/2023}{https://doi.org/10.54499/UID/50014/2023}). A.R. acknowledges support from FCT under PhD grant 2022.12332.BD.
    \end{acknowledgments}

\appendix

\section{Bayesian inference}
\label{app:Bayesian_inference}

Bayesian statistics provide a powerful framework for inference, earning it widespread use in the scientific community, namely for the characterization of quantum systems \cite{higgins2007,wang2017,craigie2021,huszar2012}. It can also be applied to the problem of amplitude estimation.

This type of inference relies on the systematic application of Bayes' rule:

\begin{equation}
    \label{eq:bayes}
    \mathbf{P}(\theta \mid D)= \frac{\mathbf{L}(\theta \mid D; m)\mathbf{P_0}(\theta)}{\mathbf{P}(D ; m)}.
\end{equation}

Here $\theta$ is the parameter of interest, $m$ is an experimental control (here the number of Grover iterations), and  $D$ is a datum (experimental outcome). To lighten notation, $m$ may alternatively be included in $D$. We use ";" to separate random variables such as $\theta$ and $D$ from non-random variables such as $m$.

The left hand side of equation (\ref{eq:bayes}) represents the posterior probability as a function of the parameter $\theta$. For sufficiently informative datasets, the posterior probability distribution should converge to a sharp peak centered in the real value of $\theta$. 

An evaluation of this function for a specific $\theta$ quantifies its merit as an estimate of the true value given experimental data. The right hand side is the likelihood of the parameter given the observation (defined as the probability that the former would have generated the latter), times our prior degree of belief in $\theta$, divided by a normalizing constant. This constant is termed the marginal likelihood or evidence, and can be used to evaluate models. 

Figure~\ref{fig:inference} summarizes the inference process applied to QAE. We use vector notation to emphasize sequences. This does not provide a way of choosing $\vec{m}$, which is the main problem in iterative QAE. 

Our scheme relies on substancial enhancements as compared to Figure~\ref{fig:inference}, with the goal of choosing good values of $m$ while keeping processing costs down. Figure~\ref{fig:update} presents a high level view of our scheme.

For amplitude estimation, the likelihood model follows from equation (\ref{eq:p1}). The prior distribution $\mathbf{P}(\theta)$ encodes previous knowledge, which can simply mean enforcing the domain $a \in [0,1]$ via a uniform distribution with support limited to that region. Finally, the denominator can be calculated numerically. 

For all experiments in this work, we use the aforementioned uninformative prior, as the comparison to other algorithms would otherwise be unfair. Still, in general, this ability to consider pre-existing information can be advantageous; in some applications, the amplitude may, for instance, be known to reside within a smaller range of values. 

Expression (\ref{eq:bayes}) can be used to extract empirical knowledge from a dataset; the data can be considered sequentially or via a batch update. However, it does not provide a direct estimate of the parameter, but rather a function quantifying the merit of any parameter value. To obtain a numerical single point estimate, further processing is required. One possibility is to maximize the probability to find the mode. This is the approach of MLAE \cite{suzuki2020}. 

This was one of the first proposals for simplified amplitude estimation, and the performance is remarkable, even when the circuits are chosen based on uncomplicated pre-determined heuristics. This approach is simple, parallelizable, and can easily accommodate noise models. However, it has some downsides. First, it falls short of Heisenberg-limited scaling, and offers no straightforward way of improving performance. Second, it does not scale well with the model complexity. Complex noise models with many parameters live in high dimensional spaces, where optimization does not fare well.

One alternative is to find the mean, which requires integrating over the distribution. Integration methods are discussed in Appendix~\ref{app:integration}. But before, we note that this extra step arises due to one of Bayesian inference's strongest features. The mean is just a particular case of an expected value we can compute based on data. More generally, we can calculate the expectation for any function of the parameter(s),

\begin{equation}
    \label{eq:expectation}
    \mathbb{E}_{\mathbf{P}( \theta)}\left[f(\theta)\right]
    = \int f(\theta) \mathbf{P}(\theta)\mathrm{d}\theta.
\end{equation}

Doing $\mathbf{P}( \theta) = \mathbf{P}( \theta \mid D)$ gives us posterior expectations. We can then obtain measures of uncertainty, like the variance, which are useful for quantifying our confidence in the final results. But the usefulness goes beyond that: we can evaluate utility functions in a \textit{look-ahead} scheme. For instance, we can estimate how much a given experimental control, or sequence thereof, is expected to reduce the variance. We can thus sweep over potential experimental controls and choose the ones expected to produce better results; for instance, a lower uncertainty, or higher information gain. This is called Bayesian experimental design, and can be used to attain Heisenberg-limited estimation \cite{higgins2007}. 

In Figure~\ref{fig:inference}, it would take the form of an \textit{a priori} optimization of $\vec m$, the sequence of controls, considering expectations over the prior. However, the quality of the predictions is heavily dependent on the accuracy of the distribution; that is, how close it is to reality (a Dirac delta function on the real value). Thus, this choice may be unsatisfactory, especially in the case of uninformative priors. 

Instead, we can unfold the inference process into several steps, considering intermediary distributions $\mathbf{P}( \theta \mid D_0)$, $\mathbf{P}( \theta \mid D_0, D_1)$, and so on. The freedom to do so stems from the structure of Bayes' rule: one may do a bulk update based on a complete dataset, or consider the data incrementally. In the second case, the posterior of one iteration becomes the prior for the next, allowing empirical knowledge to inform the latter's controls. The inference process can then occur \textit{online} and adaptively, exploiting all available information at each point. There are additional benefits to this approach (Appendix~\ref{app:expdesign}).

We adopt this adaptive approach for QAE, as reflected in Figure~\ref{fig:update}. We consider a single datum per iteration, but the process can be done with arbitrary granularity, dividing the measurements into blocks of several and considering a chunk of data per step. There are multiple trade-offs involved. Such details on the experimental design are discussed in Appendix~\ref{app:expdesign}, where we also introduce problem-tailored tricks to cut back on the optimization cost.  

In this way, we can guide the inference process according to the most up-to-date knowledge. In BAE, noting that optimization cost per measurement is considerable, and bound to yield smaller returns in the initial measurements due to the minimal accumulated knowledge, we choose to start this adaptive optimization only after an initial phase. In this initial phase, we perform \textit{wNshots} non-amplified ($m=0$) measurements. We call this phase a \textit{classical warm-up}. Once it is completed, we begin the adaptive phase, having accrued information that will make it more gainful, before it becomes increasingly so as the learning proceeds. 

This finalizes the main outline of the algorithm whose pseudo-code is provided in algorithm \ref{alg:adaptive_inference}. The estimator is a construction mapping the posterior to an estimate - such as the mean, the mode or the median.
\section{Integration Methods}
\label{app:integration}

In section \ref{sec:BAE} we discussed the strengths of Bayesian inference, and how they are tied to the probabilistic depiction of information. It remains to discuss the specifics of this depiction, which are determinant for the realization of such strengths.

In the general case, expression (\ref{eq:bayes}) yields an arbitrary distribution, meaning that an exact representation is not tractable. The authors of \cite{wang2021} work under a Gaussian assumption that remedies this. The downside of such an approach is that the limited expressibility may lead to the inference results' being wrongly captured, skewing the experimental choices and the final estimates in the wrong direction. While this may not be apparent in simple test cases, it becomes so once the models are sufficiently intricate. 

Such a scenario is likely to occur when pursuing a thorough noise characterization, the simplicity of QAE's fundamental model notwithstanding. Capturing noise accurately requires augmenting the model with additional parameters; the higher the dimension of the parameter space, the harder it is to explore, and the more unforgiving towards simplistic methods.  This is related to what is known in statistics as the curse of dimensionality \cite{bellman2003}. For instance, high dimensional spaces are prone to extrinsic redundancy, which undermines the normality assumption that Gaussian approximations rest upon. 

This extreme example is only one of several ways an inadequate representation method may compromise the inference efforts. More nuanced oversights are even more likely to arise, misguiding the experimental design and jeopardizing accuracy. Such shortcomings cannot be overcome even by a perfect generative model for the noisy data. 

In short, the statistical backbone plays a pivotal role in the inference. It is thus critical that it be refined along with its other aspects, so as not to disrupt the efforts invested in them. 

The simplest general alternative would be to evaluate the distribution on a finite grid. However, this is blatantly inefficient. As the inference process advances, we expect the distribution to peak around the real parameter. This will mean that most - eventually all - grid points will have vanishing probability.

Another option would be Markov Chain Monte Carlo (MCMC) \cite{Andrieu2003,speagle2020,betancourt2018}. This method constructs a Markov sequence whose states converge to the intended probability distribution, in this case the posterior $\mathbf{P}(\theta \mid D)$. The samples can then be used to perform Monte Carlo integration.  

MCMC overcomes the pitfall of grids, as the samples naturally concentrate in the relevant region of parameter space. However, it presents several drawbacks: it is ill-suited for adaptivity, lacks robustness against multi-modality, and is computationally intensive.

We use Sequential Monte Carlo (SMC), a highly parallelizable method that combines a grid and MCMC to address the shortcomings of each \cite{delmoral2006, doucet2013, liu2001, neal1998, south2019, south2019, daviet2016}. This type of sampler has been applied to a multitude of tasks in quantum metrology \cite{granade2012, gentile2021, wiebe2016a,wang2017, santagati2018}, although using the Liu-West filter (LWF) \cite{liu2001} instead of MCMC. The underlying idea is to rely on a grid, but to refresh the point locations by resampling (with e.g. LWF or MCMC) when necessary. Apart from being adaptive and efficient, this method has the added benefit of computing the model evidence at barely any extra cost. This is a valuable tool for model comparison and selection, and is not directly provided by MCMC. In the simulations realized in this work, we use MCMC rather than LWF, as it is more robust and provides more accurate expectations, even e.g. under multi-modality.

Details and pseudo-code for these methods are given in Appendix~\ref{app:SMC}.

\subsection{Sequential Monte Carlo}
\label{app:SMC}

Pseudo-code for a Sequential Monte Carlo update is presented in algorithm \ref{alg:SMC_update}. Note that normalization is not necessarily required after each update, as it is only required when calculating quantities (algorithm \ref{alg:expectation}). These algorithms are one possible solution for the unspecified functions in algorithm \ref{alg:adaptive_inference}; in this case, distributions are represented by sets of points  $\vec{\theta}$ (often called particles in SMC literature) and their respective weights, $\vec{w}$. For details on the hyperparameters \texttt{Npart}, \texttt{thr}, refer to Appendix~\ref{app:hyperparameters}.

\begin{figure*}
    \centering
    \begin{minipage}{0.9\textwidth}
        \begin{algorithm}[H]
\caption{Algorithm for updating a Sequential Monte Carlo distribution.}\label{alg:SMC_update}
\begin{algorithmic}[1]
\Parameters \texttt{Npart}, \texttt{thr} \Comment{Number of particles, resampling threshold.} 
\Require $\vec{\theta}$, $\vec{w}$, $\vec{m}$, $\vec{o}$ \Comment{Particle positions and weights (length \texttt{Npart}), controls, outcomes.} 
\Ensure  $\vec{\theta}$', $\vec{w}$' \Comment{Updated positions and weights.} 
\Function{batch\_bayesian\_update}{$\vec{\theta}$, $\vec{w}$, $\vec{m}$, $\vec{o}$}
    \State \Getst {N} {\Callt{length}{$\vec{m}$}} 
    \For{$i \in \{0, \dots, \ttt{N}\}$}
        \State \Getst{$\vec{\theta}$, $\vec{w}$} \Call{bayesian\_update}{$\vec{\theta}$, $\vec{w}$, \ttt{$\vec{m}$}[i], $\vec{o}$[i]}
    \EndFor
    \State \Return $\vec{\theta}$, $\vec{w}$
\EndFunction
\Function{bayesian\_update}{$\vec{\theta}$, $\vec{w}$, m, o}
    \For{$i \in \{1, \, \dots, \texttt{Npart}\}$}
        \State \Gets {\vec{w}[i]} $ \vec{w}[i]* \mathcal L(\vec{\theta}[i] \mid \ttt{o}; \texttt{m})$ \Comment{Bayes' rule.}
    \EndFor
    \State \Gets {\text{ESS}}  $\left( \sum_{i=1}^N  \ttt{w[i]} \right)^2 \bigg/
    \left( \sum_{i=1}^N \ttt{w[i]}^2 \right)$ \Comment{Effective sample size.}
    \If{\Call{ESS}{$\vec{w}$} $>$ \texttt{thr}}
        \State \Getst{$\vec{\theta}$', $\vec{w}$'} $\vec{\theta}$, $\vec{w}$
    \Else
        \State \Getst{$\vec{\theta}$, $\vec{w}$} \Call{resample}{\ttt{$\vec{\theta}$, $\vec{w}$}}  
    \EndIf 
    \State \Return \ttt{$\vec{\theta}$'}, $\vec{w}$'
\EndFunction
\end{algorithmic}
\end{algorithm}
\end{minipage}
\end{figure*}

\begin{figure*}
    \centering
    \begin{minipage}{0.9\textwidth}
        \begin{algorithm}[H]
\caption{Algorithm for resampling an SMC distribution.}\label{alg:resample}
\begin{algorithmic}[1]
\Parameters \texttt{kernel} \Comment{Function to perturb particles.}
\Require $\vec{\theta}$, $\vec{w}$, \ttt{info} \Comment{Particle positions and weights, information for the kernel.} 
\Ensure  $\vec{\theta}$', $\vec{w}$' \Comment{Updated positions and weights.} 
\Function{resample}{$\vec{\theta}$, $\vec{w}$}
    \State \Getst N \Callt{length}{$\vec{\theta}$}
    \State \Gets {\vec{\theta}'} empty list
    \For{$i \in \{1, \, \dots, N\}$}
        \State \Gets \theta $\vec{\theta}$[j] sampled with probability $\vec{w}$[j] \Comment{Multinomial sampling.}
        \State \Gets {\theta'} \Callt{kernel}{$\theta$, \ttt{info}}
        \State append $\theta'$ to \ttt{$\vec{\theta}$'}
    \EndFor
    \State \Gets {\vec{w}'} array of ones with length N
    \State \Return $\vec{\theta}$', $\vec{w}$'
\EndFunction
\end{algorithmic}
\end{algorithm}
\end{minipage}
\end{figure*}








{\begin{figure*}
    \centering
    \begin{minipage}{0.9\textwidth}
        \begin{algorithm}[H]
\caption{Algorithm for calculating expectation values given a Sequential Monte Carlo representation.}\label{alg:expectation}
\begin{algorithmic}[1]
\Require $\vec{\theta}$, $\vec{w}$, \ttt{f} \Comment{Particle positions and weights, and a function.} 
\Ensure  E \Comment{Expectation value of \ttt{f}.}
\Function{expectation}{$\vec{\theta}$, $\vec{w}$, \ttt{f}} 
        \State \Getst S  $\sum_{i}  \vec{w}$[i] \Comment{Normalization factor.}
        \State \Getst E E + $\sum_i$ \ttt{f}($\vec{\theta}$[i]) * $\vec{w}$[i]
        \State \Return E/S
\EndFunction
\end{algorithmic}
\end{algorithm}
\end{minipage}
\end{figure*}

\begin{figure*}
    \centering
    \begin{minipage}{0.9\textwidth}
        \begin{algorithm}[H]
\caption{Algorithm for calculating average expectation values given a Sequential Monte Carlo representation.}\label{alg:average_expectation}
\begin{algorithmic}[1]
\Require $\vec{\theta}$, $\vec{w}$, \ttt{f}, $m$ \Comment{Particle positions and weights, a function, and a potential control.} 
\Ensure  E \Comment{Average expectated value for \ttt{f} for $m$.}
\Function{average\_expectation}{$\vec{\theta}$, $\vec{w}$, \ttt{f}, $m$}
    \State \Getst E 0
    \For{each possible outcome \ttt{o}}
        \State \Getst P \Call{expected\_probability}{$\vec{\theta}$, $\vec{w}$, $m$,\ttt{o}}
        \State \hspace{-.5em} \Getst E $\ttt{E} + \Call{cond\_expectation}{\vec{\theta}, \vec{w}, \ttt{f}, $m$,\ttt{o}} * \ttt{P}$
    \EndFor
    \State \Return E
\EndFunction

\Function{cond\_expectation}{$\vec{\theta}$, $\vec{w}$, \ttt{f}, \ttt{o},  $m$} 
    \State \Getst {$\vec{\theta}$', $\vec{w}$'} \Call{bayesian\_update}{$\vec{\theta}$, $\vec{w}$, $m$, \ttt{o}}
    \State \Getst {cE} \Callt{expectation}{$\vec{\theta}$', $\vec{w}'$, f}
    \State \Return \ttt{cE}
\EndFunction

\Function{expected\_probability}{$\vec{\theta}$, $\vec{w}$, $m$,\ttt{o}}
    \State \Getst S  $\sum_i \vec{w}[i]$ \Comment{Normalization factor.}
    \State \Getst P 0
    \For{$i \in \{1, \, \dots, N\}$}
        \State \Getst P P + $\vec{w}[i] * \mathcal L(\vec{\theta}[i] \mid \ttt{o}; $m$)$
    \EndFor
    \State \Return P/S
\EndFunction
\end{algorithmic}
\end{algorithm}
\end{minipage}
\end{figure*}}



\subsection{Liu-West filter}

The Liu-West filter is a resampling kernel that preserves the two first moments of the distribution \cite{liu2001}. Pseudo-code is presented in algorithm \ref{alg:LW}. It may perform well enough for simple problems, while being more lightweight than Markov Chain Monte Carlo. 

\begin{figure*}
    \centering
    \begin{minipage}{0.9\textwidth}
        \begin{algorithm}[H]
        \caption{Algorithm for a Liu-West resampling kernel.}\label{alg:LW}
        \begin{algorithmic}[1]
        \Parameters {$\alpha$}
        \Require $\theta$, \ttt{mean}, \ttt{std} \Comment{Old particle location and first moments of the SMC distribution.} 
        \Ensure  $\theta'$ \Comment{New particle location.} 
        \Function{LiuWest}{$\theta$, mean, std}
            \State \Gets \mu $\alpha * \theta + (1-\alpha) * \ttt{mean}$
            \State \Gets \sigma $\sqrt{1-\alpha^2} * \ttt{std}$
            \State $\theta'$ $\sim \mathcal N(\mu, \sigma)$ \Comment{Gaussian sampling.}
            \State \Return $\theta'$
        \EndFunction
        \end{algorithmic}
        \end{algorithm}
\end{minipage}
\end{figure*}

\subsection{Markov chain Monte Carlo}
Markov chain Monte Carlo can be used as a resampling kernel that asymptotically preserves the distribution \cite{betancourt2018}. We use random walk Metropolis, where the proposals are sampled from a Gaussian distribution centered at the previous particle location. When using MCMC within SMC, the particle set can be used to inform the proposal distribution \cite{ramoa2024}. The ideal proposal variance for Metropolis is proportional to the target distribution variance ($2.38/\sqrt d$, where $d$ is the dimension) \cite{Gelman__1996}, which is estimated at each step by the current SMC distribution. We thus use the variance of the SMC distribution to tune the variance of the Metropolis proposals. 

Pseudo-code for random walk Metropolis is presented in algorithm \ref{alg:Metropolis}. Note that unlike the Liu-West filter, which requires only the SMC particle locations and weights, this resampler needs to access the full dataset. This is the cost associated with the full preservation of the distribution.

\begin{figure*}
    \centering
    \begin{minipage}{0.9\textwidth}
        \begin{algorithm}[H]
        \caption{Algorithm for a Markov (Metropolis) resampling kernel.}\label{alg:Metropolis}
        \begin{algorithmic}[1]
        \Parameters k
        \Require $\theta$, $\vec D$ \Comment{Old particle location and complete dataset.} 
        \Ensure  $\theta'$ \Comment{New particle location.} 
        \Function{Markov}{$\theta$, $\vec D$}
            \State $\theta' \sim \mathcal N(\theta, k)$
            \State \Getst p $\frac{\mathcal L(\theta \mid \vec D)}{\mathcal L(\theta' \mid \vec D)}$
            \Indent \textbf{with} probability $1-\ttt{p}$ \textbf{do}
                 \State \Getst {$\theta'$} $\theta$ \Comment{Reject new location.}
            \EndIndent
            \State \Return $\theta'$
        \EndFunction
\end{algorithmic}
\end{algorithm}
\end{minipage}
\end{figure*}

\subsection{Calculating the model evidence}
\label{subapp:model_evidence}

One may be interested in calculating the model evidence, given by:

\begin{equation}
    \mathbf{P}(\vec{D}) = \int 
    \mathbf{L}(\theta \mid \vec{D})
    \mathbf{P}(\theta) \mathrm{d}\theta.
\end{equation}

This works as a a Bayesian quantification of the merit of a model, which can be used for model selection and tuning. One interesting perk of SMC is that it allows one to calculate this quantity at a small added cost.

Let's say that we use SMC to sample from a sequence of probability distributions of length $T+1$ given by:
\begin{equation}
    \mathbf{P}_t(\theta) = 
    \frac{\eta_t(\theta)}{Z_t} \quad, \ t \in \{0,\dots,T\},
\end{equation}

\noindent having defined the normalizing constant:
\begin{equation}
    Z_t \equiv \int \eta_t(\theta) \mathrm{d}\theta.
\end{equation}

The reweightings of SMC will abide by:
\begin{equation}
    w_i^{(t)} \propto w_i^{(t-1)}* 
    \frac{\eta_{t}(\theta_i^{(t)})}
    {\eta_{t-1}(\theta_i^{(t)})},
\end{equation}

\noindent since the $Z_t$ constants are rendered irrelevant by normalization. More explicitly,
\begin{gather}
    \label{eq:reweighting_eta}
    W_i^{(t)} = w_i^{(t-1)}* 
    \frac{\eta_{t}(\theta_i^{(t)})}
    {\eta_{t-1}(\theta_i^{(t)})}\\
    w_i^{(t)} = \frac{W_i^{(t)}}{\sum_{j=1}^M W_j^{(t)}}.
\end{gather}

With this, expectations can be computed at each step as:
\begin{gather}
    \label{eq:expectation_eta}
    \mathbb{E}_{\mathbf{P}_t(\theta)}\left[f(\theta)\right]
    = \int f(\theta) \mathbf{P}_t(\theta)\mathrm{d}\theta
    \nonumber \\
    \approx \sum_{i=1}^{M} f(\theta_i^{(t)}) * w_i^{(t)}.
\end{gather}

If a resampling stage does not take place at the end of the $t$-th iteration,the $(t+1)$th particle set is unchanged: $\{\theta_i^{(t+1)}\}_{i=1}^M=\{\theta_i^{(t)}\}_{i=1}^M$. If on the contrary a resampling mechanism is activated, $\{\theta_i^{(t)}\}_{i=1}^M$ is recast to get $\{\theta_i^{(t+1)}\}_{i=1}^M$. This preparation aims to introduce variety for the following iteration, and preserves $\mathbf{P}_t(\theta)$ (or a selection of relevant attributes). 

Expectation \ref{eq:expectation_eta} can be rewritten as:
\begin{equation}
    \label{eq:expectation_eta_res}
    \mathbb{E}_{\mathbf{P}_t(\theta)}\left[f(\theta)\right]
    \approx \sum_{i=1}^{M} f(\theta_i^{(t+1)}) * w_i^{(t)}.
\end{equation}

Our interest lies in the normalizing constants $Z_t$. If the final distribution $P_T(\theta)$ is the posterior $\mathbf{P}(\theta \mid \vec{D})$, $Z_T$ is the marginal, or the model evidence.

At the other end, $Z_0$ is just the integral of the prior, which is 1 for any valid distribution:
\begin{equation}
    Z_0 = \int \mathbf{P}(\theta) \mathrm{d}\theta = 1.
\end{equation}

Thus we can write:
\begin{equation}
    \mathbf{P}(\vec{D}) = \frac{Z_T}{1} = \frac{Z_T}{Z_0},
\end{equation}

which in turn can be written as:
\begin{equation}
    \label{eq:ratio_evidence}
    \mathbf{P}(\vec{D}) = \frac{Z_T}{Z_0} = \prod_{t=1}^{T} \frac{Z_{t}}{Z_{t-1}}.
\end{equation}

\noindent, since all the intermediate terms cancel out when expanding.

Now we need to get the individual terms, or some estimate thereof:
\begin{equation}
    \widehat{\left( 
    \frac{Z_{t}}{Z_{t-1}}
    \right)}.
\end{equation}

This is written in terms of two consecutive distributions. Developing the expression further, we can write the term corresponding to the $t$-th iteration as an average over the $(t-1)$th distribution:
\begin{gather}
    \left( 
    \frac{Z_{t}}{Z_{t-1}}
    \right)
    = \frac{\int \eta_t(\theta) \mathrm{d}\theta}{Z_{t-1}}
    = \int \frac{\eta_t(\theta)}{Z_{t-1}} \mathrm{d}\theta
    \nonumber \\
    = \int \frac{\eta_t(\theta) }{\eta_{t-1}(\theta)}
    * \frac{\eta_{t-1}(\theta)}{Z_{t-1}} \mathrm{d}\theta
    \nonumber \\
    = \int \frac{\eta_t(\theta)}{\eta_{t-1}(\theta)}
    \mathbf{P}_{t-1}(\theta)\mathrm{d}\theta.
\end{gather}

At this point we fall back on \ref{eq:expectation_eta_res},
\begin{gather}
    \left( 
    \frac{Z_{t}}{Z_{t-1}}
    \right)
    = \int \frac{\eta_t(\theta)}{\eta_{t-1}(\theta)}
    \mathbf{P}_{t-1}(\theta)\mathrm{d}\theta
    \nonumber \\ 
    \approx \sum_{i=1}^{M} \frac{\eta_t(\theta_i^{(t)})}
    {\eta_{t-1}(\theta_i^{(t)})} 
    * w_i^{(t-1)}.
\end{gather}

Recognizing \ref{eq:reweighting_eta}, this becomes:
\begin{equation}
    \label{eq:ratio_estimator}
    \left( 
    \frac{Z_{t}}{Z_{t-1}}
    \right)
    \approx \sum_{i=1}^{M}  W_i^{(t)}.
\end{equation}

The finalized estimate is obtained by plugging \ref{eq:ratio_estimator} into \ref{eq:ratio_evidence}.
\begin{equation}
    \label{eq:marginal_smc}
    \mathbf{P}(\vec{D}) 
    \approx \prod_{t=1}^{T} \left( \sum_{i=1}^{M}  W_i^{(t)} \right).
\end{equation}

\section{Experimental design}
\label{app:expdesign}

As mentioned in section \ref{sec:BAE}, a Bayesian distribution can be used to choose the best measurements to be performed. More specifically, we can use it to evaluate how informative a potential experiment (or sequence thereof) is given available knowledge, and thus optimize it.  It does so by considering the possible scenarios that could result: how likely they are, and how useful they would be.  This ability can be used to optimize the control $m$, by choosing the one with the highest expected utility. In this work, we choose to minimize the expected variance. The target function is called a utility function. Other utility functions may be considered, such as the expected information gain \cite{rainforth2024}.

Let us consider a prospective control $m$ whose benefit we want to assess. That is, we want to forecast how executing the associated circuit would improve our knowledge. We may be at any stage of the inference process, and denote the current distribution $\mathbf{P}(\theta)$, which may consider previous data.

We begin by computing the utility conditional on each possible outcome $D$ for control $m$, by doing a hypothetical Bayesian update on the current distribution (equation \ref{eq:bayes}) and then calculating the expectation of the utility function over the resulting distribution (equation \ref{eq:expectation}). 

We then calculate the probability of each outcome, by integrating the likelihood over the distribution,

\begin{equation}
    \label{eq:outcome_prob}
    \mathbb{E}_{\mathbf{P}(\theta)}\left[P(D;m)\right]
    = \int \mathbf{L}(\theta \mid D; m) \mathbf{P}(\theta)\mathrm{d}\theta,
\end{equation}

since $\mathbf{L}(\theta \mid D; m) = \mathbf{P}(D \mid \theta; m)$ by definition. 

Finally, we calculate a weighted average of the former according to the latter,

\begin{equation}
    \label{eq:expected_utility}
    U(m) = \sum_{D}  \mathbb{E}_{\mathbf{P}(\theta)}\left[P(D;m)\right]
    *
    \mathbb{E}_{\mathbf{P}(\theta \mid D;m)}
    \left[ 
    U(\theta,D;m)  .
    \right] 
\end{equation}

Here we suppress vector notation for simplicity, but both $D$ and $m$ can be sequences of arbitrary length.

\begin{figure}[!htb]
    \centering
    \includegraphics[width=\linewidth]{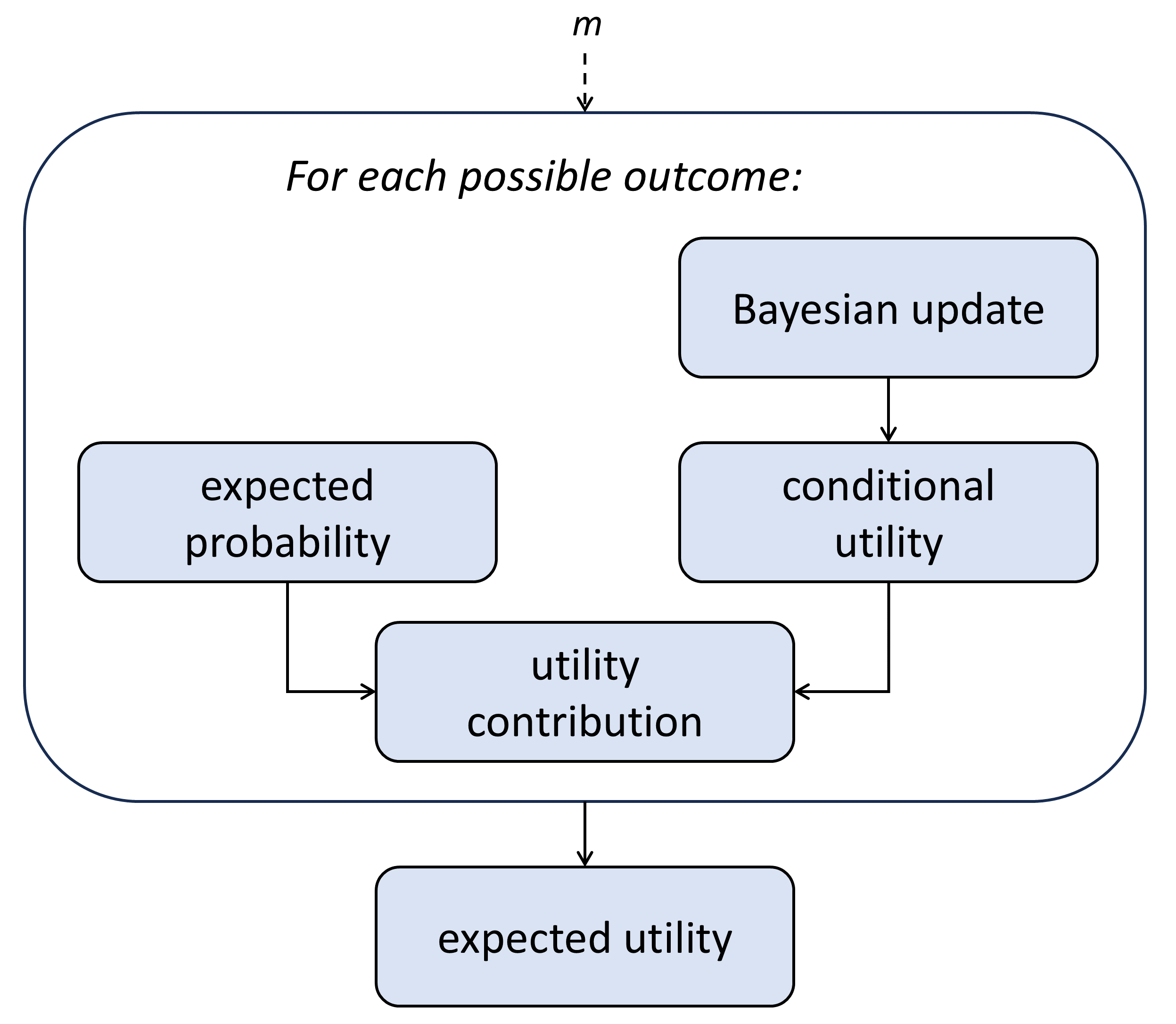}
    \caption{Diagrammatic representation of a utility calculation given an experimental control $m$. }
    \label{fig:utility}
\end{figure}

Figure~\ref{fig:utility} shows how to evaluate the utility for an experiment in this framework. Here the Bayesian update serves as an auxiliary calculation for the utility contribution of an outcome: we want to estimate its potential effect on the distribution, and weigh it according to the estimated probability of it happening. 

Being able to calculate the utility of a potential $m$, we can then optimize over a set of $M$ possible values of $m$ to choose the best option. Let us consider the case of Figure~\ref{fig:inference}, where we optimize from the outset the full sequence of controls based on the prior.  If one intends to do $N>1$ experiments, as is usually the case, the control to be optimized is a sequence of $N$ values for $m$, and "outcomes" are binary words of length $N$. 

The cost of evaluating the utility for a single sequence of controls $\vec{m}$ grows exponentially with $N$, since there are $2^N$ possible outcomes whose contributions must be calculated. Let $C$ represent the cost of a full iteration of the inference algorithm without optimization, consisting essentially of a Bayesian update. The cost of calculating the utility contribution of each outcome for each control is larger than $C$. Denoting by $M$ the number of possibilities for $m$ (integers), there are $M^N$ possible sequences $\vec{m}$, for each of which $2^N$ contributions must be calculated. As such, this optimization strategy will increase the algorithm's cost from $N*C$ to over $(M^N*2^N+N)C$. Clearly, the complexity of this approach is not favourable.

This is alleviated by a unitary look-ahead, adopting a locally optimal, or \textit{greedy}, strategy. The adaptive reformulation greatly simplifies the task, reducing the cost of evaluating the utility for a sequence of $N$ experiments to linear in $N$, $(M+1)*N*C$. 

A greedy approach is not guaranteed to be globally optimal, but has been shown to perform well \cite{ferrie2012}. Between the globally optimal and greedy strategies, there are multiple intermediate possibilities, where we consider a-aheads of length $L$. For instance, for $L=3$, we would consider the $2^3$ possible scenarios arising from three consecutive binary experiments. The optimization cost is exponential in the look-ahead size \cite{ramoa2024}; the expressions we derived are special cases for $L=1$ and $L=N$. 

The choice of a look-ahead size $L$ is ultimately a case-dependent negotiation between costs and benefits. The larger $L$, the greater the classical optimization cost (as the possible outcomes branch out exponentially); the greater the parallelism in the quantum component (since $L$ experiments can be performed at once); and the smaller the information usage, as the knowledge used to choose the controls for the second to $L$th experiments in a block is outdated the moment the first one is performed, and increasingly so as we advance in the block. In other words, a look-ahead of length $L$ will not make use of any of the $L$ data to inform any of the $L$ experiments (as they have been pre-determined).

This approach is then an adaptive one, requiring online processing: after each measurement, a Bayesian update must be performed considering the outcome, so that the following experimental control can be chosen based on all available information. We note that this type of adaptivity has been used to achieve quantum-enhanced precision in metrology, as an alternative to entangled state preparation \cite{higgins2007}. 

Even with this simplification where the cost of one utility evaluation does not scale exponentially, the cost of optimizing it over possible choices does. To curb this, we employ problem tailored heuristics to cut down the processing costs while retaining the ability to negotiate cost to benefit trade-offs. We do this by optimizing within an exponentially expanding window that grows adaptively to suit each particular execution. We choose the utility evaluations so that the cost does not depend on the window's width. Details are presented in subsection \ref{subapp:window}.

We additionally employ extra cost-sparing measures, whose effect is demonstrated in section \ref{sec:results}. First, the updates for the utility are not as critical as those of the main inference process. They may misguide effectiveness, but should not affect formal rigor. As such, we do not resample in the utility calculations. Second, one may perform multiple shot measurements for each optimized control, despite the greedy optimization. This further lowers the optimization costs. While this is not necessarily close to the optimum, good performance is observed in practice, while decreasing the classical costs and increasing the quantum parallelism. 

\subsection{Expanding window strategy}
 \label{subapp:window}

In this subsection we present the problem-tailored strategy we use to adaptively define the search range for the optimization, drastically reducing optimization costs while preserving the quantum advantage. 

Choosing the optimal control can be done in a window. The simplest method would be brute force optimization over all possible controls up to a high maximum value. As mentioned in section \ref{sec:BAE}, the utility of each can be calculated in a look-ahead fashion, after which the one with the maximal value is chosen. The utility function could be the negative variance, the expected information gain, or other. 

However, this process is computationally expensive. Evaluating the utility for a single experiment has twice the cost of an entire iteration otherwise, plus the cost of integrating twice over the distribution.

Furthermore, it is expected that the experimental control scales exponentially with the iteration number for an optimal strategy \cite{ferrie2012, suzuki2020}. Clearly, the search range must contemplate this optimum. As a result, the optimization cost will increase exponentially with the number of measurements to be performed. 

For this type of problem, the likelihood takes the form of a squared sinusoidal wave whose frequency increases with the experimental control - or equivalently, the number of queries/probing time. The higher frequencies achieved by large controls bring more informative data, in that the likelihood peaks are sharper. However, they also bring redundancy, due to the periodicity of the  measured function. 

As a result, the inference process should start with low controls that bring modest but unambiguous information, and use progressively higher ones as knowledge accumulates. This pacing is also beneficial for the Sequential Monte Carlo representation. Too abrupt changes in the successive distributions cause particle depletion and low effective sample sizes, i.e. poor statistical significance. In statistics, that is the primary motivation for the use of SMC, and the sequences of distributions are carefully fabricated for that purpose. Although is not the case here - metrological concerns being the main driver -, how smoothly the distribution evolves is still of concern.

With this in mind, we search for the optimal control in a dynamic window that follows the exponential increase. We periodically multiply the maximum of the window by a factor. This period could be fixed beforehand. However, the inference algorithm is probabilistic, and hard to pre-tune in such a rigid way. In some cases, the learning will be faster than average, in which case the pre-determined schedule will slow down the learning. In less lucky runs, it may be slower, and computational resources will be unnecessarily used to evaluate the higher controls.

We obtained better results using an adaptive criterion where the expansion schedule is tailored to each run. The key idea is to expand the search window when its maximum control has been chosen and used for a measurement. As the optimal probing time tends to be exponential with the acquired information, this signals that the algorithm has captured enough information that it would benefit from a window expansion, and thus we double the maximum. 

At the same time, we choose to increase the previous window minimum to the previous maximum. This will mean that phases separated by a window expansion never have overlapping controls. This further reduces the processing cost. 

In practice, this scheme may be aggressive, in particular due to the approximations in other domains of the algorithm. The largest control could sometimes be chosen too early due to particularities of the data and statistical representation. If that happens, the irreversible expansion may cause the algorithm to fail, creating redundancies that cannot be solved without access to lower experimental controls. 

On the other hand, the opposite issue might arise: due to numerical artifacts, the highest control may never be chosen, but rather e.g. the penultimate or some other.

To avoid both of these problems, we use a more robust trigger for the expansion: the chosen controls must be among the $R$ highest available options for $T$ iterations (not necessarily consecutive) ; $R$, $T$ are hyperparameters.

\begin{figure}[!htb]
    \centering
    \includegraphics[width=\linewidth]{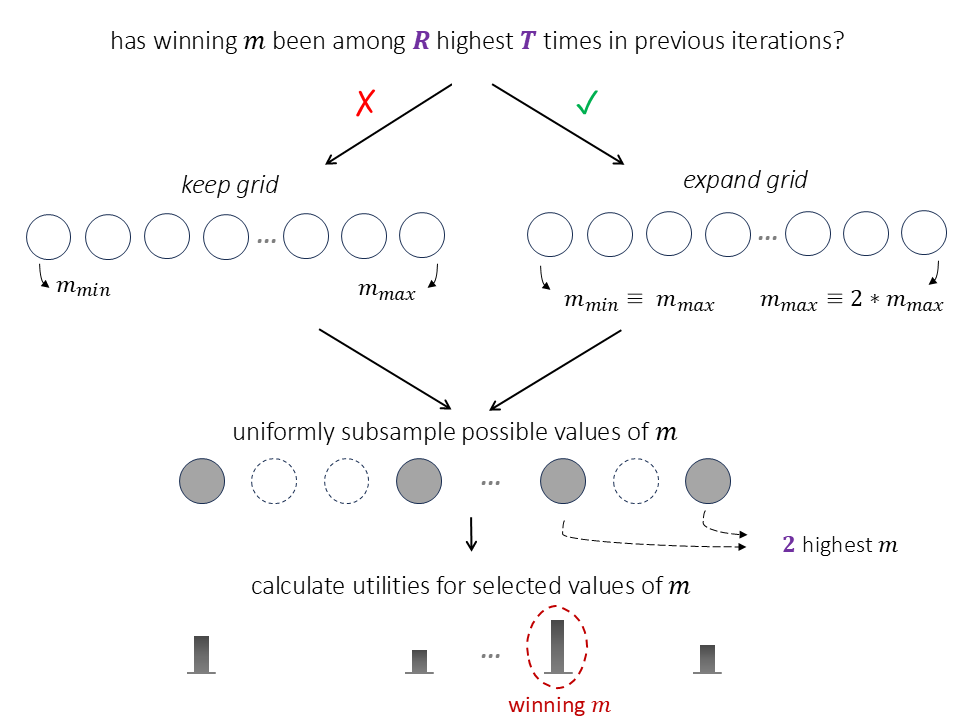}
    \caption{Grid expansion strategy for dynamic experimental design. Purple labels represent algorithm hyperparameters to be chosen by the user.}
    \label{fig:grid_exp}
\end{figure}

Figure~\ref{fig:grid_exp} illustrates one occurrence of this process. In the represented case, an expansion would be triggered for the next iteration if $R \geq 2$ and one of the highest $R$ subsampled values for $m$ had already been picked in $T-1$ previous iterations since the last expansion (not necessarily consecutive).

Using an expanding window reduce the optimization cost, but its exponential progression still holds, as the window width itself is expected to grow exponentially. To remedy this, we subsample a constant number of possible values of $m$ within the window. As the search range increases, we may conjecture that the importance of choosing the exact optimum (as compared to e.g. the second best) decreases. In the beginning we have only a few options to choose from, and the trade-off between sharpness and ambiguity is particularly strong due to the absence of insight. However, once we have acquired more information and have more options to choose from, we expect the specific choice to be less critical. 

This change accomplishes a constant cost per iteration: the optimization cost increases linearly - rather than exponentially - with the number of iterations/queries. In other words, the number of potential controls whose utilities are evaluated in each optimization is a constant \ttt{Nevals} throughout execution, regardless of the window expansions. This is crucial for the scalability of the algorithm. We determine the initial window maximum by multiplying \ttt{Nevals} by a constant $k_0$.

 As mentioned in section \ref{sec:BAE}, we opt for a greedy strategy (unitary look-ahead). This has two advantages: leveraging all available information and minimizing the optimization cost, which would increase exponentially with the size of the look-ahead.

 Pseudo-code for the discussed strategies is presented in algorithm \ref{alg:optimize_control}. Specifics of the utility calculations can be found in Appendix~\ref{app:SMC}.

\begin{figure*}
    \centering
    \begin{minipage}{0.9\textwidth}
        \begin{algorithm}[H]
\caption{Algorithm for optimizing an experimental control $m$.}\label{alg:optimize_control}
\begin{algorithmic}[1]
\Parameters \ttt{Nevals}, $k_0$, $R$, $T$, \ttt{utility\_fun}
\Require \ttt{d} \Comment{Probability distribution.} 
\Ensure $\ttt{m}_\ttt{opt}$ \Comment{Optimized experimental control.}
\Function{optimize\_control}{\ttt{d}}
    \State \textit{global } $m_\ttt{min}$, $m_\ttt{max}$ \Comment{Persistent across invocations.}
    \If {first call}
        \State \Gets {m_\ttt{min}}{$0$}
        \State \Gets {m_\ttt{max}} {$k_0*\ttt{Nevals}$}
    \ElsIf {controls $\ttt{m}_\ttt{opt}$ of prior iterations among $R$ highest $T$ times} \Comment{See text.} 
        \State \Gets {m_\ttt{min}}{$m_\ttt{max}$}
        \State \Gets {m_\ttt{max}}{$2*m_\ttt{max}$}
    \EndIf 
    \State \Getst {grid} \ttt{Nevals} samples from $\{m_\ttt{min}, ..., m_\ttt{max}\}$ 
    \State \Getst {utilities} [\Call{average\_expectation}{\ttt{d}, \ttt{utility\_fun}, \ttt{m}} for \ttt{m} in \ttt{grid}]	
    \State \Getst {$\ttt{m}_\ttt{opt}$} \Call{argmax}{\ttt{utilities}}
    \State \Return $\ttt{m}_\ttt{opt}$
\EndFunction
\end{algorithmic}
\end{algorithm}
\end{minipage}
\end{figure*}

The \textit{else if} case is the case where a window expansion is triggered. The condition "controls of prior iterations among $R$ highest $T$ times" falls outside the scope of this function, as it refers to information from previous invocations. That is, if in $T$ of all previous invocations of the function  \texttt{optimize\_control} since the last window expansion the returned value was among the top $R$ possibilities, this case applies and the search window is enlarged. 

For details on the hyperparameters involved in the experimental design process, refer to Appendix~\ref{app:hyperparameters}.
\section{Noise tailoring}
\label{app:noise}

One advantage of the Bayesian approach is its natural ability to accommodate noise models by adapting the likelihood function. This brings the ability to study the undesirable properties alongside the parameters of interest. As the former undermine the characterization of the latter, this offers a largely customizable path to robust characterization, without any structural modifications to the algorithm. On the one hand, we can account for the fact that the system's behavior is affected by confounding factors, and correct the estimates accordingly. On the other, we are in a position to pursue realistic measurement selection under less than ideal conditions. 

Both of these advantages are inhibited by optimization based approaches. While they technically still allow for modeling at least static noise, they are ill-equipped to deal with high-dimensional probability distributions. Optimization blindly seeks probability density without heeding volume; whereas it is their product, the probability mass, what best conveys information \cite{betancourt2018, ramoa2024}. As the parameter space's dimension grows, volume moves away from the highest probability region(s), rendering their contribution to the distribution's behavior negligible. Integration weighs the density according to volume, making it a more sound approach.

Secondly, optimization restricts adaptivity. In \cite{suzuki2020}, heuristic pre-defined measurements are used instead, but they do not attain the Heisenberg limit. An adaptive strategy could be adopted where the interim mode is found in between measurements and used to inform the following ones. However, this information is very limited, and can become unreliable for the reasons described above.  

To enjoy the flexibility, robustness and scalability of integration, it is vital that one employs efficient integration methods. Furthermore, to enable adaptivity, these methods must be compatible with sequential data processing. Appendix~\ref{app:integration} discusses this. 

Incorporating a description of noise into Bayesian inference requires a noise model.  We consider that the noise model is given, but it can be learned from the data \cite{gentile2021}. If the parameters are known, inference can proceed directly without significant modifications. If they are not, one must switch to a multi-parameter estimation regime, augmenting the likelihood model with these extra parameters. They can then be estimated alongside the amplitude, and dynamically adapt to the quantum device in real-time. 

Alternatively, if the noise does not have strong time fluctuations, it can be learned in a pre-processing phase. This can be done once for a given quantum device, and periodically calibrated. Even if there are strong time fluctuations, a pre-processing phase may be used to establish a prior for the estimation, which is then fine tuned and updated during the main estimation process. For some noise parameters, such as the coherence time, it may be possible to perform estimation independently from the amplitude, by observing the system decohere. Several phases of single-parameter estimation can be more efficient than one phase of multi-parameter estimation. 

In the simulations performed for this paper, we consider an exponential damping model for decoherence (subsection \ref{sub:qsim}), and a constant coherence time. BAE learns this time in a pre-processing phase as described above. The process is similar to the approach used to learn the amplitude. The main difference is that for amplitude measurements, longer evolutions (higher $m$) bring more informative but more ambiguous data, and thus $m$ tends to increase through the iterations. This is beneficial not only for the learning rate, but also for the numerical approximations, as our adaptive discretization has the opportunity to progressively focus on the correct region of the parameter space. For the coherence time estimation, the opposite occurs \cite{ramoa2024}. As such, we consider them in descending order when performing the Bayesian updates. Adaptive experimental design could also be applied to noise estimation, but we choose to use pre-determined uniformly spaced evolution times - between 0 and a provided maximum value for the coherence time. This maximum value is the only information that BAE requires to learn the coherence time, as a finite prior must be established. 
\section{Processing adaptive data}
\label{app:proc_adapt}

Previous QAE proposals present numerical results for a single execution and a single chosen amplitude  \cite{grinko2021, wie2019}. As discussed in section \ref{sec:methods}, to obtain a more robust benchmark, we instead average the results over randomized amplitudes. These averages must be taken over several runs for each data point. 

For algorithms with fixed amplification schedules, this is straightforward. Since these executions always use the same sequences of experiment controls and thus the same sequence of query numbers, we simply take the average of all the estimation errors for each query number. For any query number, we are sure to have as many summands as there are runs.

However, most algorithms we have analyzed, including BAE, have adaptive amplification schedules. This complicates the averaging, as the sequences of query numbers are bound to vary between runs. What is more, they tend to increase exponentially, which means that the data will become exponentially sparse as the iteration number grows. Owing to this, it would be difficult to obtain averages by brute force. That is, we could run many simulations and hope that for many distinct and spaced out numbers of queries, there exist coincidental occurrences in a significant proportion of these runs. We could then represent the data points corresponding to these lucky query numbers for which we have sufficient information. However, this is extremely inefficient, especially for large iteration numbers; note that in the algorithms we have seen, the number of queries grows exponentially with the iterations. 

We thus seek alternative processing methods. Ideally, we would make use of all the data, condensing it into a handful of representative data points. The aim is for the resulting graph to relay the underlying tendency correctly and intelligibly.

A sensible first step is to bin the data, splitting them into groups depending on their numbers of queries. The remaining question is how to process each subset of data to get a single point representing the bin in a way that does not over or under-estimate the learning rate, nor introduce artifacts that hinder visual analysis.

To test possible approaches without having to actually run simulations, which require longer processing and may introduce confounding factors, we generated dummy data observing the Heisenberg Limit. We did so as follows:
\begin{enumerate}
    \item Sample an $x$ (representing $N$) coordinate uniformly at random on a log scale. 
    \begin{equation}
        x \sim \text{unif}_{\log} ([x_\text{min}, x_\text{max}])
    \end{equation}
    \item Sample auxiliary variables $z$ depending on $x$: 
    \begin{equation}
        \label{eq:auxvars}
        z \sim \mathcal N (\mu, \sigma(x)),
    \end{equation}
    where 
    \begin{equation}
        \sigma(x) \propto 1/x.
    \end{equation}
    The choice of the mean $\mu$ is arbitrary, whereas the constant of proportionality is determined by fixing a point.
    \item Calculate $y$ (representing $\epsilon^2$) as $y=(x-\mu)^2$.
\end{enumerate}

Clearly, points thus generated will emulate ideal Heisenberg-limited behavior: 
\begin{equation}
    \mathbb{E} (y) = \mathbb{V}(x) \propto 1/x^2
\end{equation}

The goal is to find a processing strategy that will portray them accordingly.

Alternatively to steps 2-3, we could assign $y=\sigma(x)$ to get “noiseless” points standing directly on the Heisenberg limit. This does not mimic the behavior of actual Heisenberg-limited estimation, to which randomness is intrinsic. It is merely meant to aid the reasoning.

\begin{figure}[!htb]
    \centering
    \includegraphics[width=.8\linewidth]{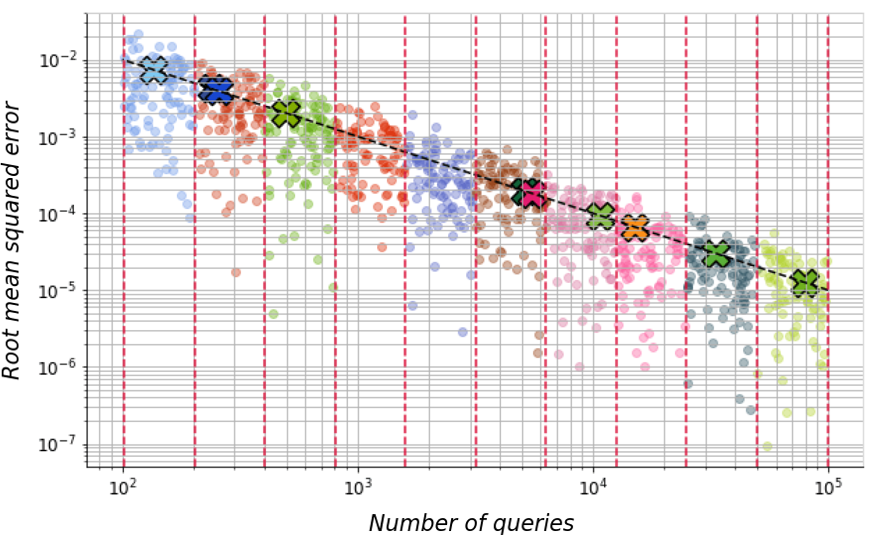}
    \caption{Graphical representation of "Heisenberg-limited" dummy data. Dashed lines represent bin edges. The 'x' markers represent "noiseless" points, which lie on the Heisenberg limit (dashed diagonal line)}
    \label{fig:dummy_HL}
\end{figure}

A plot of data generated like so should resemble Figure~\ref{fig:dummy_HL}.

 Now, if instead of each bin containing scattered points we had multiple points with the same $x$, we could simply average their $y$ coordinates to get a well-behaved representation. However, since the $x$ coordinate is distributed on a continuum, only with vanishing probability will any two points get the same $x$. 
 
 In this case, the naive approach would be to average their $x$ and $y$ coordinates independently, but this does not seem justified. We need only think of points lying on a straight line. Clearly, averaging over $y$ for constant $x$ will produce another point on the line; but averaging over each coordinate will not.

We tried subjecting the groups (datapoints in each bin) to the following treatments:

\begin{enumerate}
    \item \textbf{Independent $x$/$y$ averaging.} Average their $x$ coordinates and their $y$ coordinates separately - the naive approach. After averaging, take the square root of $\bar y$ to get a quantity akin to the estimation error (RMSE, or in this case standard deviation). Plot $\bar x, \sqrt{\bar y}$.
    \item \textbf{Average "log-"coordinates.} Same as the previous point, but doing the averages in the log-domain, i.e. taking logarithms before averaging, then exponentiating the averages.
    \item \textbf{Average slopes.} Average their $x$ coordinates to get $x$. Separately, computing the `` log-slope'' of each point in the group relative to the fixed point, and take the average slope $\bar m$. Compute the image  of the mean $x$ under the power function given the average slope, and the fixed point: $f_{\bar m}(\bar x)$.
\end{enumerate}

The reason for trying the two latter options is the intuition that in a noiseless case the processed points should all exactly overlapped with a line, as do the original ones. The obvious way to fix it is to work in the log domain, since an affine function $f(x)=mx+b$ applied to an average,  $f(\bar x)$, is the average of the images of the individual elements, $\overline{f(x)}$. E.g., $f((x_1+x_2)/2)=m(x_1+x_2)/2+b=[(mx_1+b)+(mx_2+b)]/2$. 

We further tried two strategies without prior binning: 
\begin{enumerate}
    \setcounter{enumi}{3}
    \item \textbf{Curve fitting}.
    \item \textbf{Spline interpolation}.
\end{enumerate}

The motivation for these is clear: they are standard data treatment strategies. In this case the results are functions, which can then be evaluated at the bin midpoints to produce evenly spaced sample points comparable to those of the other strategies.

\begin{figure*}[p]
\centering
\captionsetup[sub]{font=footnotesize, skip=2pt}
\begin{subfigure}[t]{0.42\textwidth}
  \includegraphics[width=\linewidth]{figs/11a.average_xy_noisy.png}
  \captionsetup{font=footnotesize, skip=2pt}
  \caption*{a) Independent averaging (noisy).}
  \label{fig:average_xy_noisy}
\end{subfigure} \hspace{2em}
\begin{subfigure}[t]{0.42\textwidth}
  \includegraphics[width=\linewidth]{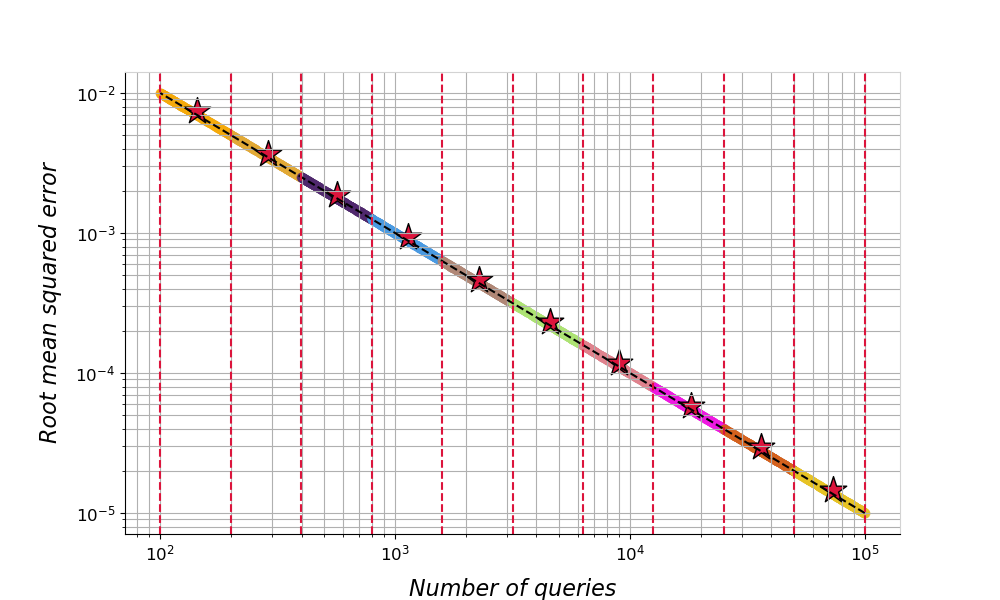}
  \captionsetup{font=footnotesize, skip=2pt}
  \caption*{b) Independent averaging (noiseless).}
  \label{fig:average_xy_noiseless}
\end{subfigure}

\vspace{-0.25em}

\begin{subfigure}[t]{0.42\textwidth}
  \includegraphics[width=\linewidth]{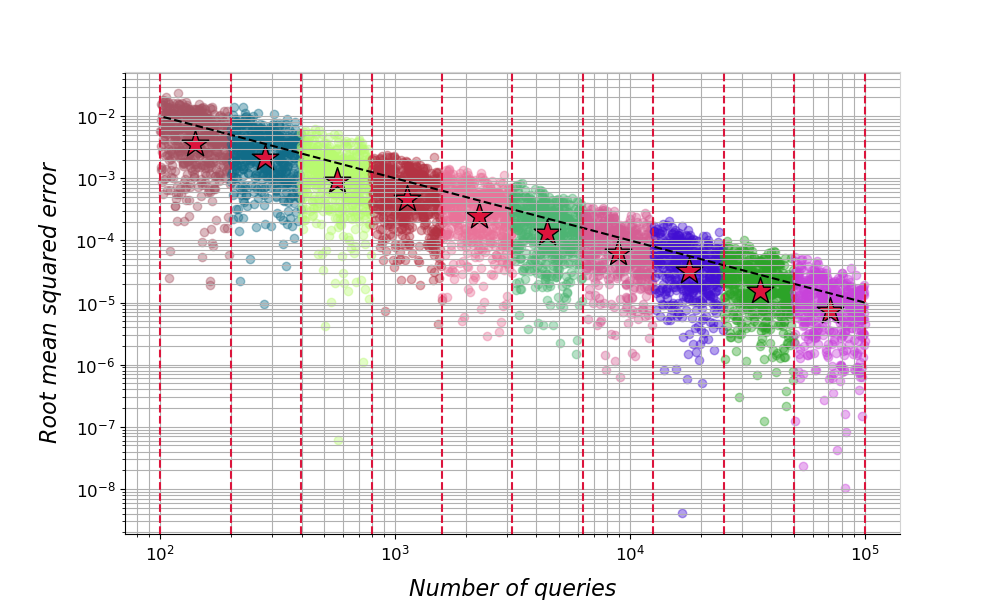}
  \caption*{c) Average log-coordinates (noisy).}
  \label{fig:average_logxy_noisy}
\end{subfigure}\hspace{2em}
\begin{subfigure}[t]{0.42\textwidth}
  \includegraphics[width=\linewidth]{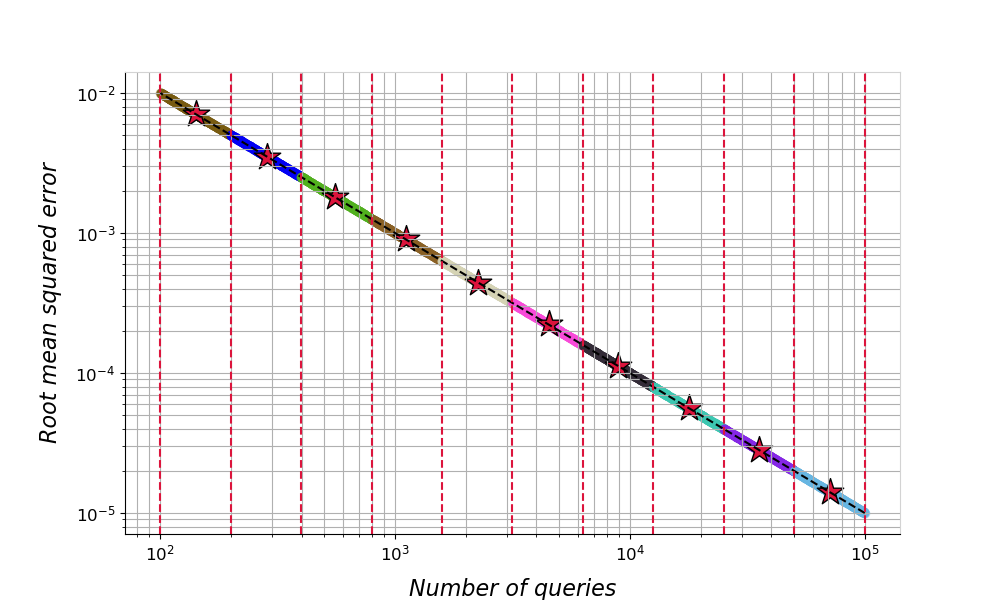}
  \caption*{d) Average log-coordinates (noiseless).}
  \label{fig:average_logxy_noiseless}
\end{subfigure}

\vspace{-0.25em}

\begin{subfigure}[t]{0.42\textwidth}
  \includegraphics[width=\linewidth]{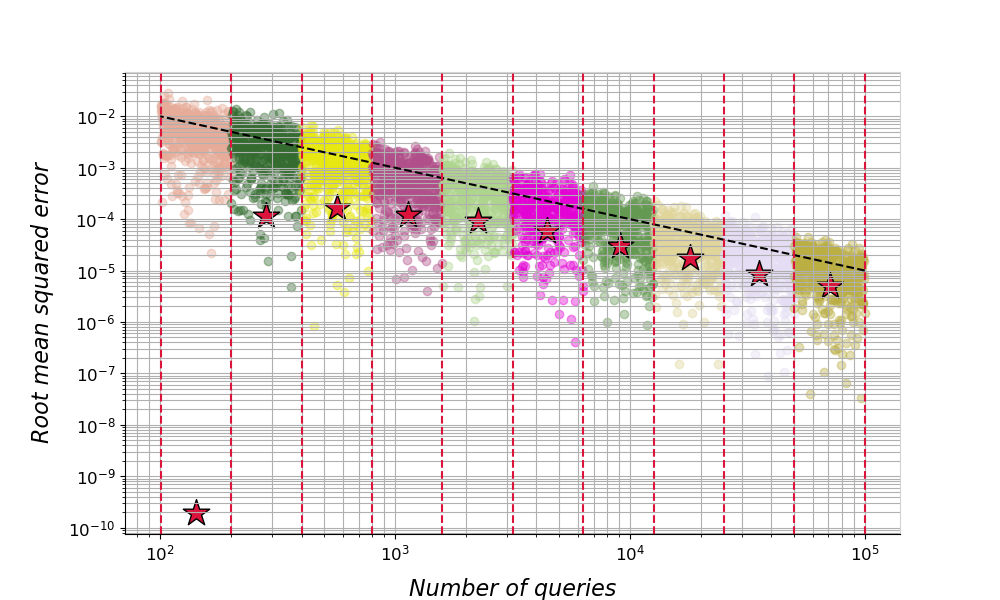}
  \caption*{e) Average slopes (noisy).}
  \label{fig:average_m_noisy}
\end{subfigure}\hspace{2em}
\begin{subfigure}[t]{0.42\textwidth}
  \includegraphics[width=\linewidth]{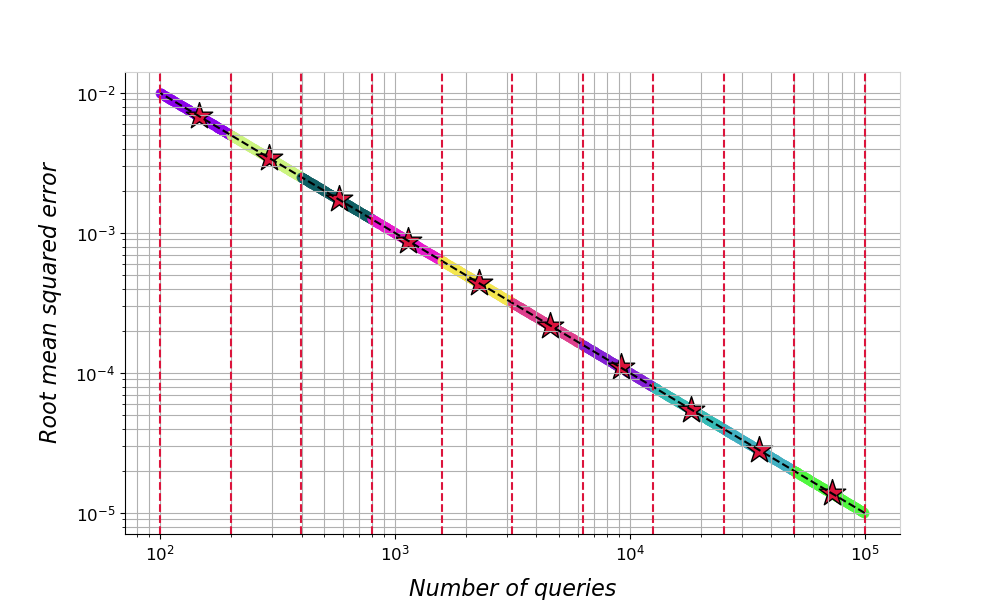}
  \caption*{f) Average slopes (noiseless).}
  \label{fig:average_m_noiseless}
\end{subfigure}

\vspace{-0.25em}

\begin{subfigure}[t]{0.42\textwidth}
  \includegraphics[width=\linewidth]{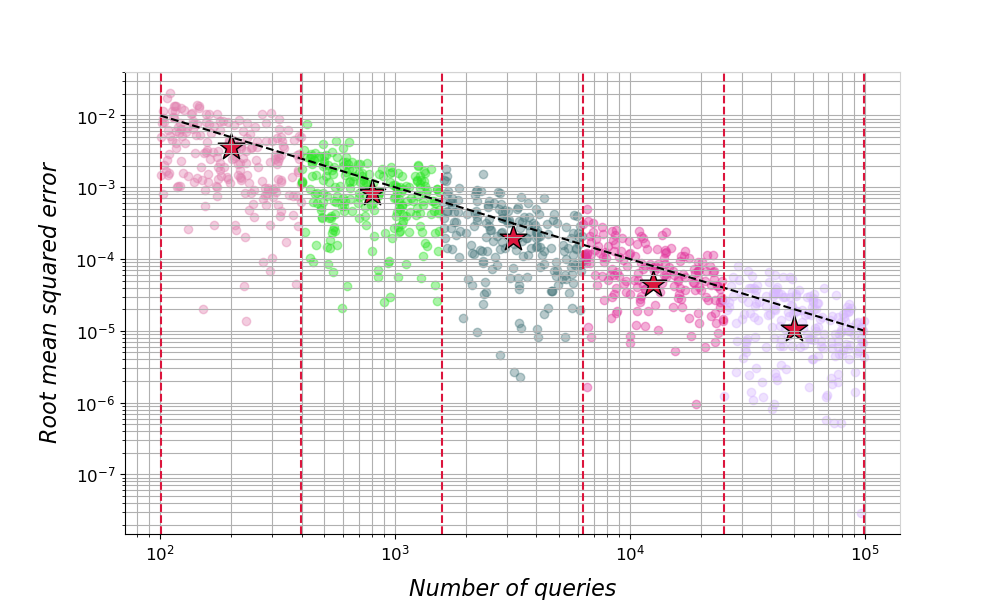}
  \caption*{g) Curve fitting (noisy).}
  \label{fig:fit_noisy}
\end{subfigure}\hspace{2em}
\begin{subfigure}[t]{0.42\textwidth}
  \includegraphics[width=\linewidth]{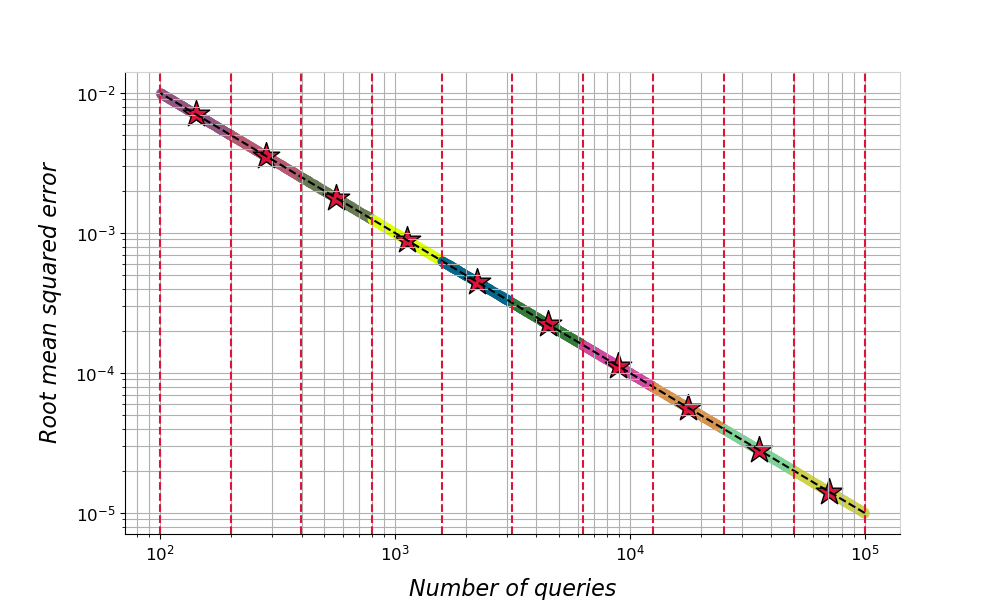}
  \caption*{h) Curve fitting (noiseless).}
  \label{fig:fit_noiseless}
\end{subfigure}

\vspace{-0.25em}

\begin{subfigure}[t]{0.42\textwidth}
  \includegraphics[width=\linewidth]{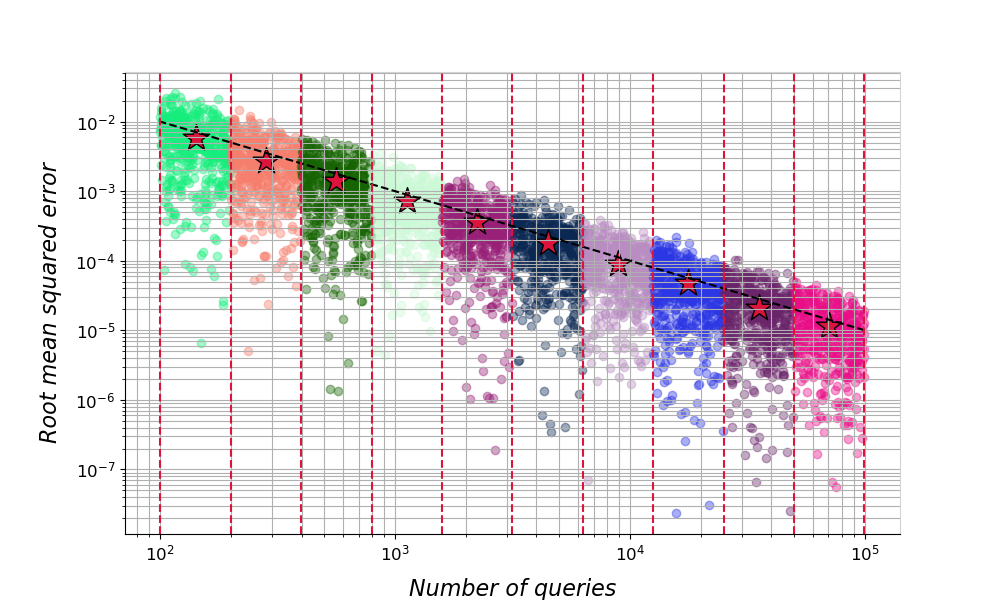}
  \caption*{i) Spline interpolation (noisy).}
  \label{fig:spline_noisy}
\end{subfigure}\hspace{2em}
\begin{subfigure}[t]{0.42\textwidth}
  \includegraphics[width=\linewidth]{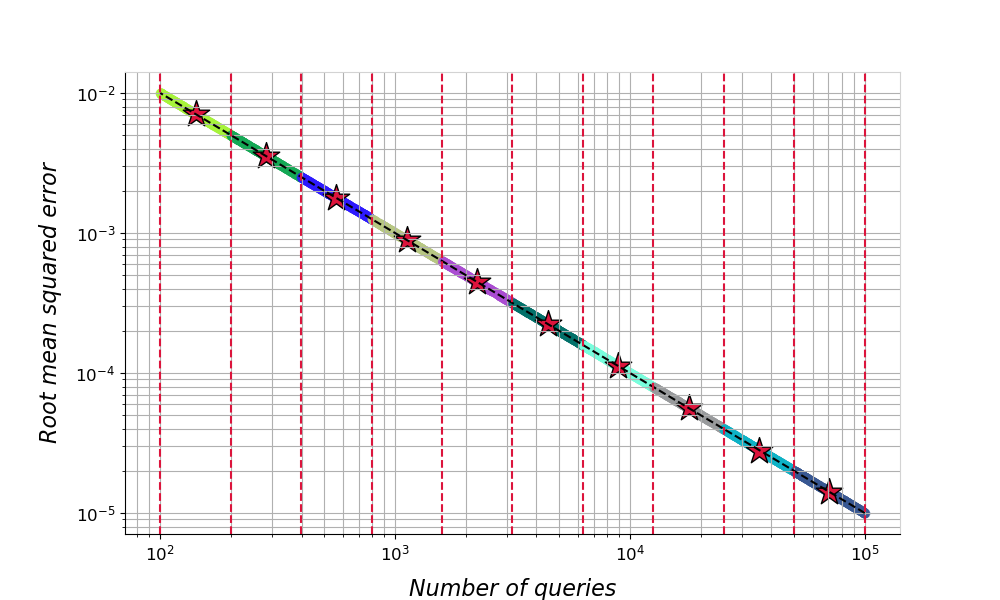}
  \caption*{j) Spline interpolation (noiseless).}
  \label{fig:spline_noiseless}
\end{subfigure}

\caption{Comparison of summary points obtained by various strategies for noisy (left) and noiseless (right) data.}
\label{fig:processing_dummy_all}
\end{figure*}

Figure~\ref{fig:processing_dummy_all} showcases the performance of each of these approaches. An ideal candidate is one who translates each group of binned data points into a single point lying on the dashed line.

Interestingly, the method that seems to work best for our purposes is the naive one - despite the fact that most other methods work near perfectly with the noiseless dataset, which would intuitively seem like a good property. When operating over actual points whose images are probabilistic, they seem to introduce bias. 

Since independent $x$/$y$ averages seem to yield the best results, we adopt this strategy where necessary in the main text. However, we note that a processing techniques working well for these dummy data is a necessary but not sufficient condition for them to work in the general case, as the noise model of equation \ref{eq:auxvars} may have properties that do not hold for real data. However, it is correct in the limit of infinite bins, and we expect it to be a good approximation  for high enough numbers of bins and data. Tests for various settings, as well as comparisons with single executions and analytical results, support this. 

To further demonstrate the robustness of the results independently of processing strategies, we present in Appendix~\ref{app:otherstat} the results for each of the processing strategies that convey the Heisenberg-limited scaling correctly. These are: average log coordinates, curve fitting, and spline interpolation. The results demonstrate consistency across processing strategies.

\section{Numerical results for other statistics}
\label{app:otherstat}

In the joint plots of section \ref{sec:results} (figures \ref{fig:noiseless_median} and \ref{fig:noisy_median}), we presented median errors using the "binning followed by independent x-y averaging" strategy described in Appendix~\ref{app:proc_adapt}. The mean errors tend to follow a less stable trend for most algorithms, especially for randomized amplitudes and/or in the presence of decoherence, making them harder to read.

\begin{figure*}[ht]
  \centering

  \begin{subfigure}[t]{0.48\linewidth}
    \centering
    \includegraphics[width=\linewidth]{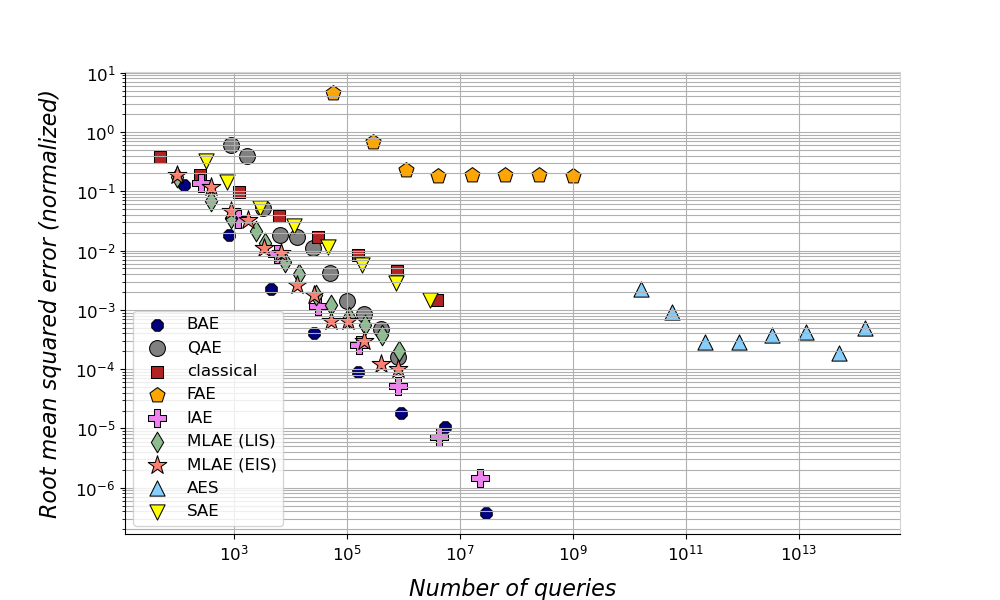}
    \caption{Coordinate averages (noiseless).}
    \label{fig:noiseless_mean}
  \end{subfigure}\hfill
  \begin{subfigure}[t]{0.48\linewidth}
    \centering
    \includegraphics[width=\linewidth]{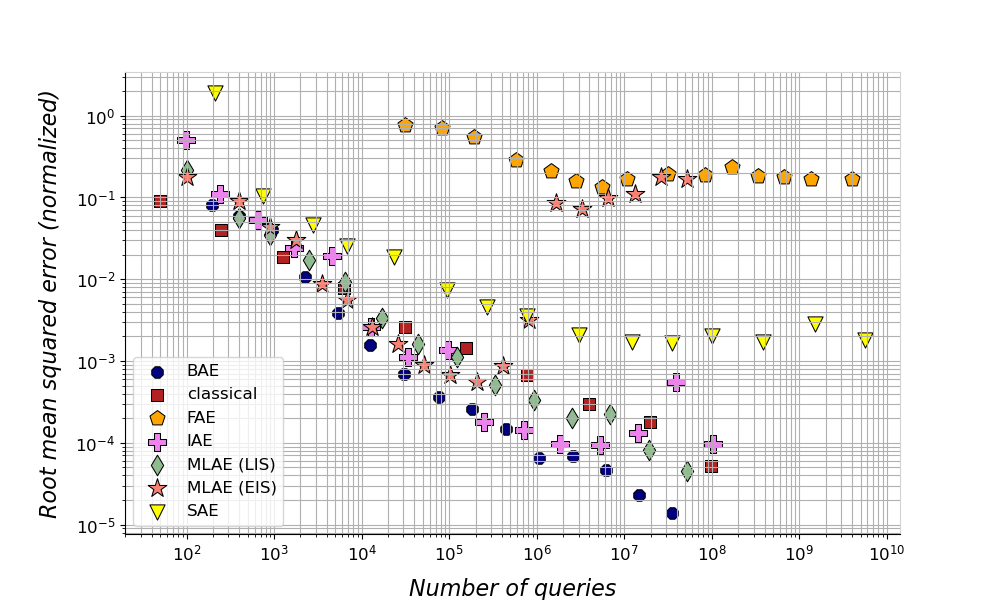}
    \caption{Coordinate averages (noisy).}
    \label{fig:noisy_mean}
  \end{subfigure}

  \vspace{1em}

  \begin{subfigure}[t]{0.48\linewidth}
    \centering
    \includegraphics[width=\linewidth]{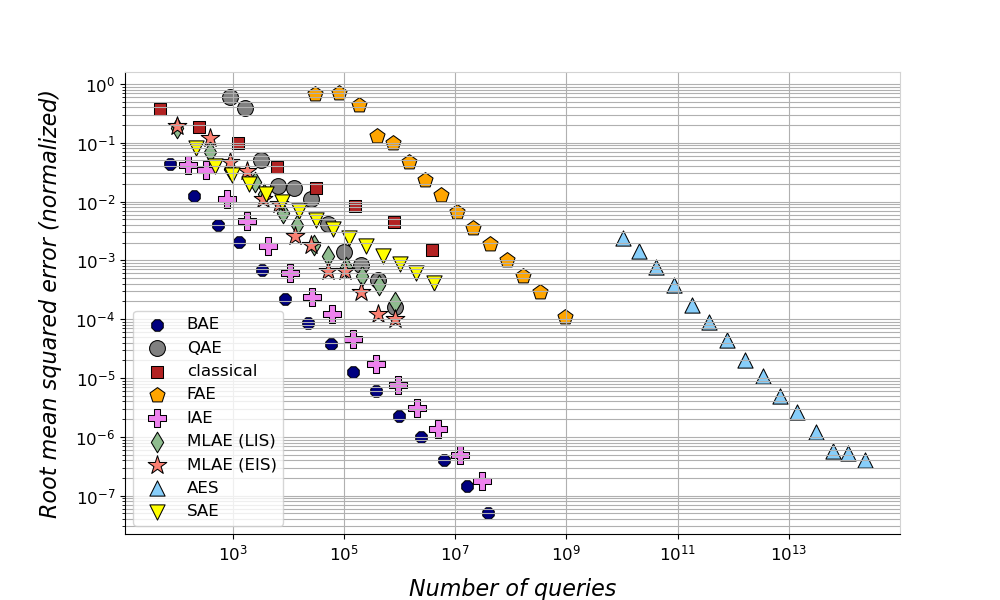}
    \caption{Log-coordinate averages (noiseless).}
    \label{fig:noiseless_mean_log}
  \end{subfigure}\hfill
  \begin{subfigure}[t]{0.48\linewidth}
    \centering
    \includegraphics[width=\linewidth]{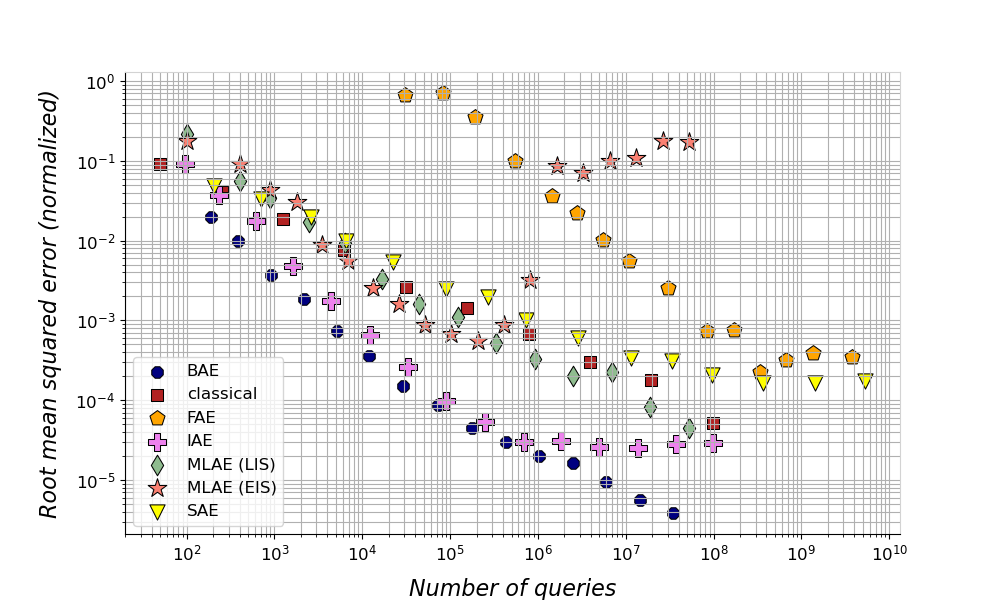}
    \caption{Log-coordinate averages (noisy).}
    \label{fig:noisy_mean_log}
  \end{subfigure}

  \caption{Simulation results (RMSE evolution) for a selection of QAE algorithms in the presence and absence of noise, using different strategies to process adaptive data: independent averaging of the coordinates (top) and log-coordinates (bottom).}
  \label{fig:rmses_combined}
\end{figure*}

For completeness, we additionally present the evolution of the root mean squared errors associated with the same results. This is shown in Figures~\ref{fig:noiseless_mean} and \ref{fig:noisy_mean} for the noiseless and noisy cases respectively. BAE continues to rank better than alternatives for this statistic. 

We additionally present the results of using other statistics described in Appendix~\ref{app:proc_adapt} for completeness. Figures~\ref{fig:noiseless_mean_log} and \ref{fig:noisy_mean_log} show the results of independent averaging of the logarithms of the coordinates. Interestingly, the log-mean results are closer to the non-log-medians presented in the main text, which suggest that the averages over logarithms have more regularizing properties. 

Finally, we consider curve fitting and spline interpolation. We do not use these two to fit the noisy performances, as they are expected to have less regular behavior, and the exhaustive analysis in the noiseless case shows that the medians and log-averages are solid evaluations. 

\begin{figure*}
  \begin{minipage}[t]{0.48\linewidth}
    \includegraphics[width=\linewidth]{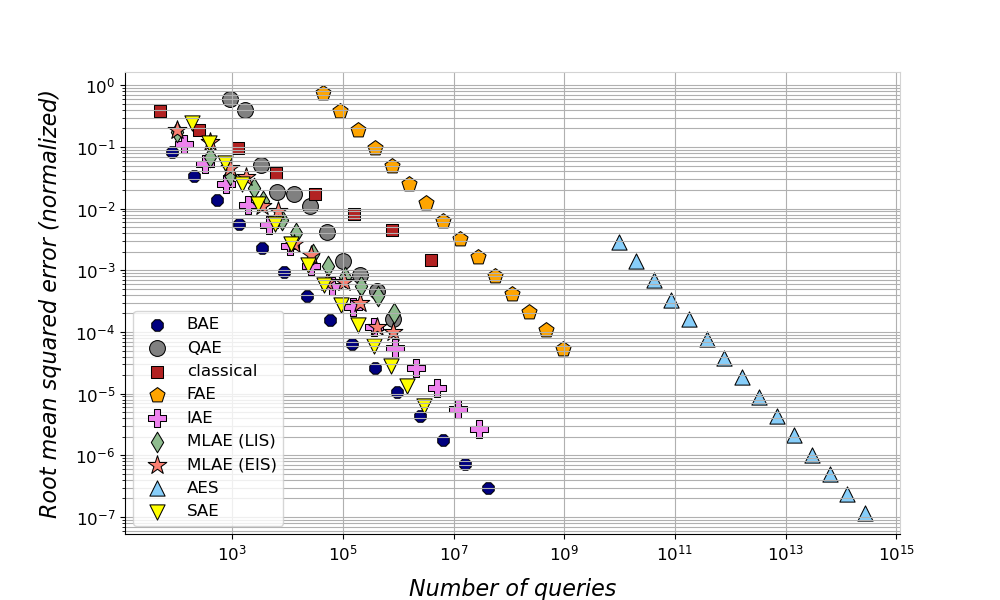}
  \text{a) Curve fitting.}
  \label{fig:noiseless_fit}
  \end{minipage}\hfill%
  \begin{minipage}[t]{0.48\linewidth}
  \includegraphics[width=\linewidth]{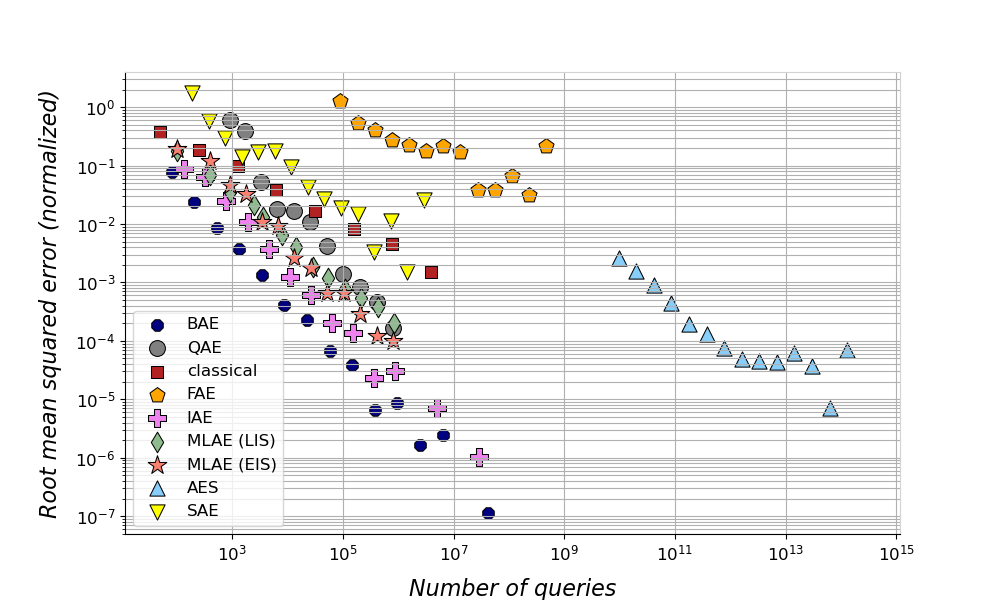}
  \text{b) Spline interpolation.}
  \label{fig:noiseless_spline}
  \end{minipage}%
  \caption{Simulation results (RMSE evolution) for BAE and a selection of other QAE algorithms in a noiseless scenario, using different strategies to process adaptive data: curve fits and interpolation.}
  \label{fig:rmses_fit_spline}
  \end{figure*}

  These results demonstrate the robustness of BAE endently of the data processing and statistics adopted: it is the best performing algorithm in all cases. 
\section{Defining the intercept}
\label{app:intercept}

To plot data along with the fundamental limits of section \ref{sub:quantum_enhanced}, we must define the y-intercept. This is not given by these limits alone: they describe only the slope (in a log-scale), which represents the complexity. 

To define the vertical offset of the graphs, we used the following strategy:
\begin{itemize}
    \item Do a curve fit $Be^m$ with parameters $B$, $m$:
    \begin{gather*}
        y = Bx^m \leftrightarrow \log(y) = m\log(x)+\log(B)\\
    \end{gather*}
    \item Calculate the image $y_0$ of the the x-coordinate $x_0$ of the first datapoint, under the model from the previous point;
    \item Make the straight lines of the standard quantum and Heisenberg limits pass by this point.
\end{itemize}

This seems to produce sensible results. 
\section{Numerical treatment of QAE}
\label{app:num_qae}

\subsection{Original algorithm (QFT)}
\label{sub:num_orig}

The probabilistic behaviour of the original QAE algorithm \cite{brassard2002} can be described analytically. This allows us to simulate the algorithm efficiently, without actually running the circuits, significantly lowering the cost of testing the algorithm. 

We start by considering the outcome distribution of QPE. It estimates $\phi$ in $\exp (i2\pi\phi)$, but the actual measurement outcomes target $K\phi$, with $K$ the order of the Fourier transform and the number of possible outcomes. If that is an integer, the outcome is deterministically $x=K\phi$, and we calculate $\phi$ as $\phi = x/K$. If it is not, the probability of each of the $K$ outcomes, which we'll call $x$, increases with its accuracy, and observes the following expression:

\begin{equation}
\label{eq:p_QPE} 
    P(\text{measuring } x \mid QPE(\phi)) = \frac{\sin^2(K \Delta\pi)}{K^2\sin^2(\Delta\pi)},
    \end{equation}

\noindent where $\Delta$ is a circular distance, and also the error in the estimate produced by $x$:
\begin{equation}\Delta =\left\lvert \phi-\frac{x}{K}\right\rvert \mod 1.\end{equation}

This is an angular distance, divided by $2\pi$.

In QAE, when we perform QPE, we measure one of two eigenvalues of the Grover operator, which changes the looks of the expressions above:

\begin{itemize}
    \item One eigenvalue is $\exp (i2\theta )$, meaning that $2\pi\phi = 2\theta \leftrightarrow \phi = \theta/\pi$. In this case:
    \begin{itemize}
        \item The exact case outcome is $x_0=M\theta/\pi$.
        \item We would calculate $\theta$ from an outcome $x$ as $\pi x/M$.
        \item $\Delta$ is defined as $\Delta = \lvert \theta/\pi-x/M\rvert \mod 1$, which measures the distance between $\theta/\pi$ (the phase-estimated angle divided by $2\pi$) and $x/M$ (the exact phase encodings representable by the QPE auxiliary register).
    \end{itemize}
    \item The other eigenvalue is $\exp (-i2\theta )$, meaning $2\pi\phi = 2\pi - 2\theta \leftrightarrow \phi = 1 - \theta/\pi$. We're basicallly replacing $\theta \rightarrow \pi - \theta$ as compared to the previous case:
    \begin{itemize}
        \item The exact case outcome is $x_0'=M-M\theta/\pi=M(1-\theta/\pi)=M-x_0$ using the $x_0$ definition from the first case.
        \item We would calculate $\theta$ from an outcome $x$ as $\pi(M-x)/M=\pi - \pi x/M$. 
        \item $\Delta$ is defined as $\Delta = \lvert 1-\theta/\pi-x/M\rvert  \mod 1$, which measures the distance between $1-\theta/\pi$ and $x/M$; or equivalently, between $\theta/\pi$ and $1-x/M$.
    \end{itemize}
\end{itemize}

The second points in each of the two cases, which tell us how to calculate the unknown parameter $\theta$ from an outcome $x$, may seem problematic: in practice, we have no way of knowing which eigenphase was behind a measurement. However, as luck would have it, the object of our interest is not actually $\theta$, but rather $a=\sin^2(\theta)$. Due to the symmetry of the sine function about $\pi/2$ in $[0,\pi]$, $a$ can always be calculated as $a=\sin^2( \pi x/M)$, regardless of the underlying eigenvalue. Equivalently, it can be calculated as $a=\sin^2( \pi - \pi x/M)$. 

Since we measure each eigenvalue with probability $1/2$, the probability of measuring some specific outcome $x$ is given by the average:
\begin{equation}
\label{eq:qae_dist_formula1}
\begin{aligned}
P( x \mid QAE(\theta)) & =  \frac{P( x \mid QPE(\theta/\pi))}{2} \\
& + \frac{P(x \mid QPE(1-\theta/\pi))}{2}
\end{aligned}
\end{equation}

Or alternatively, as can be seen from formula (\ref{eq:p_QPE}) and the previous considerations on $\Delta$:

\begin{equation}
\begin{aligned}
\label{eq:qae_dist_formula2}
P(x \mid QAE(\theta)) &= \frac{P(x \mid QPE(\theta/\pi))}{2} \\
&+ \frac{P(M-x \mid QPE(\theta/\pi))}{2} .
\end{aligned}
\end{equation}

These equations give the exact probabilities of each possible QAE output. By sampling from a multinomial distribution defined by these probabilities, we can accurately reproduce the ideal behavior of the canonical quantum amplitude estimation algorithm under shot noise. This emulates the performance of a perfect quantum device while requiring a much shorter runtime, since it relies on a small number of simple analytical and sampling operations. 

Note that this does not mean that this construction can replace the QAE algorithm, because it requires knowing the solution we're seeking: the calculations require specifying the amplitude parameter. In other words, these considerations are useful to study the behaviour of QAE, but not to solve the problem it tackles. 

\subsection{Maximum likelihood post processing}
\label{sub:qae_mle}

If we execute the original QAE algorithm for one of the few cases allowing an exact result, we will always get the same outcome. But for most cases, we will get a distribution of outcomes, according to section \ref{sub:num_orig}. One must then pick one estimate for the amplitude.

One option would be to pick the value with the highest relative frequency. In practice, that is one of the few values that we are \textit{absolutely sure} is wrong. Such an amplitude could never have generated the distribution, since it finds an exact representation in our discretization. As such, it would have generated only 2 outcomes, both corresponding to the same amplitude. The same goes for any other of the amplitude values corresponding to an observed outcome, for the very reason they are a possible outcome.

Out of those values, the mode is indeed our best choice, as it falls closer to the real value than any other. But we can actually do better than choosing on the grid, by availing ourselves of equation (\ref{eq:qae_dist_formula1}) - or (\ref{eq:qae_dist_formula2}) -, which describes how likely a given amplitude would be to generate an observed outcome. This allows us to consider amplitudes on a continuum: given any value, we can calculate how likely it would have been to produce the list of outcomes we observed. We do so by multiplying how likely it would have been to produce each of them. 

We can then sweep over a much denser grid of amplitudes, without expending any extra quantum resources. All we need is the previously gathered data. This maximum likelihood estimation approach was suggested in \cite{grinko2021} for this problem. 

Even though no additional quantum resources are required, classical processing is. However, the optimization can be quite agile; more so than increasing the number of qubits to achieve the same precision. Even still, it becomes harder as the number of measurement repetitions increases. One easy trick that can aid it is to make a more localized search, by reducing the search range to the vicinity of the mode. 

Since the most likely outcome should be the one with a smaller estimation error, the mode should be closer to the true value than any other grid point. Thus, it should suffice to search in a range centered at the mode, and limited to within the halfway distance between itself and its nearest "grid points" (exactly representable amplitudes). This is also suggested in \cite{grinko2021}.

We use this strategy to numerically test the original QAE algorithm. Another advantage is smoothing out its behavior, avoiding abrupt changes in the error or aberrant cases (namely those admitting a discrete representation, for which the error becomes a constant after a sufficient number of qubits is reached).

\subsection{Grover measurements}
\label{sub:num_grov}

Similar to section \ref{sub:num_orig}, measurement data for Grover measurements can be generated efficiently. In this case, the probabilities are straightforward and given by equation (\ref{eq:p1}). To add shot noise, one need only draw from the Bernoulli distribution defined by this parameter. 

The Hadamard tests of \cite{wie2019} allow an identical treatment; one need only change the mathematical expressions for each probability. 

\section{Quantum and classical parallelism of BAE}
\label{app:parallelism}

The main operations involved in our algorithms are highly parallelizable. On the classical processing side, this is known to be a strength of the sequential Monte Carlo algorithm \cite{doucet2013,delmoral2006,south2019}. The key operations of SMC are the reweighting and resampling steps, which can be performed in parallel for each particle, giving \texttt{Npart} threads, where \texttt{Npart} is often in the order of the hundreds or thousands. 

Furthermore, for the adaptive experimental design, the optimization process itself is parallelizable, as the expected utility must be calculated for each possible control whose utility is to be evaluated, resulting in \texttt{Nevals} parallel computations. We use \texttt{Nevals} in the order of the tens. 

On the quantum side, parallelization comes from the possibility, explored in Appendix~\ref{app:expdesign}, of using $\texttt{Nshots}>1$ shots for each optimized control $m$. This can be executed in parallel in $\texttt{Nshots}$ quantum devices. We have used values of $\texttt{Nshots}$ in the order of the hundreds. 

For details on the parameters \texttt{Npart}, \texttt{Nevals}, \texttt{Nshots}, refer to Appendix~\ref{app:hyperparameters}.   
\section{Tuning the hyperparameters}
\label{app:hyperparameters}

In this appendix we provide additional information on the hyperparameters of our algorithm, how to tune them, and the trade-offs they can negotiate. We additionally illustrate their impact on performance via numerical simulations where we vary one parameter while keeping the others constant. We show that BAE is not overreliant on specific configurations, further demonstrating its robustness. 

For each hyperparameter, we specify the default values used in our simulations. Thus, the plots presented here for various parameter values can additionally be compared with e.g. Figure~\ref{fig:res2}, which uses these defaults. This comparison demonstrates that we can have large (often ten-fold) variations in the hyperparameters with negligible impact on performance; but extra numerical tests show that after a point the performance begins to suffer significantly. 

Cost offsets refer to \textit{quantum} cost offsets, i.e. vertical position in the plots. 

\begin{figure*}[!htbp]
  \centering
  \begin{subfigure}[t]{0.38\linewidth}
    \includegraphics[width=\linewidth]{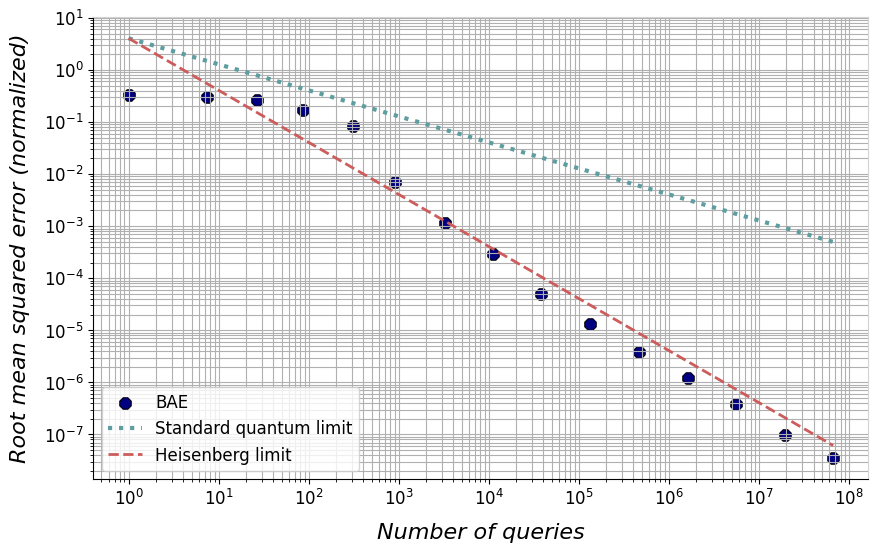}
    \caption{$\texttt{wNshots}=1$.}
    \label{fig:wNshots_a}
  \end{subfigure}\hspace{5em}
  \begin{subfigure}[t]{0.38\linewidth}
    \includegraphics[width=\linewidth]{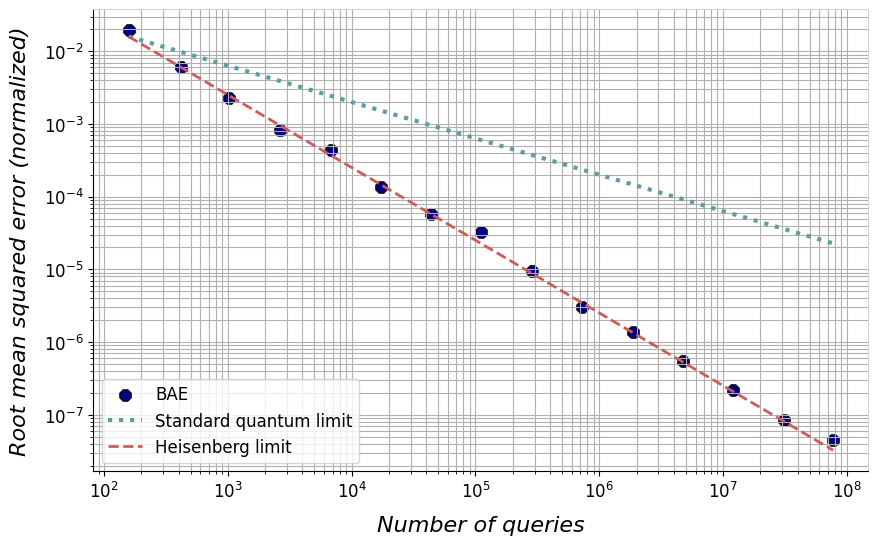}
    \caption{$\texttt{wNshots}=100$.}
    \label{fig:wNshots_b}
  \end{subfigure}
  \caption{Simulation results (RMSE evolution) for BAE with different values of \texttt{wNshots}. The case $\texttt{wNshots}=1$ has less stable performance, but a lower offset than $\texttt{wNshots}=10$. Both are Heisenberg-limited.}
  \label{fig:wNshots}
\end{figure*}

\textbf{\texttt{wNshots}: number of shots for the classical warm-up phase.} \textit{Default value: $50$.} This parameter defines how many initial experiments with $m=0$ are performed. These measurements have a regularizing effect on performance, as they convey slow and reliable information, at a stage where the adaptive strategy wouldn't be surgical due to the lack of information in the early stages of the inference process. For too low values, performance may be slightly less stable; for too high values, the query offset will be worse, as quantum effects are not exploited in these measurements. Figure~\ref{fig:wNshots} illustrates this. 

\begin{figure*}[!htbp]
  \centering
  \begin{subfigure}[t]{0.38\linewidth}
    \includegraphics[width=\linewidth]{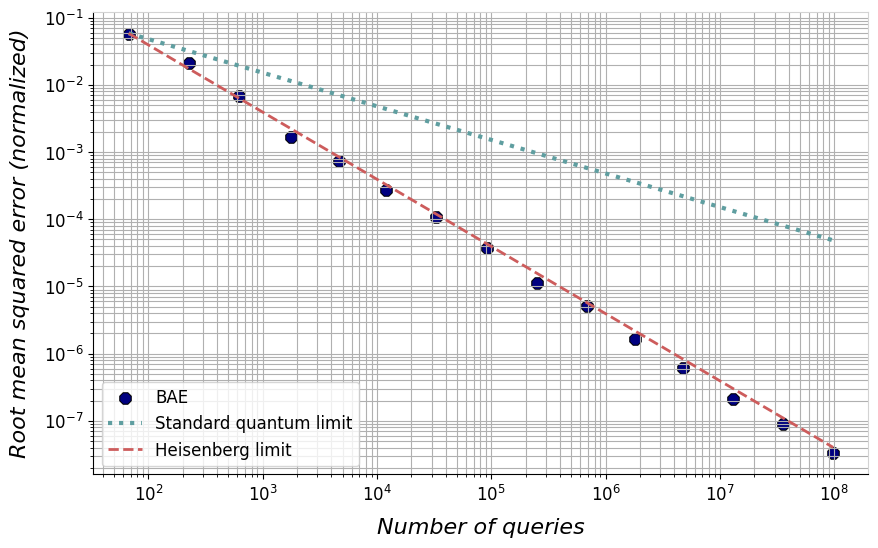}
    \caption{$\texttt{Nshots}=10$.}
    \label{fig:Nshots_a}
  \end{subfigure}\hspace{5em}
  \begin{subfigure}[t]{0.38\linewidth}
    \includegraphics[width=\linewidth]{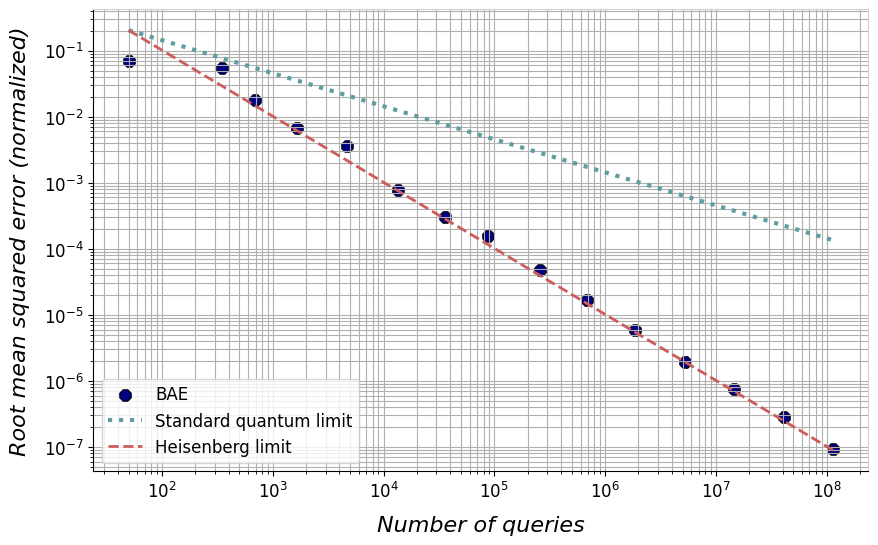}
    \caption{$\texttt{Nshots}=100$.}
    \label{fig:Nshots_b}
  \end{subfigure}
  \caption{Simulation results (RMSE evolution) for BAE with different values of \texttt{Nshots}. The case $\texttt{Nshots}=10$ has lower offset, but higher classical optimization cost and lower quantum parallelism. Both are Heisenberg-limited.}
  \label{fig:Nshots}
\end{figure*}

\textbf{\texttt{Nshots}: number of shots per measurement in the adaptive phase.} \textit{Default value: $1$.} This parameter defines how many shots are performed for the optimized control. This does not change the optimization process; BAE always optimizes the control as if a single shot were to be performed. While this is not formally correct, it performs well in numerical simulations, and avoids the exponential cost in \texttt{Nshots} that would come otherwise. This parameter controls the degree of parallelism in the quantum component, as the \texttt{Nshots} measurements can be performed in parallel. Lower values have higher optimization cost per measurement, while higher values allow higher parallelism but may have a worse cost offset, as can be seen in Figure~\ref{fig:Nshots}.

\begin{figure*}[!htbp]
  \centering
  \begin{subfigure}[t]{0.38\linewidth}
    \includegraphics[width=\linewidth]{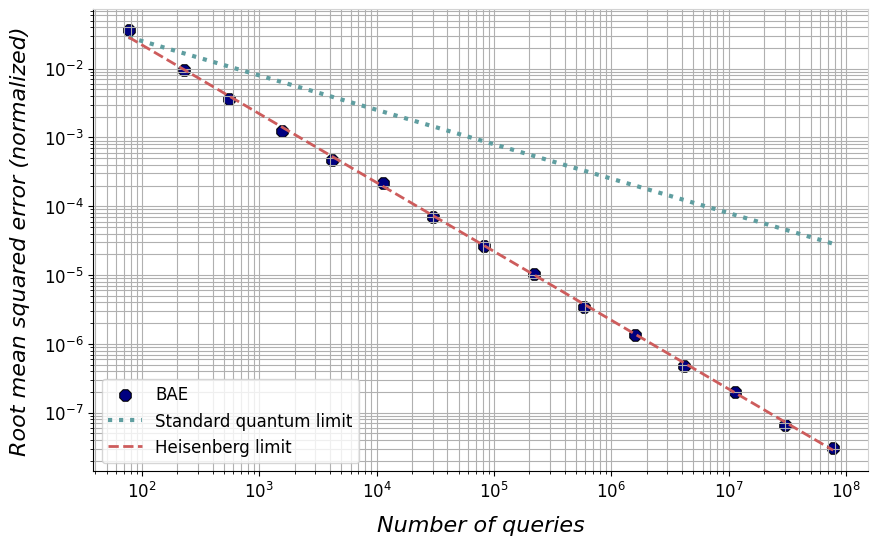}
    \caption{$\texttt{Nevals}=25$.}
    \label{fig:Nevals_a}
  \end{subfigure}\hspace{5em}
  \begin{subfigure}[t]{0.38\linewidth}
    \includegraphics[width=\linewidth]{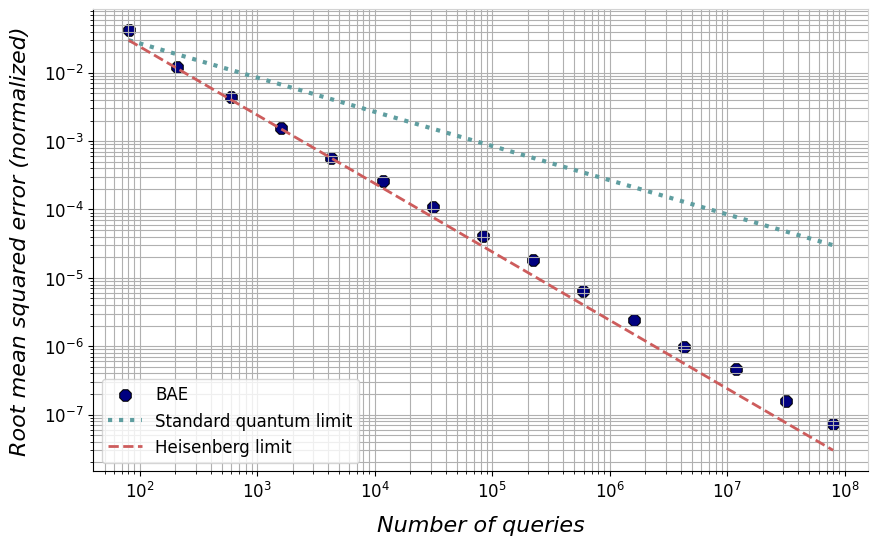}
    \caption{$\texttt{Nevals}=10$.}
    \label{fig:Nevals_b}
  \end{subfigure}
  \caption{Simulation results (RMSE evolution) for BAE with different values of \texttt{Nevals}. The case $\texttt{Nevals}=25$ has higher classical optimization cost than $\texttt{Nevals}=10$, but a better slope.}
  \label{fig:Nevals}
\end{figure*}

\textbf{\texttt{Nevals}: number of potential experimental controls contemplated in each optimization.} \textit{Default value: $50$.} This parameter determines how many potential controls $m$ are sampled from the search window of Appendix~\ref{app:expdesign}. Higher values have higher optimization costs, as the expected utility must be evaluated for each contemplated control. Lower values may have worse performance, as the optimization is coarser. This is shown in Figure~\ref{fig:Nevals}.

\begin{figure*}[!htbp]
  \centering
  \begin{subfigure}[t]{0.38\linewidth}
    \includegraphics[width=\linewidth]{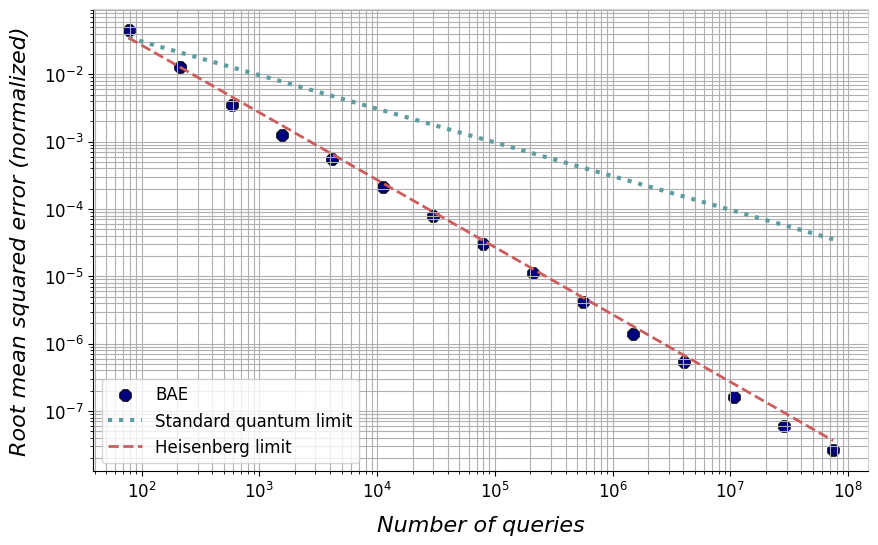}
    \caption{$k_0=2$.}
    \label{fig:k0_a}
  \end{subfigure}\hspace{5em}
  \begin{subfigure}[t]{0.38\linewidth}
    \includegraphics[width=\linewidth]{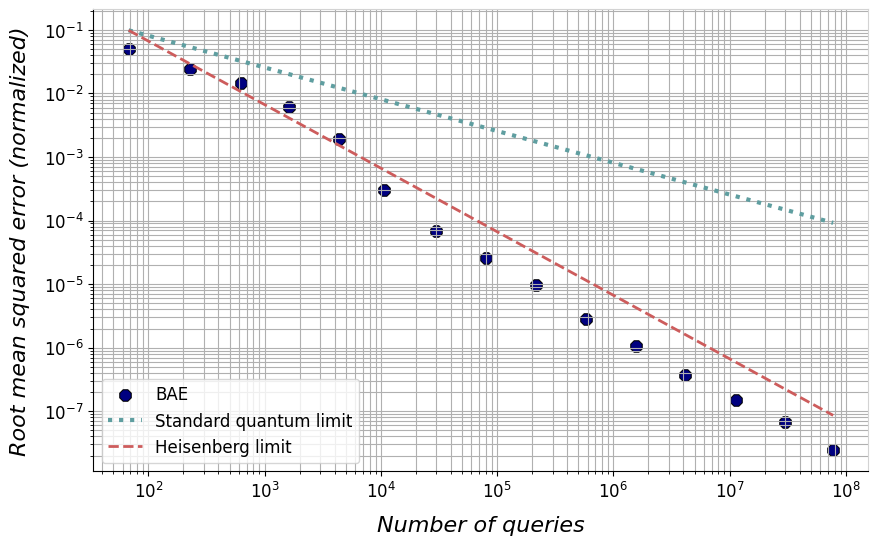}
    \caption{$k_0=10$.}
    \label{fig:k0_b}
  \end{subfigure}
  \caption{Simulation results (RMSE evolution) for BAE with different values of $k_0$. The case $k_0=2$ has more regular behavior than $k_0=10$. Both are Heisenberg-limited.}
  \label{fig:k0}
\end{figure*}

\textbf{$k_0$: factor for the initial window width.} \textit{Default value: $1$.} Along with \texttt{Nevals}, this value determines how high the initial upper bound for the control $m$ is: $k_0*\texttt{Nevals}$. This affects later iterations, as the window expansions depend on the starting point, with the upper bound doubling for each expansion. This parameter may be increase to leave the window bounds unchanged as \texttt{Nevals} is reduced to spare costs. For low combined values of \texttt{Nevals} and $k_0$, the expansion may be too slow, falling short of the quadratic speed-up and/or leading to a worse offset. For too high values, it may be too fast, causing less stable behavior. This is shown in Figure~\ref{fig:k0}.

\begin{figure*}
  \centering
  \begin{subfigure}[t]{0.38\linewidth}
    \includegraphics[width=\linewidth]{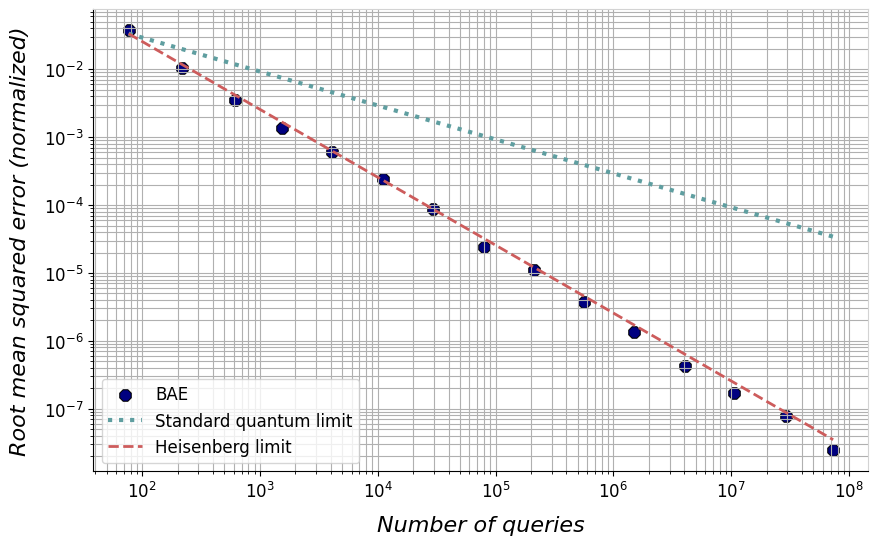}
    \caption{$R=1$.}
  \end{subfigure}\hspace{5em}
  \begin{subfigure}[t]{0.38\linewidth}
    \includegraphics[width=\linewidth]{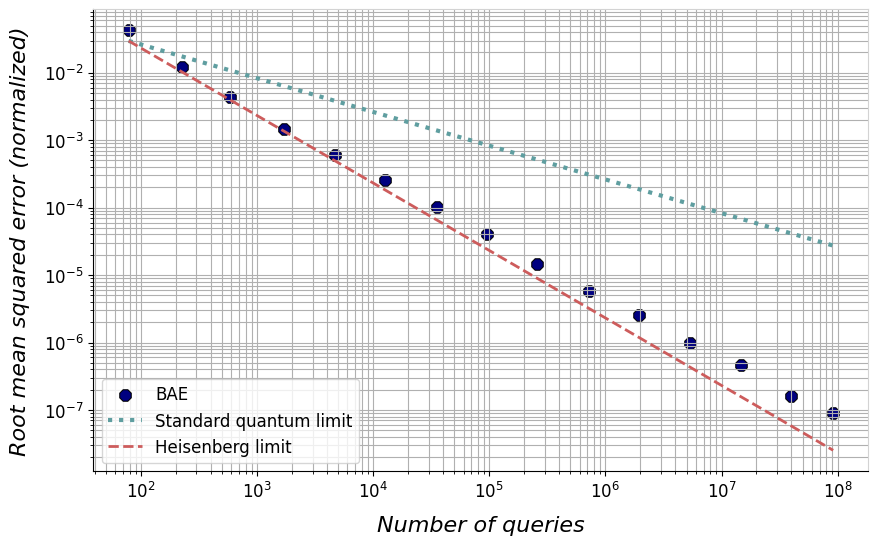}
    \caption{$R=25$.}
  \end{subfigure}
  \caption{Simulation results (RMSE evolution) for BAE with different values of $R$. The case $R=1$ has a better offset and slope than $R=25$, but the impact on performance is limited for a 25-fold increase.}
  \label{fig:R}
\end{figure*}

\textbf{$R$: number of top values to be used as references for window expansions.} \textit{Default value: $3$.} This parameter determines how many of the highest values are used as references for the window expansions. A hit occurs when the chosen control $m$ is among the top $R$ values that could have been selected. When $T$ (another hyperparameter, see below) hits have occurred since the last expansion, a window expansion is triggered, taking the search range to higher values of $m$. A low value of $R$ may hamper window expansions, leading to a smaller speed-up. A high value of $R$ may be too unspecific, leading to  premature expansions and thus a worse performance. Although both extremes, combined with poor choices of $T$, may affect the degree of quantum advantage in the average case (Figure~\ref{fig:R}), there is a wide range of values of $R$ for which the performance is consistent. 

\begin{figure*}
  \centering
  \begin{subfigure}[t]{0.38\linewidth}
    \includegraphics[width=\linewidth]{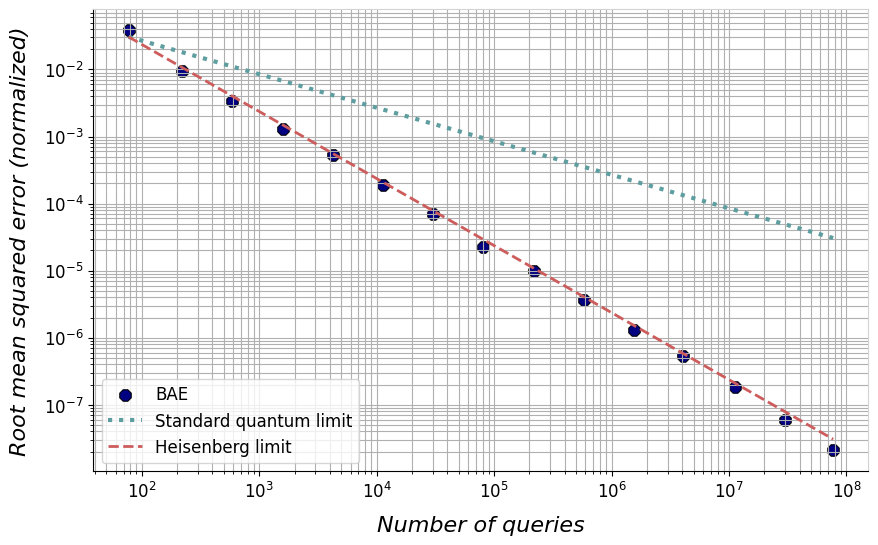}
    \caption{$T=1$.}
  \end{subfigure}\hspace{5em}
  \begin{subfigure}[t]{0.38\linewidth}
    \includegraphics[width=\linewidth]{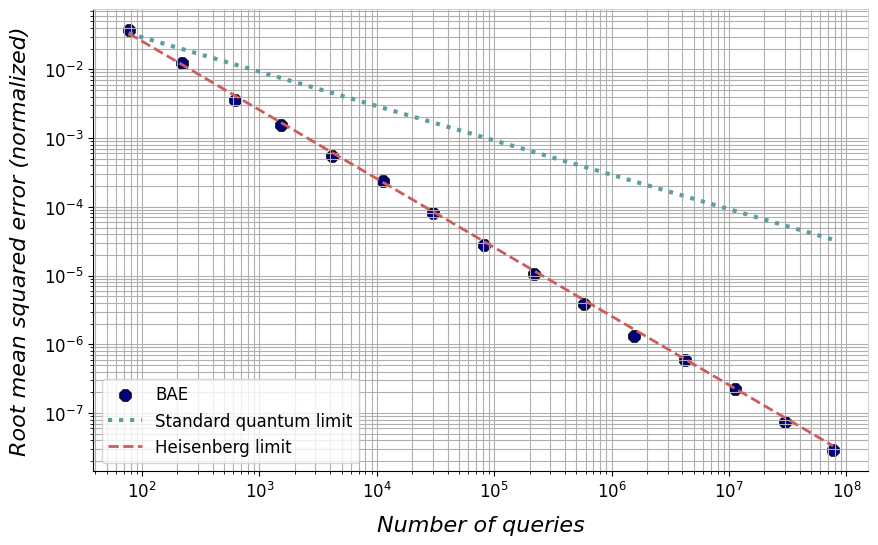}
    \caption{$T=10$.}
  \end{subfigure}
  \caption{Simulation results (RMSE evolution) for BAE with different values of $T$. The case $T=1$ has a better offset than $T=10$, but the difference is slight. Both are Heisenberg-limited.}
  \label{fig:T}
\end{figure*}

\textbf{$T$: threshold for the window expansions.} \textit{Default value: $3$.} This parameter determines how many times the chosen control must fall among the top $R$ values for the window to expand, i.e. the number of hits necessary to trigger a window expansion. Inversely to $R$, too low values may cause unstable behavior due to fast expansion, while low values may be too slow; both cases may harm performance. However, as with $R$, there is a large range of values for which the algorithm works similarly well. In fact, it is even more so the case, as we find numerically that higher values of $T$ do not result in significant deviations from the Heisenberg limit. Figure~\ref{fig:T} illustrates this. However, the offset and classical runtime do increase for higher $T$.

\begin{figure*}
  \centering
  \begin{subfigure}[t]{0.38\linewidth}
    \includegraphics[width=\linewidth]{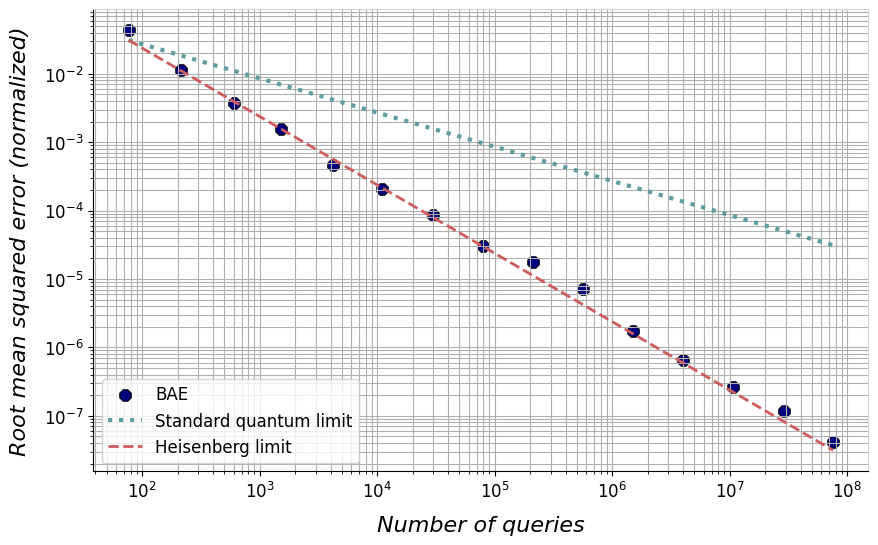}
    \caption{\texttt{Npart} = 500.}
  \end{subfigure}\hspace{5em}
  \begin{subfigure}[t]{0.38\linewidth}
    \includegraphics[width=\linewidth]{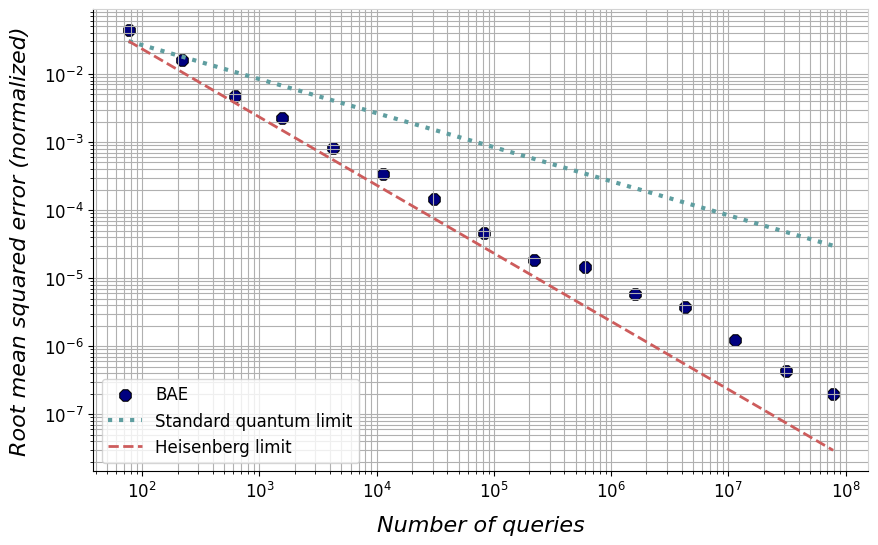}
    \caption{\texttt{Npart} = 100.}
  \end{subfigure}
  \caption{Simulation results (RMSE evolution) for BAE with different values of \texttt{Npart}. The case with \texttt{Npart} = 500 has higher cost per iteration, but better performance than \texttt{Npart} = 100. The median results are identical.}
  \label{fig:Npart}
\end{figure*}

\textbf{\ttt{Npart}: number of particles/grid points for the sequential Monte Carlo (SMC) approximation.} \textit{Default value: $5 000$.} This parameter determines how many discrete points are used in the numerical representation. It is one of the 2 main parameters controlling the quality of the approximations of the Bayesian distributions (the other being the \ttt{thr} below), which is determinant for performance. This affects the error in the estimate of the amplitude, as well as the errors in the utility computations for the experimental design. High numbers of particles have higher computational cost, with quickly saturating gains; low numbers of particles can lead to unstable behavior, larger errors and poor average performance. This is reflected in Figure~\ref{fig:Npart}. Additionally, very low particle numbers may counterproductively lead to higher total costs despite lower costs per iteration, because the estimation process becomes inefficient. 

This parameter is not original to BAE; it is a key parameter in the sequential Monte Carlo algorithm \cite{delmoral2006, doucet2013, south2019}. However, it is determinant for performance, and may cause one to wonder if the complexity of the SMC algorithm could cancel out the quantum advantage. We argue that this is not the case. In SMC, due to the resampling mechanism, $N$ does not necessarily scale with $\epsilon$: the particles' successive relocations should allow the precision to increase indefinitely until we run into issues due to limited machine precision. If this were a problem, we could replace SMC by e.g. the trapezoid rule, where the number of particles  would scale as $N \in \mathcal O (\sqrt e)$ for an error $\epsilon$ in the estimate (in our case, of the amplitude). However, SMC is more cost efficient. 
    
This can also be seen by looking at tests for varied numbers of particles, as in Figure~\ref{fig:Npart}. Decreasing the number of particles from $5 000$ to $500$ didn't appreciably impact the results, even for RMSE $10^{-1}$ to $10^{-7}$; and we never observed a deterioration of performance for lower RMSE given a constant number of particles.

We also note that the cost per particle is that of relatively cheap operations: in most cases, evaluating a single trigonometric function per update or expected value calculation. This may be significantly lower than that of a query in amplitude estimation problems, and thus in practice the prefactor in the complexity could be meaningfully smaller. For instance, in option pricing, one query would involve stochastic simulation of a trajectory for the underlying assets' time evolution until maturity, often for hundreds of times steps \cite{egger2020}.

\textbf{\texttt{thr}: resampling threshold for the SMC algorithm.} \textit{Default value: $0.5$.} This parameter determines how easily a resampling step should be triggered in the face of increased correlation between samples. The resampling corrects this, increasing statistical efficiency. Higher values of \ttt{thr} entail higher costs (more resampling steps, e.g. Markov transitions), but enforce stricter lower bounds on the effective sample size, resulting in better approximations. This hyperparameter is again intrinsic to sequential Monte Carlo, rather than BAE. Its value should be chosen jointly with \ttt{Npart}; along with it, it is determinant for the quality of the discretization of the distributions, and thus crucial for good performance. A higher \ttt{thr} may compensate for a lower \ttt{Npart}, depending on the efficiency of the resampling kernel; conversely, a higher \ttt{Npart} allows a lower \ttt{thr} for the same performance standards.

\begin{figure*}
  \centering
  \begin{subfigure}[t]{0.38\linewidth}
    \includegraphics[width=\linewidth]{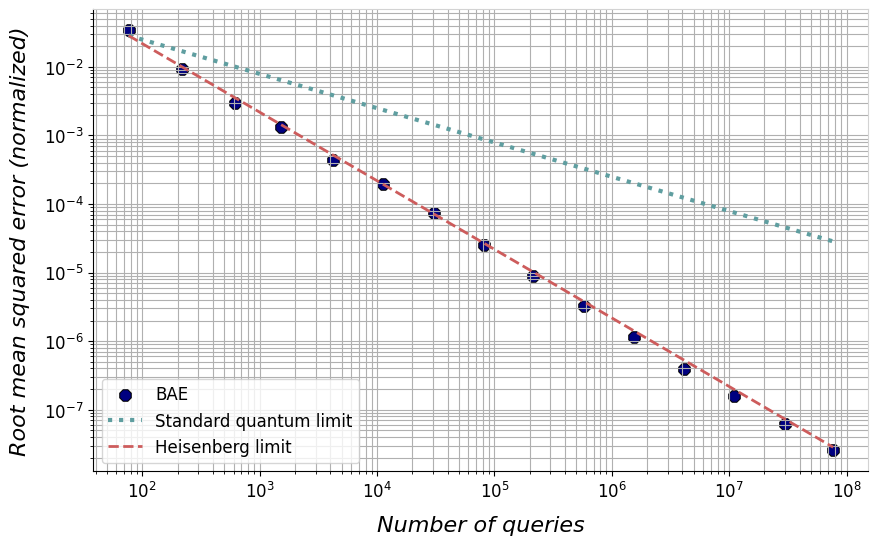}
    \caption{\texttt{thr} = 0.25.}
  \end{subfigure}\hspace{5em}
  \begin{subfigure}[t]{0.38\linewidth}
    \includegraphics[width=\linewidth]{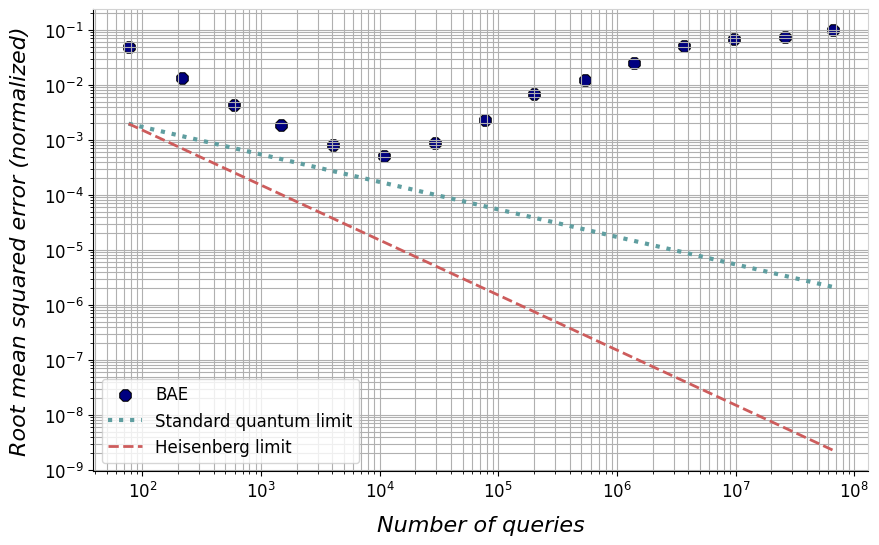}
    \caption{\texttt{thr} = 0.}
  \end{subfigure}
  \caption{Simulation results (RMSE evolution) for BAE with different values of \texttt{thr}. The algorithm works well for \texttt{thr} = 0.25 and even lower values, but a too low value is detrimental to performance, as demonstrated by the disastrous results for \texttt{thr} = 0. This highlights the importance of resampling.}
  \label{fig:thr}
\end{figure*}

\bibliographystyle{plainnat}
\bibliography{QAE}
\end{document}